\documentclass[final,3p,11pt]{elsarticle}
\usepackage{geometry}
\usepackage[table]{xcolor}
\usepackage[utf8]{inputenc}
\usepackage{graphicx}
\usepackage{tikz}
\usepackage{soul}
\usepackage{multirow}
\usepackage{array}
\usepackage{makecell} 
\usepackage{caption}
\usepackage{subcaption}
\usepackage{lineno}
\usepackage{pgfplotstable}
\usepackage{dsfont}
\usepackage{amsmath,amsfonts}
\usepackage{algpseudocode}
\usepackage{upgreek}
\usepackage{algorithm}
\usepackage{amssymb}
\usepackage{amsthm}
\usepackage{mwe}
\usepackage{MnSymbol}
\usepackage{rotating}
\usepackage{pdflscape}
\usepackage{colortbl}
\usepackage[hidelinks]{hyperref}
\usepackage[none]{hyphenat}
\usepackage{float}
\usepackage{colortbl}
\usepackage{adjustbox}
\usepackage{longtable}
\emergencystretch=\maxdimen
\makeatletter
\def\ps@pprintTitle{%
  \let\@oddhead\@empty
  \let\@evenhead\@empty
  \let\@oddfoot\@empty
  \let\@evenfoot\@oddfoot
}
\makeatother

\begin{document}

\begin{frontmatter}

\title{\textbf{Understanding Heterogeneity of Automated Vehicles and Its Traffic-level Impact: A Stochastic Behavioral Perspective}}
\author[add1]{Xinzhi Zhong}
\author[add2]{Yang Zhou} 
\author[add1]{Soyoung Ahn\corref{cor1}}
\author[add3]{Danjue Chen} 

\address[add1]{Civil and Environmental Engineering, University of Wisconsin-Madison}
\address[add2]{Zachry Department of Civil and Environmental Engineering Texas A $\&$ M, College Station\\}
\address[add3]{Civil, Construction, and Environmental Engineering at NC State University}
\cortext[cor1]{Corresponding author: sue.ahn@wisc.edu}

\begin{abstract}
This paper develops a stochastic and unifying framework to examine variability in car-following (CF) dynamics of commercial automated vehicles (AVs) and its direct relation to traffic-level dynamics. The asymmetric behavior (AB) model by \citet{chen2012behavioral} is extended to accommodate a range of CF behaviors by AVs and compare with the baseline of human-driven vehicles (HDVs). The parameters of the extended AB (EAB) model are calibrated using an adaptive sequential Monte Carlo method for Approximate Bayesian Computation (ABC-ASMC) to stochastically capture various uncertainties including model mismatch resulting from unknown AV CF logic. The estimated posterior distributions of the parameters reveal significant differences in CF behavior (1) between AVs and HDVs, and (2) across AV developers, engine modes, and speed ranges, albeit to a lesser degree. The estimated behavioral patterns and simulation experiments further reveal mixed platoon dynamics in terms of traffic throughout reduction and hysteresis.

\end{abstract}
\begin{keyword}
Automated Driving \sep Extended Asymmetric Behaviour \sep Stochastic Calibration \sep Traffic Hysteresis \sep Uncertainty \sep Mixed Traffic. \\
\emph{The code supporting the findings of this study are available in  \href{https://github.com/CeciZhong}{Github page}}. 
\end{keyword}

\end{frontmatter}

\section{Introduction}
\label{S:1}

Vehicles on the road today have various automation features considered SAE Level 2-4 \citep{shladover2015cooperative}. However, no uniform standards for these automation features currently exist. This can give rise to highly heterogeneous mixed traffic, consisting of automated vehicles (AVs) with wide-varying behaviors by design and human-driven vehicles (HDVs), which are highly variable by nature. For instance, adaptive cruise control (ACC) is now a basic automation function that regulates vehicle car-following (CF) maneuvers. The lack of uniform standards has evidently led to varying CF behaviors as observed by recent field studies \citep{gunter2020commercially,he2020energy,makridis2020empirical,li2021car,makridis2021openacc}. The variation of CF manifests itself in the variations of the response time \citep{makridis2020empirical}, string instability \citep{gunter2020commercially,makridis2021openacc} and energy consumption \citep{he2020energy} across different ACC systems.

While the variation of CF partly stems from customizable settings (e.g., headway) \citep{li2021car, makridis2021openacc}, it could stem from a wide range of factors such as ACC algorithms and different elements of vehicle design (e.g., engine). Specifically, ACC algorithms are prolific in academic literature with many design options for the spacing policy and control logic. The spacing policy governs the steady-state CF behavior. Existing policies include constant spacing policy \citep{swaroop1999constant}, constant time gap policy \citep{milanes2013cooperative}, and other, less prevalent policies \citep{besselink2017string}. The control logic influences how ACC responds to traffic disturbances (e.g., a leading vehicle's acceleration and deceleration). Various paradigms of control logic exist in the literature, largely classified into: (1) linear feedback-based controller \citep{ploeg2011design,milanes2013cooperative,ploeg2013controller,zhou2020stabilizing}; (2) constrained optimization-based controller (e.g., model predictive control) \citep{wang2014rolling,wang2018delay,zhou2017rolling,yu2019managing}; (3) data-driven or artificial intelligence (AI)-based controller \citep{qu2020jointly,shi2021connected,jiang2022reinforcement}. The detailed designs of these approaches can also differ significantly, depending on the control objectives, parameters, and penalties/constraints. Besides these differences in formulation, various uncertainties (e.g., in vehicle dynamics) and stochasticity (e.g., stemming from user setting) further add complexity. All these differences present overwhelming challenges to systematically analyze to what extent the CF dynamics can vary and how the variation impacts traffic dynamics. Further, the myopic and adaptive nature of existing controllers, dictating the vehicle action (e.g., acceleration) in the next control time step based on the current system state, makes it challenging to gain insights at the traffic flow level.   

In light of these challenges, the objective of the present paper is two-fold: (1) develop a unifying and stochastic approach to approximate the CF behaviors of different AVs in a more tractable manner and (2) analyze the variation in ACC CF behavior and its traffic-level impact, in comparison to HDVs. 
To this end, we extend the Asymmetric Behavior (AB) model by \citet{chen2012behavioral} to approximate the CF behavior of various types of AVs. The AB model captures asymmetric reaction patterns, describing the evolution of driver response time with respect to the original value, while experiencing a traffic disturbance. The reaction pattern in this model gives direct insight into the disturbance evolution (e.g., amplifying, decaying, and duration). Further, these behavioral patterns are directly linked to important traffic phenomena, including the `capacity drop' phenomenon (i.e., a drop in traffic throughput) \citep{chen2012microscopic}. 

While the AB model was originally developed to describe the CF behavior of HDVs, its flexible structure lends a unifying framework to analyze the CF behavior of different types of AVs \citep{kontar2021multi,srivastava2021modeling}. Notably, \citet{kontar2021multi} adopted the AB model to analyze the physical mechanisms of several well-known ACC algorithms. However, they observed more complex reaction patterns (e.g., concave followed by convex), suggesting the need for more general reaction patterns to effectively capture the behavior of AVs. Further, the control logic of commercially available AVs is mostly unknown due to its proprietary nature. A stochastic approach is thus needed to cope with this uncertainty.

This paper addresses these shortcomings and provides an extended AB model (EAB model hereafter) to enable more flexible descriptions of the behavioral patterns. The model parameters are estimated in a stochastic fashion to capture the inherent randomness in the CF behavior and potential model mismatch. Specifically, we apply Approximate Bayesian Computation (ABC) with an adaptive sequential Monte Carlo sampler (ABC-ASMC) to calibrate the model parameters using field data of several ACC vehicles. This method is likelihood function free and relies on simulation instead to estimate a posterior joint distribution of parameters \citep{del2012adaptive, sisson2018handbook,Zhou}. Thus, it provides a flexible platform for stochastic calibration when information about the parameter distribution is limited. The estimated behavioral patterns, confirmed by simulation experiments, elucidate platoon dynamics, particularly throughput reduction and traffic hysteresis. 

Our analysis based on the proposed stochastic and unifying framework has led to the following findings. (1) The estimated behavioral patterns and the posterior distributions of the parameters reveal marked differences in CF dynamics between AVs and HDVs, and across AV developers, albeit to a lesser degree. (2) Even for the same ACC maker, CF dynamics vary by speed range and engine modes. (3) With increasing penetration of ACC vehicles, the mixed platoon exhibits more complete hysteresis loops compared to HDVs, indicating lower throughput reduction. (4) The mixed platoon also exhibits smaller traffic hysteresis in the median-high speed range with higher penetration of ACC vehicles; however, the trend is opposite in low speed. While these specific findings could change as the technology develops, the proposed analysis framework is general and can lead to new insights as more data become available in the future.

The rest of the paper is organized as follows. Section 2 presents the EAB CF model and its direct relation to traffic-level dynamics, along with the proposed stochastic calibration method based on ABC-ASMC. Section 3 provides the calibration results and discussion on the parameter variability that represents heterogeneity in CF behavior of commercial ACC vehicles and HDVs. Section 4 establishes a method to systematically measure traffic hysteresis and presents the quantified results of how ACC CF behavior impacts throughput and hysteresis. Section 5 further investigates the impact of heterogeneity and stochasticity on disturbance propagation in mixed traffic. Section 6 provides some discussion and concluding remarks.

\section{Extended AB Model and Stochastic Calibration Method}
\label{S:2}

In this section, we first construct the EAB model to describe ACC behavioral patterns and their direct relation to the traffic-level dynamics. Then, we provide the stochastic calibration method by ABC-ASMC to calibrate the model parameters.

\subsection{EAB Model}
Here we extend the AB model \citep{chen2012behavioral} to describe a wide range of potential CF behaviors under disturbances. We start our discussion with Newell's simplified CF model \citep{newell2002simplified} and then extend to the EAB model.

Newell’s simplified CF model \citep{newell2002simplified} states that follower $i$’s position at time $t$, $x_i(t)$, can be determined by a constant temporal-spatial shift ($\tau_i$ and $\delta_i$) of leader $i-1$’s position. The $\tau_i$ and $\delta_i$ are the response time and the minimum spacing for driver $i$, respectively, and represent the travel time and distance of a traffic disturbance. Note that the ratio of the average minimum spacing, $\delta$, and average response time, $\tau$, across vehicles corresponds to the congestion wave speed, $w=-\frac{\delta}{\tau}$, in the Kinematic Wave theory (KWT) with a triangular fundamental diagram.

\begin{flalign}
    &x_i(t+\tau)=x_{i-1}(t)-\delta 
\end{flalign}

\citet{laval2010mechanism} observed that the follower would deviate from the Newell equilibrium around a disturbance. They formulated a time-dependent term, $\eta_i(t)=\frac{\tau_i(t)}{\tau}=\frac{\delta_i(t)}{\delta}$, to describe the dynamic deviations from the Newell trajectory, as shown in Eq. (2). \citet{chen2012behavioral} empirically verified the L-L model and further extended it to incorporate driver heterogeneity and asymmetric $\eta_i(t)$ evolution. They suggest five patterns for $\eta_i(t)$: concave, convex, nearly equilibrium, non-decreasing, and non-increasing when analyzing HDV CF.
 
\begin{flalign}
    &x_i(t+\eta_i(t)\tau)=x_{i-1}(t)-\eta_i(t)\delta 
\end{flalign}
 
The above-mentioned studies pertain to HDV traffic. For ACC controllers, \citet{kontar2021multi} observed composite patterns, concave followed by convex (concave-convex) and convex-concave patterns, which are beyond the AB model's scope. Hence, to make the AB model more flexible, we extend the AB model using a parsimonious generalized piece-wise function as shown in Eq. (3). The parameter vector of the EAB model is $\Vec{\theta}=[\eta^0, \eta^1, \eta^2, \eta^3, \epsilon^0, \epsilon^1, \epsilon^2,t^1]^\top$. The differences between $\eta^0$ and $\eta^1$,  $\eta^1$ and $\eta^2$,  $\eta^2$ and $\eta^3$ are denoted as $\Delta\eta^{0-1}$, $\Delta\eta^{1-2}$, and $\Delta\eta^{2-3}$, respectively. Based on the signs of $\Delta\eta^{0-1}$, $\Delta\eta^{1-2}$, and $\Delta\eta^{2-3}$, the EAB describes seven behavioral patterns, as summarized in Table \ref{eta_pattern}.

 \begin{flalign}
    &\eta_i(t)=
  \begin{cases}
      \eta_i^0 & 0<t\leq t_i^1 \\
      \eta_i^0+\epsilon_i^0(t-t_i^1) & t_i^1< t\leq t_i^2 \\
      \eta_i^1+\epsilon_i^1(t-t_i^2) & t_i^2< t\leq t_i^3 \\
      \eta_i^2+\epsilon_i^2(t-t_i^3) & t_i^3< t\leq t_i^4 \\
      \eta_i^3 & t_i^4< t
      \end{cases}
  \end{flalign}\\
  where $ \eta_i^1=\eta_i^0+\epsilon_i^0(t_2-t_1), \eta_i^2=\eta_i^1+\epsilon_i^0(t_3-t_2), \eta_i^3=\eta_i^2+\epsilon_i^0(t_4-t_3)$;
  $\eta _i^0$ is the original equilibrium state, which is a constant value before the disturbance;
  $\eta_i^1$ and $\eta_i^2$ are the critical $\eta_i(t)$ values where the follower reaches the maximum and  minimum deviations (or minimum and maximum) from the equilibrium state;
 $\eta_i^3$ is a new equilibrium state, which is a constant value after the disturbance. $\eta_i^3$ is not necessarily equal to $\eta_i^0$ due to asymmetric CF behavior;
 $t_i^1$ is the time point when the follower starts to deviate from the original equilibrium state;
 $\epsilon_i^0$, $\epsilon_i^1$ and $\epsilon_i^2$ are the average slopes of $\eta_i(t)$  between  $\eta_i^0$ and  $\eta_i^1$,  $\eta_i^1$ and  $\eta_i^2$,  and $\eta_i^2$ and  $\eta_i^3$, respectively. When $\exists j: \epsilon_i^j \rightarrow 0, j \in \{0,1,2\}$, EAB model will converge to AB model.

\begin{figure*}[h]
 \captionsetup{justification=centering} 
        \centering
        \begin{subfigure}[b]{0.49\textwidth}
            \centering
            \includegraphics[width=\textwidth]{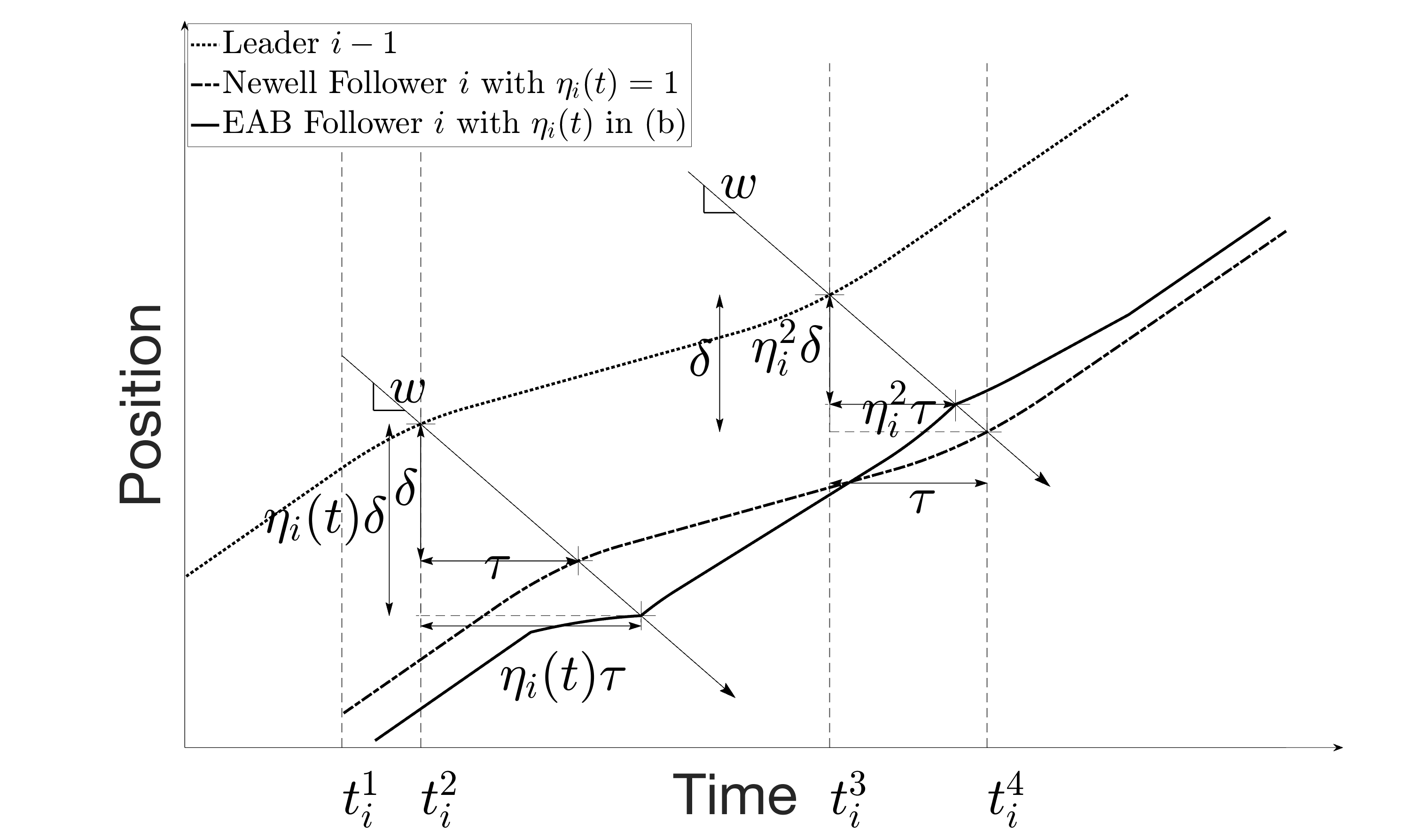}
            \caption[]%
            {{\small}}    
            \label{}
        \end{subfigure}
        \hfill
        \begin{subfigure}[b]{0.49\textwidth}  
            \centering 
            \includegraphics[width=\textwidth]{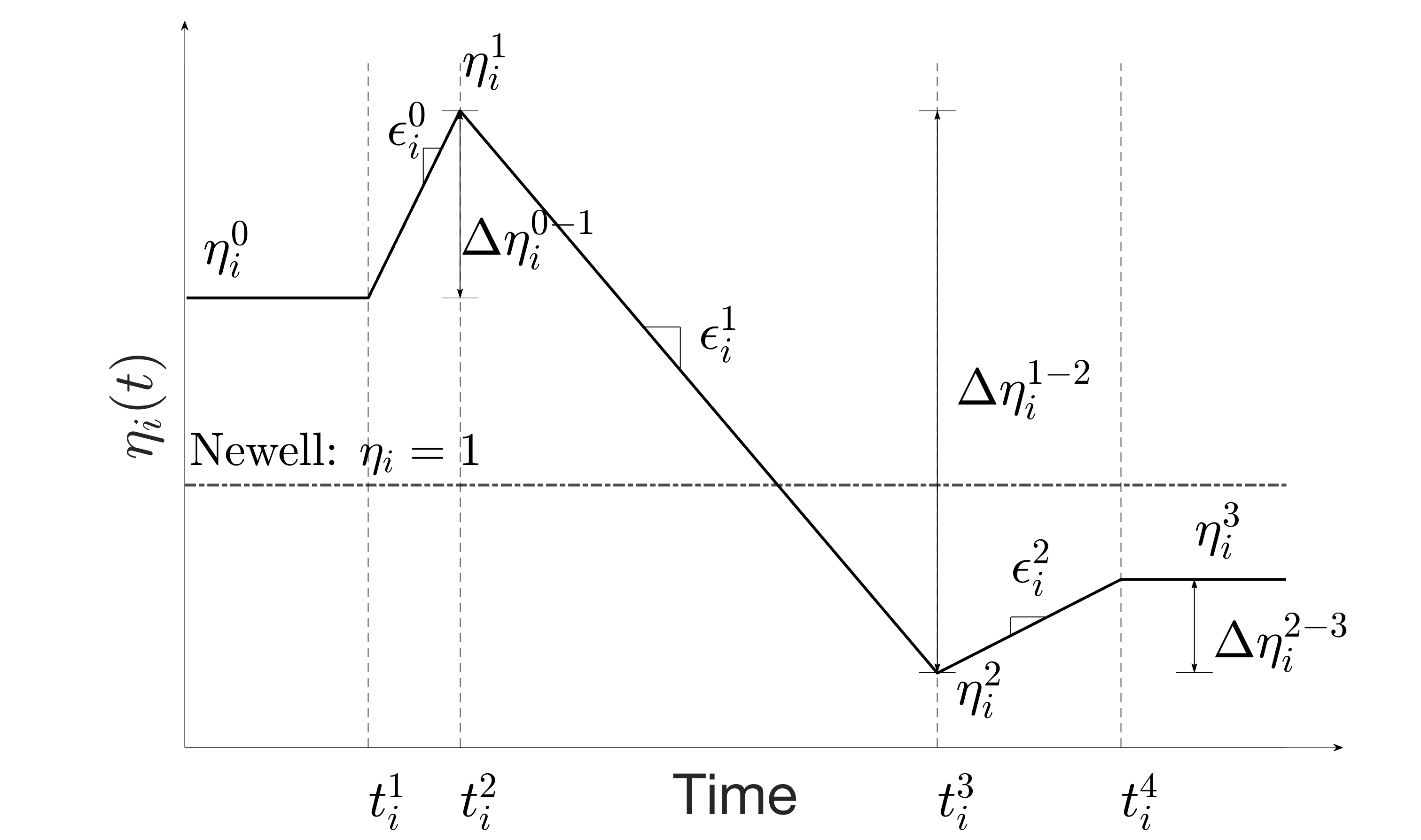}
            \caption[]%
            {{\small }}    
            \label{}
        \end{subfigure}
       \caption[{Schematic Illustration of EAB Model}]
        {\small Schematic Illustration of Extended AB Model\\(a) Trajectories of the leader, Newell follower and EAB follower with the $\eta_i(t)$ in (b)\\ 
        (b) Concave-convex $\eta_i(t)$ pattern} 
        \label{EAB}
    \end{figure*}
    
Fig. \ref{EAB} is a schematic illustration of the EAB model with a concave-convex $\eta_i(t)$ pattern as an example when a disturbance propagates upstream at the speed of $w$. In this example, $\eta_i^0 > 1$ as shown in Fig. \ref{EAB}(b), which represents conservative behavior. In Fig. \ref{EAB}(b), the $\eta_i(t)$ increases at the rate of $\epsilon_i^0$ from $t_i^1$ to $t_i^2$ and starts to decrease at the rate of $\epsilon_i^1$ when the follower reaches its maximum spacing. $\eta_i(t)$ continues to decrease till the follower reaches the minimum spacing at $t_i^3$. When the follower realizes the leader is accelerating, the follower catches up to its equilibrium trajectory, and $\eta_i(t)$ is restored to an equilibrium at $t_i^4$. In this example, $\eta_i^3 < \eta_i^0$, indicating a shorter equilibrium spacing after the disturbance. Consistent with $\eta_i(t)$, the disturbance initially amplifies and then partially decays.

Based on empirical observations, \cite{chen2012microscopic} identified a direct connection between the reaction pattern and the hysteresis in the leader's velocity, $v_{i-1}$ - $\eta_i$ evolution. The reaction pattern could capture three predominant hysteresis patterns: straight line (SL), Counter-clockwise (CCW), Clockwise (CW)). They also found that the response time (i.e., early or late response) impact the hysteresis orientation.

We extend this microscopic relation into more macroscopic hysteresis in the density-flow evolution, as shown in Table \ref{eta_pattern}. We first define different response timings: (1) \emph{early response} if the follower starts to restore to the new equilibrium during the deceleration phase of the leader, and (2) \emph{late response} after the acceleration phase of the leader. The composite pattern, concave-convex, is typically considered as the early-response concave pattern followed by the late-response convex pattern. Likewise, the convex-concave pattern is considered as the early-response convex and then the late-response concave pattern. Note that for other combinations (e.g., early-response concave followed by early-response convex, etc.) are empirically rare. In the middle column, we use the blue line to represent the initial equilibrium state and the red line to represent the new equilibrium state after disturbance. The seven hysteresis patterns are defined: (1) SL: flow and density change along the slope of the wave speed. (2) $CW^{-}$, $CCW^{-}$: the flow and density drop below the initial equilibrium and remain below the new equilibrium after disturbance in clockwise and counter-clockwise orientation, respectively. (3) $CW^{+}$, $CCW^{+}$ : the flow and density move above the line of the initial equilibrium and continue to remain above the line of the new equilibrium. (4) $CW$: flow and density drop below the initial equilibrium and then rise above the new equilibrium. (5) $CCW$: flow and density will rise above the initial equilibrium and then drop below the new equilibrium. As indicated by the respective reaction patterns and hysteresis patterns, the evolution of the disturbance can be identified in the rightmost column of the Table \ref{eta_pattern}.  Readers could refer to \citet{xinzhi}  (preparing) for a detailed proof.

\begin{table}
\small
\caption{$\eta_i(t)$ Pattern Categorization and its direct relation to traffic-level dynamics}\label{tab:versions}
	\begin{center}
		\begin{tabular}{c |c |c |c}\Xhline{1pt}
Category & Response & \makecell[c]{Hysteresis\\Orientation }& Disturbance Evolution\\\Xhline{1pt}
	\makecell[c]{Nearly Equilibrium(NE)\\ \includegraphics[width=3.5cm,height=1.5cm]{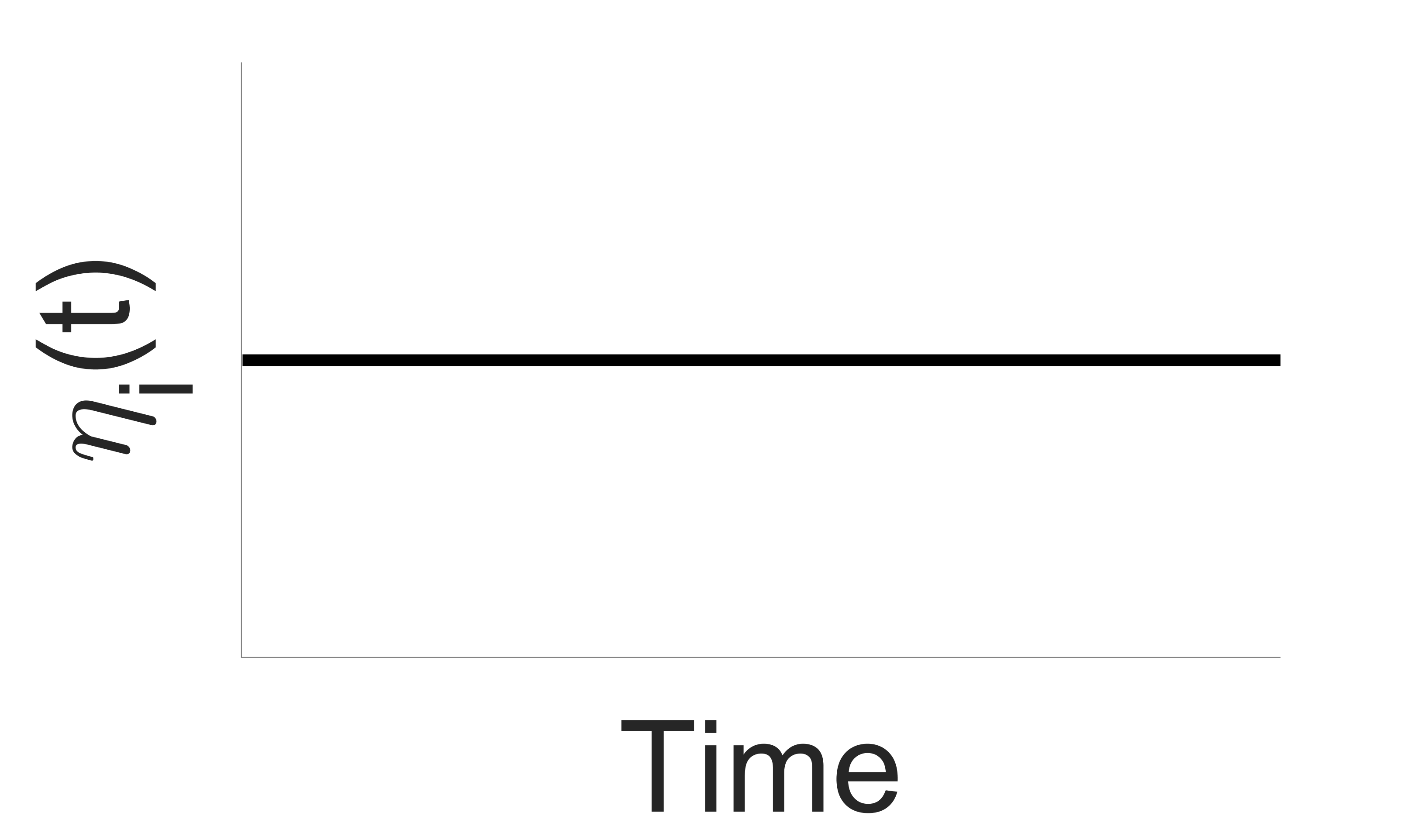}}& / &\makecell[c]{SL\\ \includegraphics[width=3.5cm,height=1.5cm]{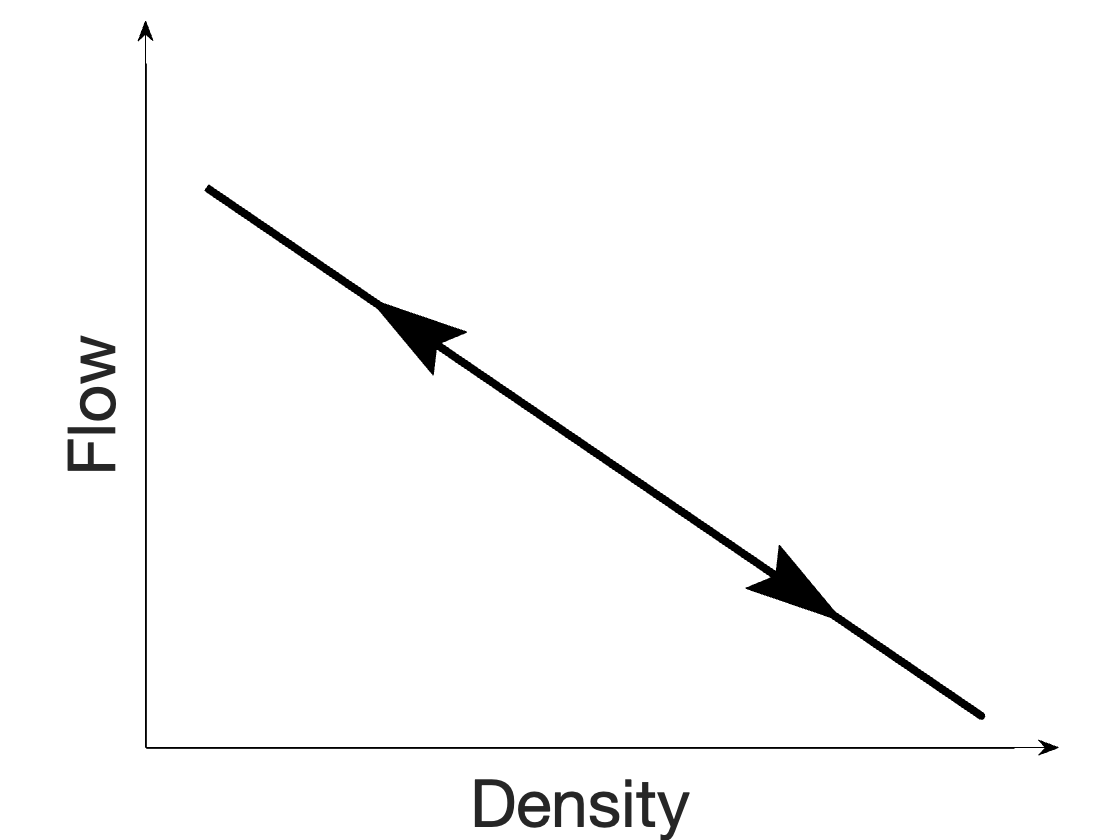}}&Disturbance will not amplify or decay.\\\hline
\multirow{2}*{\makecell[c]{Concave\\\includegraphics[width=3.5cm,height=1.5cm]{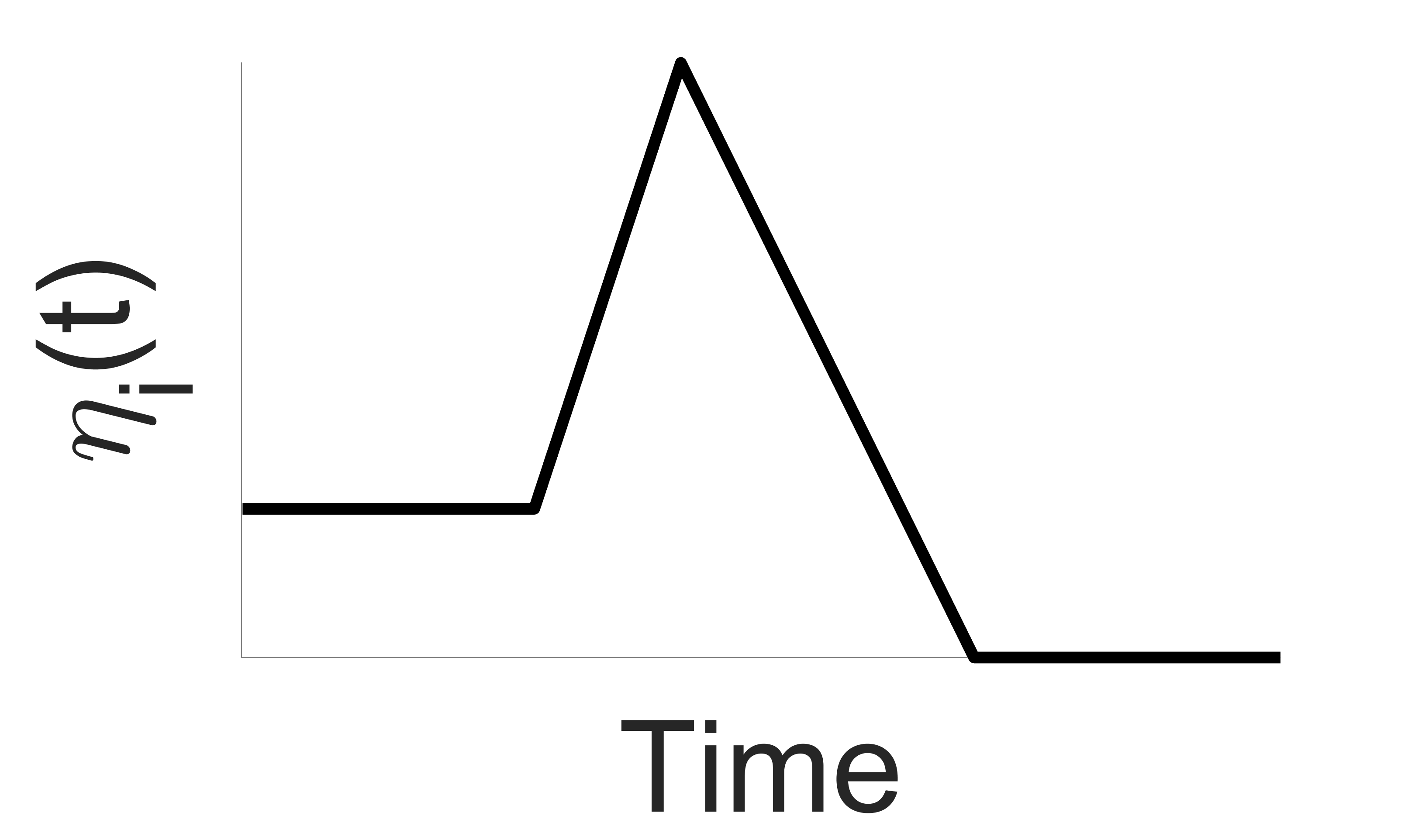}}}&early & \makecell[c]{CCW$^{-}$ \\\includegraphics[width=3.5cm,height=1.5cm]{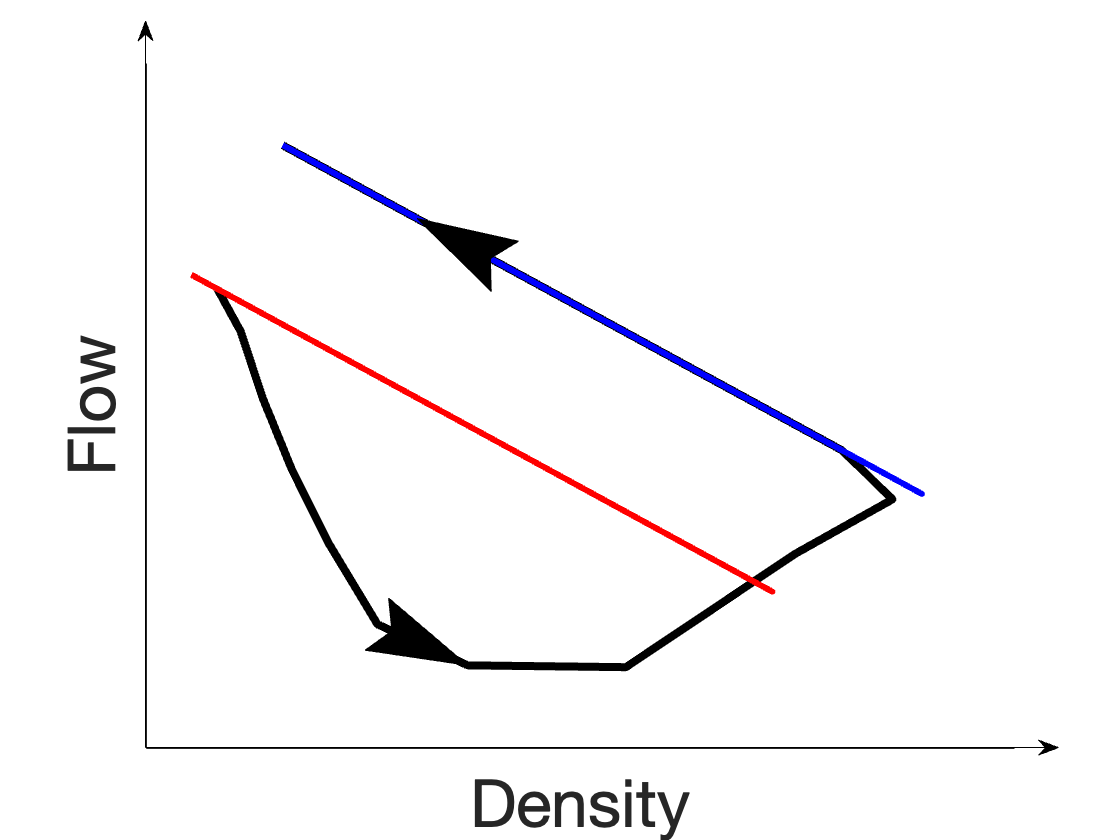}} & \multirow{2}*{ \makecell[c]{\\\\\\Disturbance will amplify.\\\\}}\\\cline{2-3}
&late&  \makecell[c]{CW$^{-}$ \\\includegraphics[width=3.5cm,height=1.5cm]{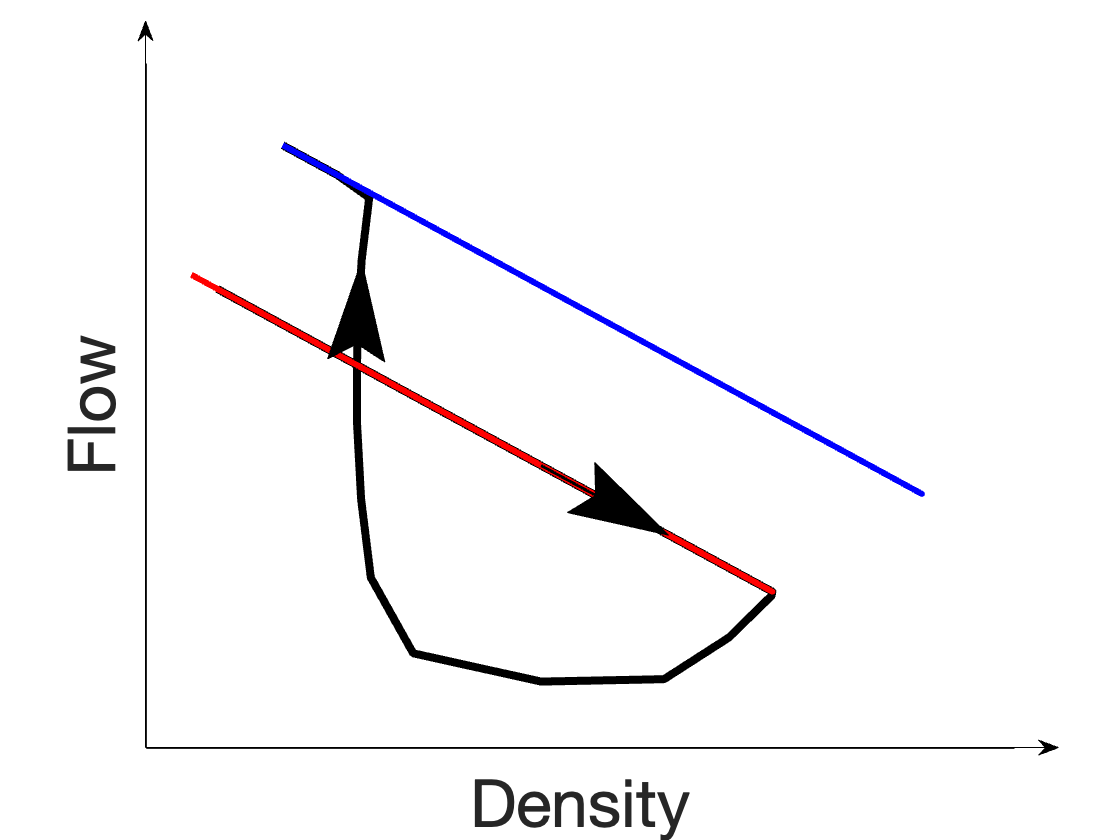}} &\\\hline
\multirow{2}*{\makecell[c]{Convex\\\includegraphics[width=3.5cm,height=1.5cm]{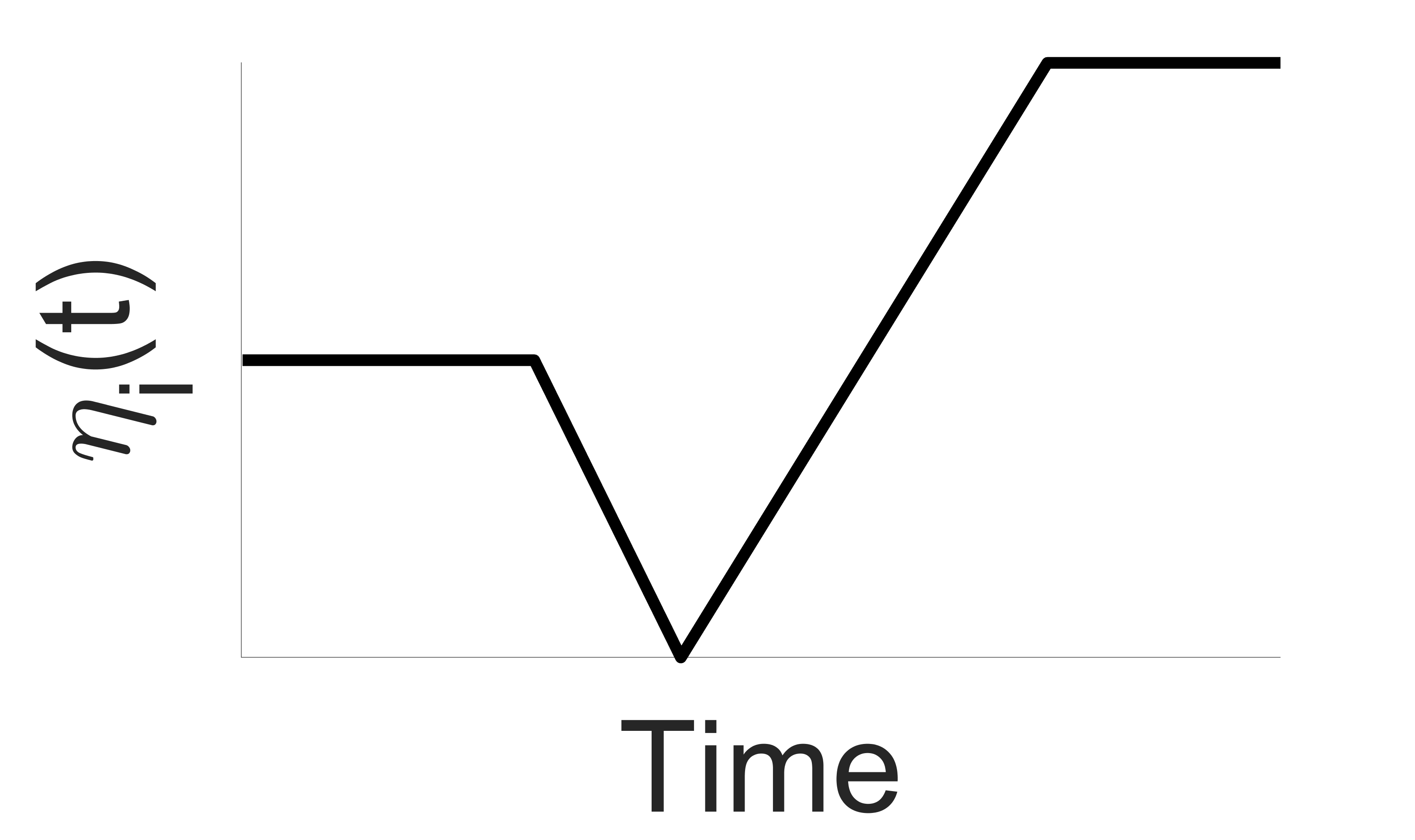}}}&early & \makecell[c]{CW$^{+}$ \\\includegraphics[width=3.5cm,height=1.5cm]{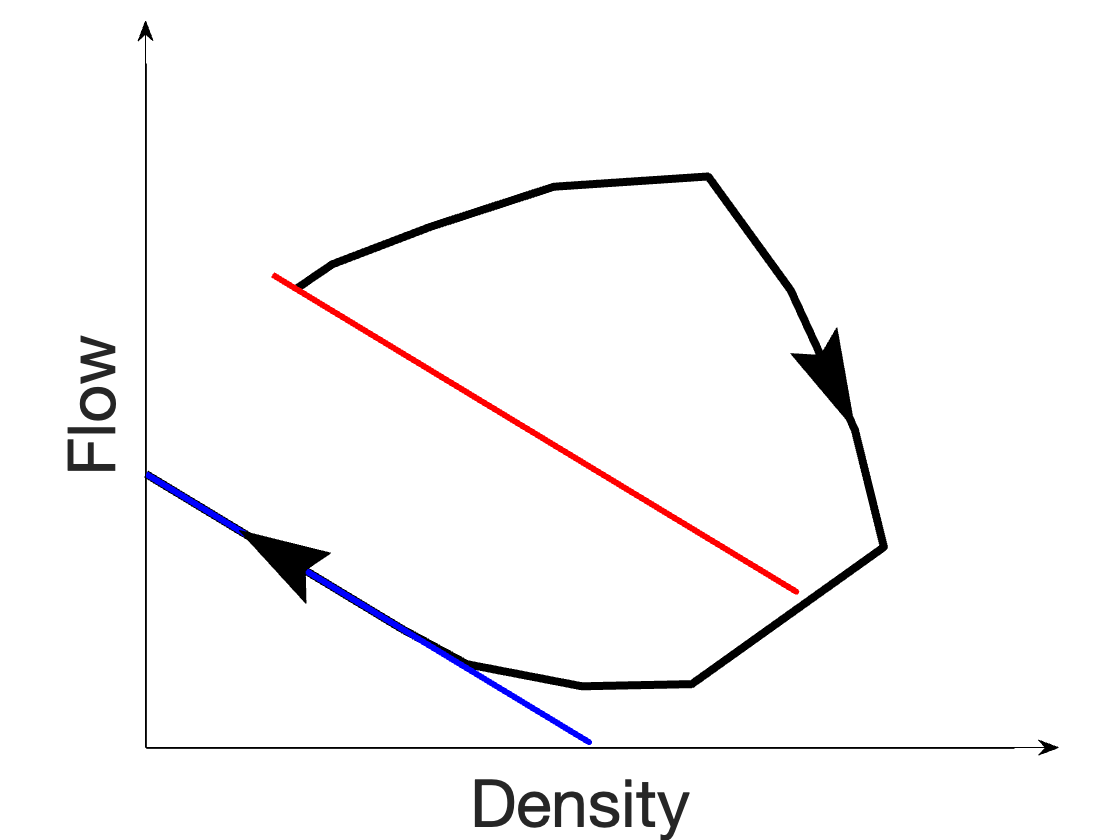}}&\multirow{2}*{\makecell[c]{\\\\\\Disturbance will decay.\\\\}}\\\cline{2-3}
& late &\makecell[c]{CCW$^{+}$ \\\includegraphics[width=3.5cm,height=1.5cm]{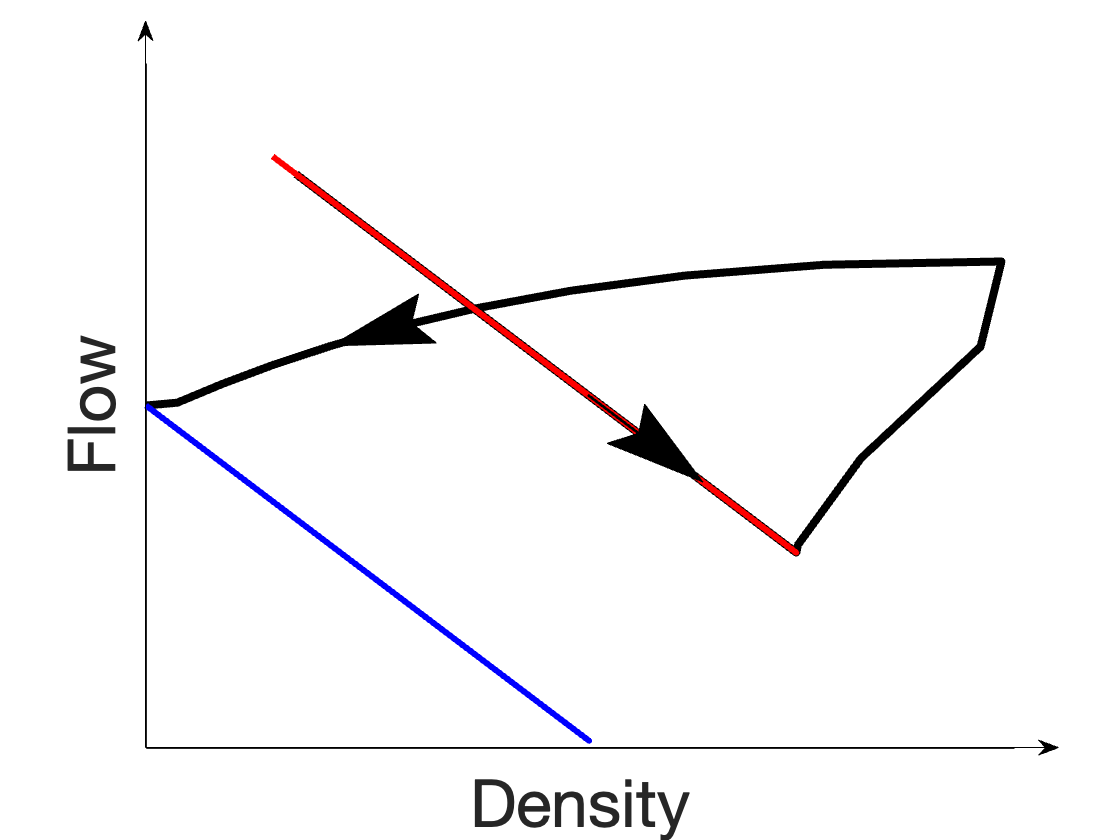}}& \\\hline
 \makecell[c]{Concave and Convex\\\includegraphics[width=3.5cm,height=1.5cm]{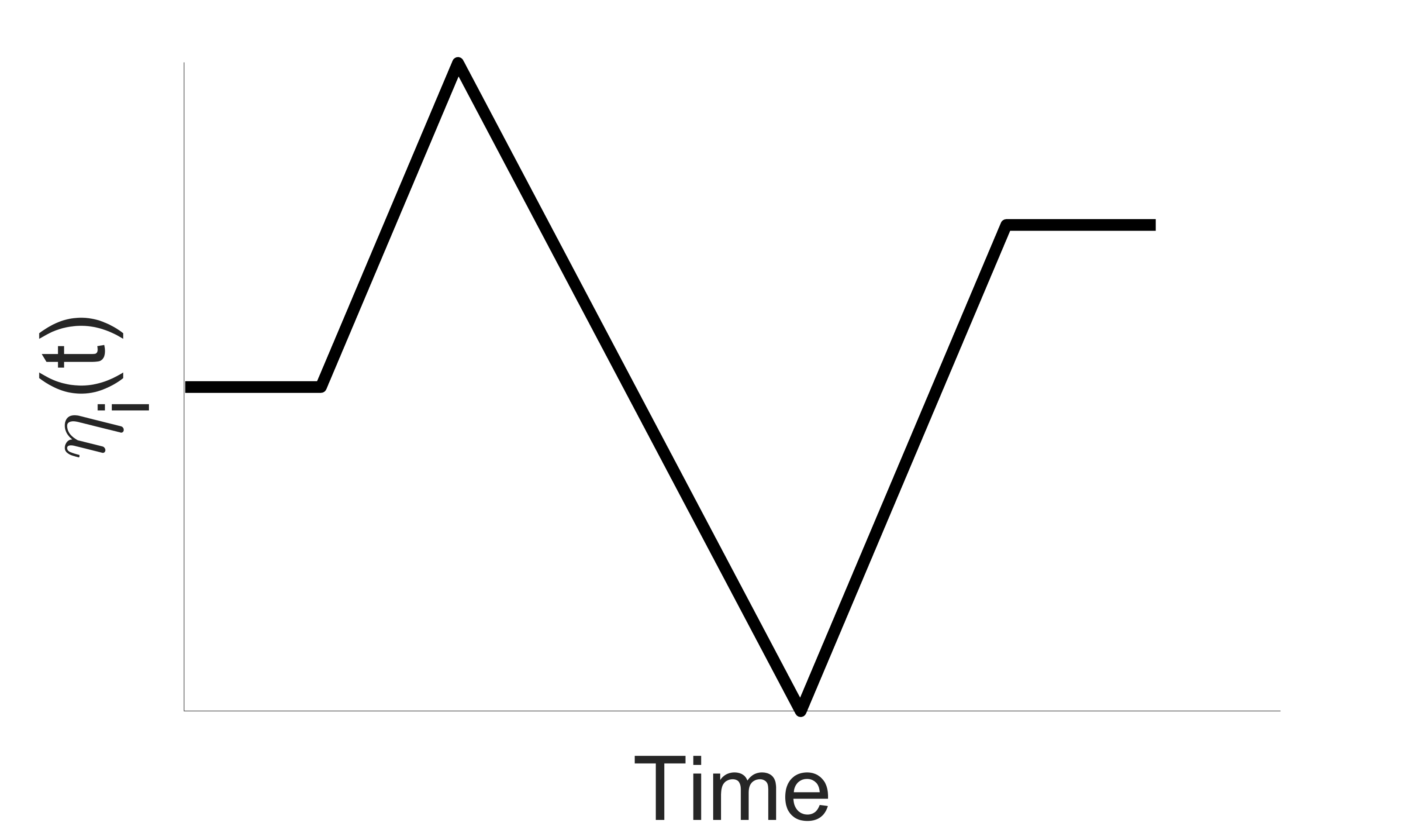}} & / &\makecell[c]{CCW \\\includegraphics[width=3.5cm,height=1.5cm]{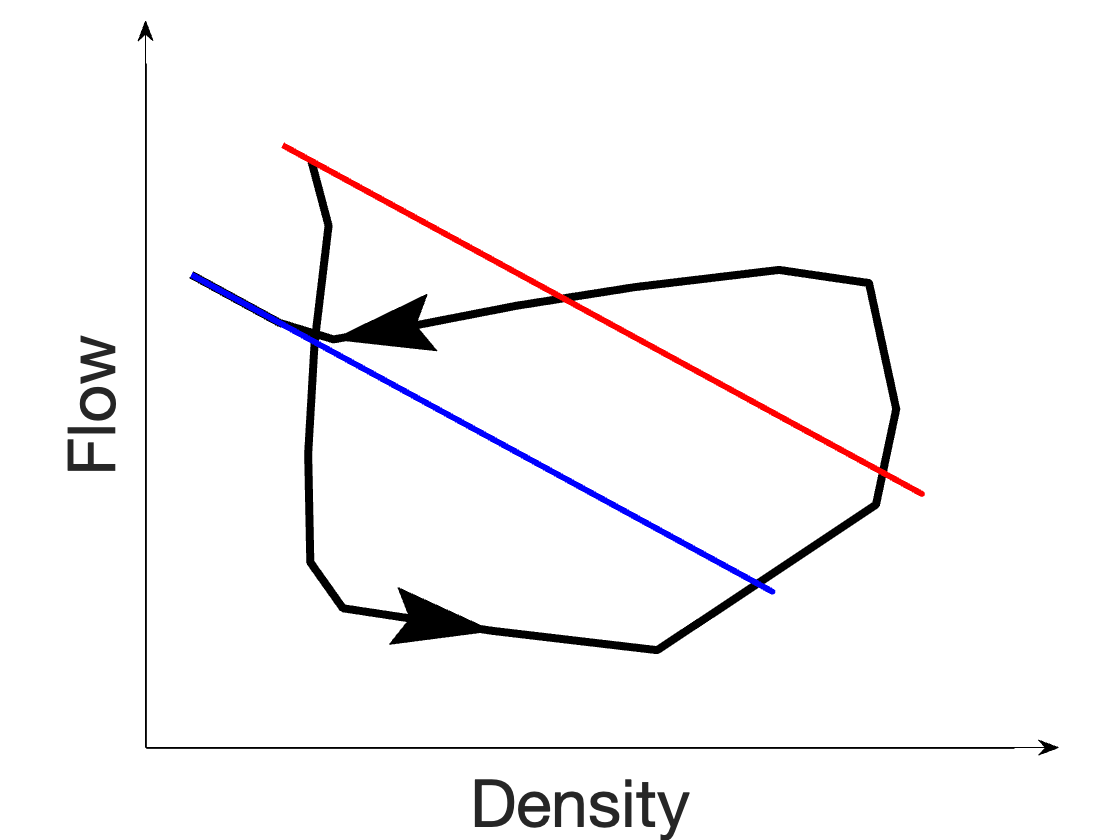}} & \makecell[c]{Disturbance will amplify and then partially decay.}\\\hline
\makecell[c]{Convex and Concave\\\includegraphics[width=3.5cm,height=1.5cm]{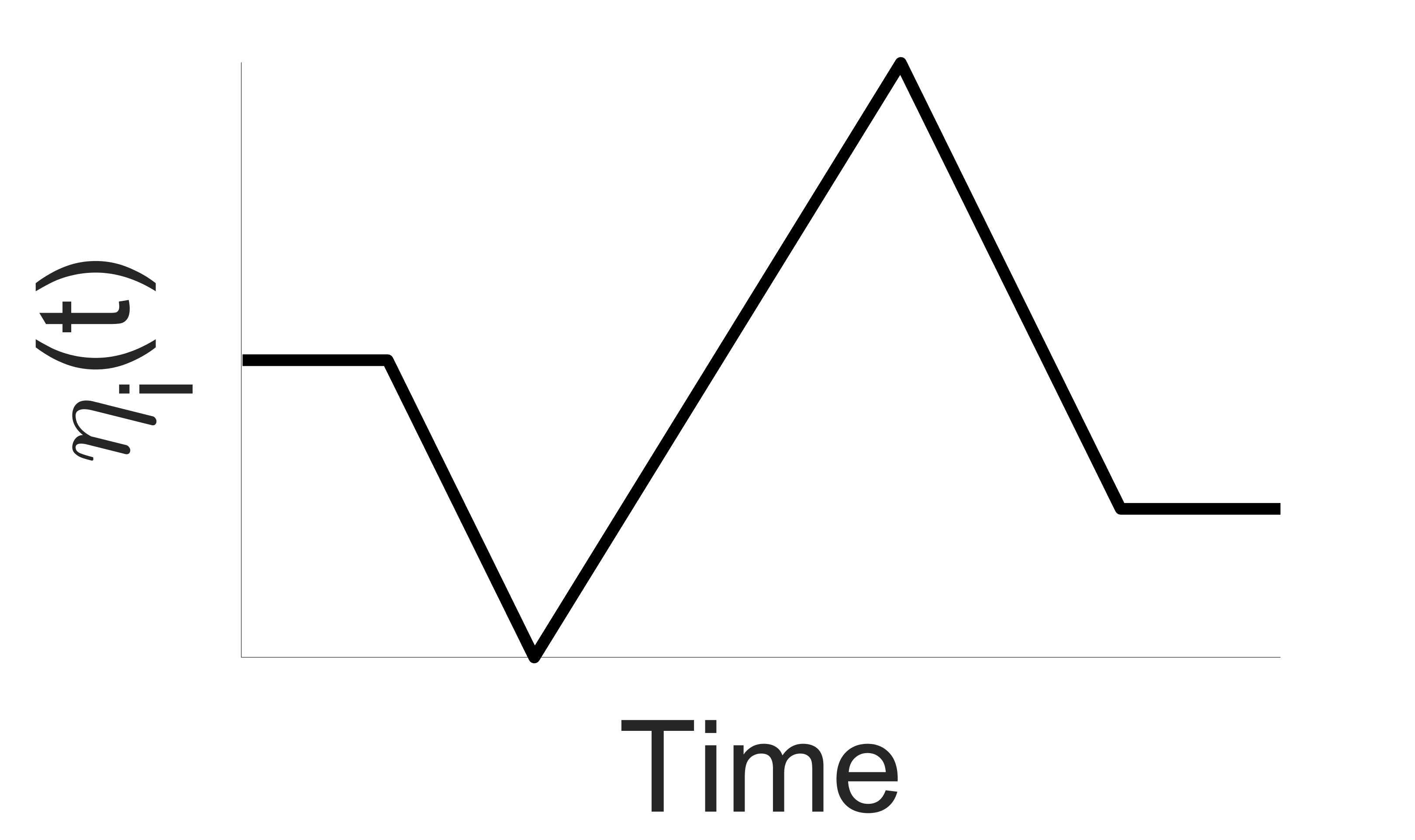}}& / &\makecell[c]{CW \\\includegraphics[width=3.5cm,height=1.5cm]{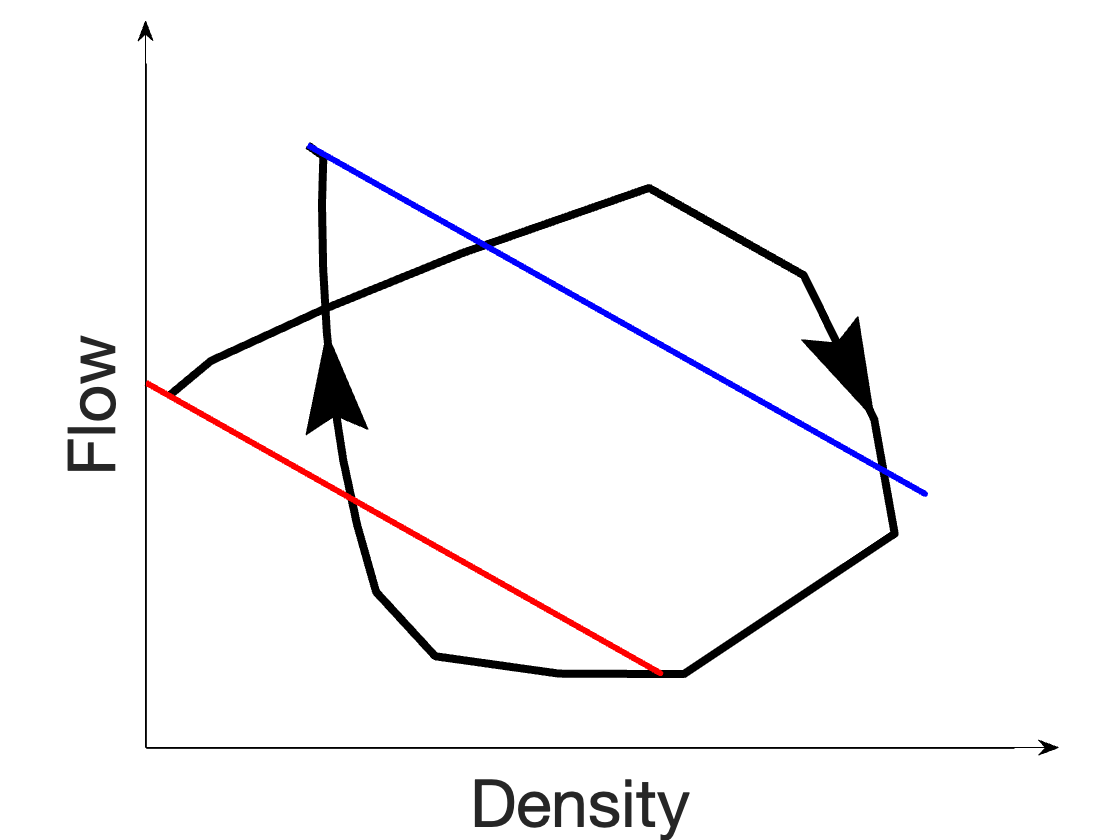}} & \makecell[c]{Disturbance will firstly decay and then amplify.}\\\hline
\multirow{2}*{\makecell[c]{Non-decreasing\\\includegraphics[width=3.5cm,height=1.5cm]{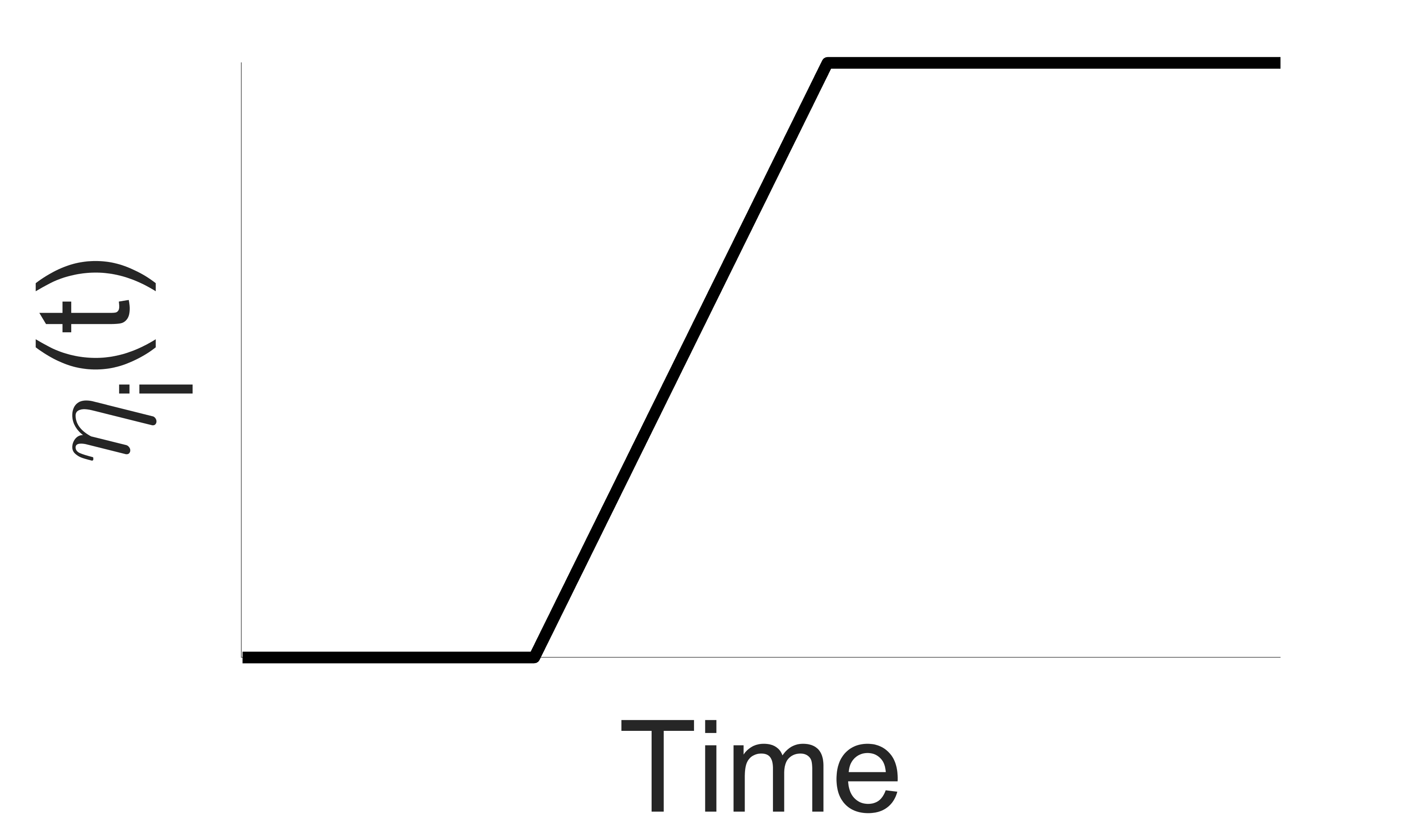}}}& \makecell[c]{\\early\\\\} & CCW$^-$ & \multirow{2}*{\makecell[c]{\\Disturbance will decay, \\likely resulting in an decreased capacity.\\}}\\\cline{2-3}
& \makecell[c]{\\late\\\\} & CW$^-$ & \\\hline
\multirow{2}*{\makecell[c]{Non-increasing\\\includegraphics[width=3.5cm,height=1.5cm]{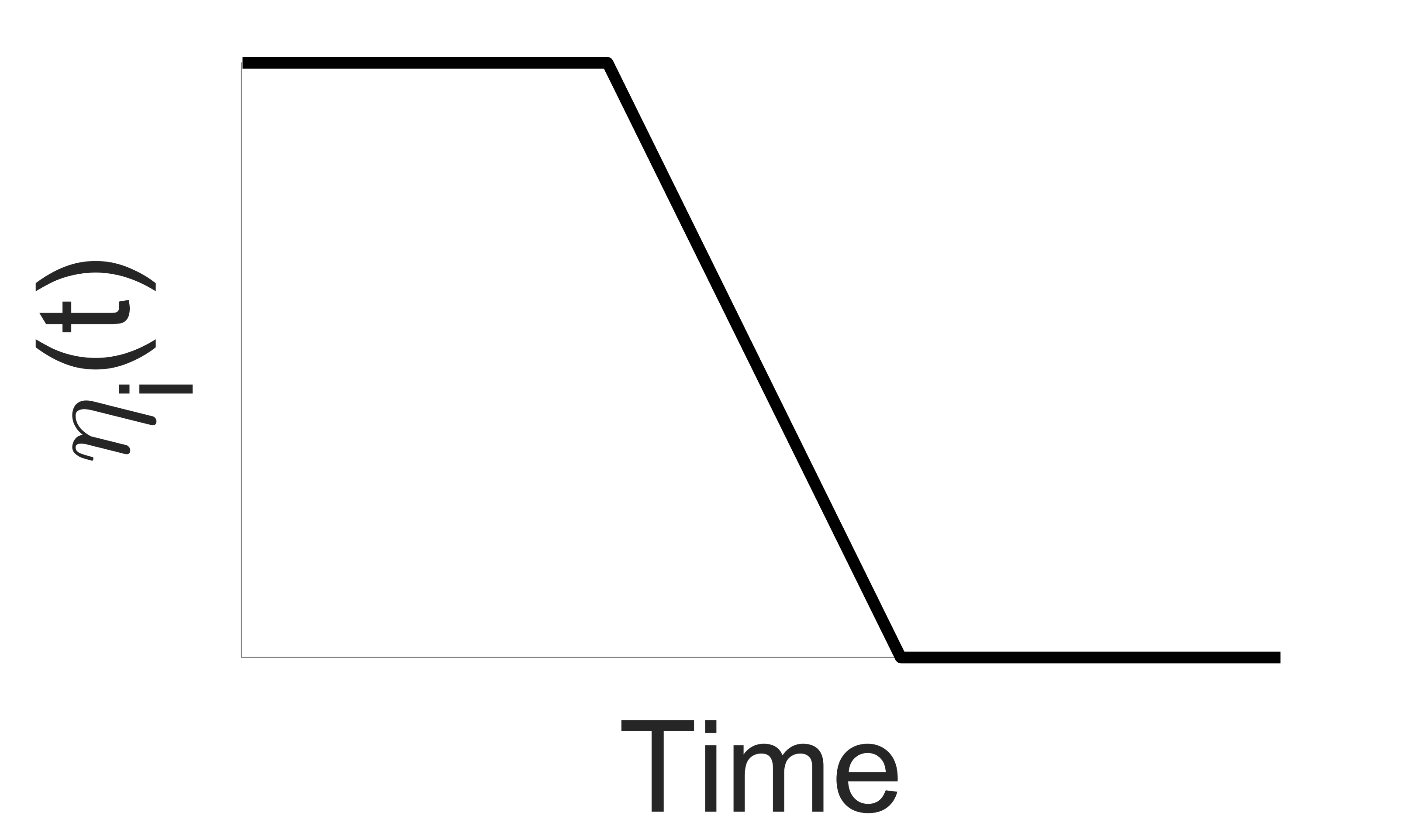}}}&\makecell[c]{\\early\\\\}& CW$^+$ & \multirow{2}*{\makecell[c]{\\Disturbance will decay, \\likely resulting in an increased capacity.\\\\}}\\\cline{2-3}
& \makecell[c]{\\late\\\\}& CCW$^+$  &
	 \\\Xhline{1pt}
		\end{tabular}
	\end{center}
 \small
 \label{eta_pattern}
\end{table}

Note that we work with the proposed EAB model to approximate the CF behaviors of various ACC vehicles. Considering potential model mismatch and stochasticities of ACCs, we take the stochastic calibration approach, ABC-ASMC, to estimate the joint distribution of the model parameters, $\tilde{\pi}(\Vec{\theta})$. Details follow. 

\subsection{EAB model calibration: ABC-ASMC}
In this subsection, we apply ABC-ASMC \citep{del2012adaptive} to approximate the joint distribution of parameters, $\tilde{\pi}(\Vec{\theta})$, building on the basic ABC-based method developed by \citet{Zhou}. The basic ABC method (ABC-rejection sampling) is a likelihood-free Bayesian inference, where the likelihood is replaced by the simulations of parameter values (called "particles") sampled from prior distributions. A goodness of fit measure (GOF) is applied to evaluate the closeness between the observed and simulated data (e.g., vehicle trajectories). The simulated data close to the observed would be accepted based on a reasonable tolerance level $\gamma$ of GOF and used to approximate the posterior distributions. Thus, $\gamma$ is the main decisive variable shaping appropriate posterior distributions. Readers could refer to \citet{Zhou} and \citet{csillery2010approximate} for a detailed review of the ABC-rejection sampling method. This earlier method has a simple structure of independently sampling particles from prior distributions until enough particles are accepted for convergence to estimate the posterior joint distribution. This independent structure makes it easy to implement but can bring very high computational burden due to the naive search process (i.e., trial and error). The proposed ABC-ASMC has a more strategic structure for sampling particles to reach quicker convergence by searching $\gamma$ in an automatic fashion, meaning a significant advantage for computational efficiency. Details of the method follow.

Given the observed CF pair, leading vehicle $i-1$ trajectory (i.e., position and velocity), $x_{i-1}, v_{i-1}$, and following vehicle $i$ trajectory (i.e., position and reaction pattern), $x_{i, obs}, \eta_{i, obs}$, the ABC-ASMC aims to approximate the posterior distribution, $\tilde{\pi}(\Vec{\theta})$, by Bayes' Theorem:
 
\begin{flalign} &\pi(\Vec{\theta}|x_{i, obs}, \eta_{i, obs},x_{i-1}, v_{i-1})=\frac{f(x_{i, obs}, \eta_{i, obs}|\Vec{\theta}, x_{i-1}, v_{i-1})\pi(\Vec{\theta})}{\int{f(x_{i, obs}, \eta_{i, obs}|\Vec{\theta}},x_{i-1}, v_{i-1})d\Vec{\theta}}
\end{flalign}
where $\pi(\Vec{\theta})$ is the prior distribution of $\Vec{\theta}$ usually set by prior knowledge and $f(x_{i, obs}, \eta_{i, obs}|\Vec{\theta},x_{i-1}, v_{i-1})$ is the likelihood to reproduce  $x_{i, obs}, \eta_{i, obs}$ given the CF (control) model, $g(\Vec{\theta},x_{i-1}, v_{i-1})$ with parameter $\Vec{\theta}$, and the leading trajectory, $x_{i-1}, v_{i-1}$ (e.g., EAB model), where  $\Vec{\theta}=[\theta_1,...\theta_n,...\theta_N]$ and $N$ is the total number of the parameters in CF (control) model (e.g., $N=8$ in EAB model). 

Since $f(x_{i, obs}, \eta_{i, obs}|\Vec{\theta},x_{i-1}, v_{i-1})$ is computationally intractable to obtain, an approximation algorithm is desired. ABC-ASMC is an effective tool to address this challenging inferential problem by replacing the likelihood function with simulations, adaptively decreasing $\gamma$ till convergence \citep{csillery2010approximate,marin2012approximate,sisson2018handbook}. 

The ABC-ASMC mainly consists of four steps: initialization, sampling, updating, and stopping criteria checking. Details are given below:\\
\hfill \break
(1) Step 1 (Initialization): For iteration $l=0$, set $\gamma_l=\gamma_0$, where $\gamma_0$ is relatively a large number. Set the sampling weight $W_l^k=\frac{1}{K}$, $k=1\dots K$, where $K$ is the total number of particles, and the prior distribution at stage $l$, $\pi_l(\Vec{\theta})=\pi(\Vec{\theta})$.\\ \hfill \break
(2) Step 2 (Sampling): Sample $K$ particles ($\Vec{\theta}$) from $\pi_l(\Vec{\theta})$ according to $W_l^k$ to get a posterior distribution,  $\hat{\pi}_l(\Vec{\theta}|x_{i, obs}, \eta_{i, obs},x_{i-1}, v_{i-1})$ satisfying  $GOF(x_{i, sim}, \eta_{i, sim},x_{i, obs}, \eta_{i, obs})<\gamma_l$, where $x_{i, sim}, \eta_{i, sim}=g(\Vec{\theta},x_{i-1}, v_{i-1})$ under $\gamma_l$.\\ 
\hfill \break
(3) Step 3 (Updating): Set $\gamma_{l+1}$ as the $\lambda$ percentile of largest GOF. We partitioned the  $\hat{\pi}_l(\Vec{\theta}|x_{i, obs}, \eta_{i, obs},x_{i-1}, v_{i-1})$ into two subsets, an alive particle set, $\hat{\pi}_{l,A}(\Vec{\theta}|x_{i, obs}, \eta_{i, obs},x_{i-1}, v_{i-1})$, and a perturbed particle set, $\hat{\pi}_{l,P}(\Vec{\theta}|x_{i, obs}, \eta_{i, obs},x_{i-1}, v_{i-1})$. The alive particle set only selects $\Vec{\theta}$ from $\hat{\pi}_l(\Vec{\theta}|x_{i, obs}, \eta_{i, obs},x_{i-1}, v_{i-1})$ satisfying $GOF(x_{i, sim}, \eta_{i, sim},x_{i, obs}, \eta_{i, obs})<\gamma_{l+1}$, to ensure that more fitted particles are kept during the iteration. The perturbed particle set consists of $(1-\lambda)K$ particles satisfying $GOF(x_{i, sim}, \eta_{i, sim},x_{i, obs}, \eta_{i, obs})<\gamma_{l+1}$, based on a component-wise independent normal zero-mean random walk according to the kernel function $\chi_l$ proposed by \cite{beaumont2009adaptive}, to further explore sampling space:

\begin{flalign}
\chi_l(\theta_n^{l,P}|\theta_n^{l,A})=(2var(\theta_n^{l,A}))^{-1/2}\varphi((2var(\theta_n^{l,A}))^{-1/2}(\theta_n^{l,P}-\theta_n^{l,A}))
\end{flalign}
where $\theta_n^{l,P}$ and $\theta_n^{l,A}$ are the marginals of $\Vec{\theta}$ in the perturbed particle set and alive particle set, respectively.
Based on the component-wise independent random walk, we can calculate the particle acceptance ratio of the perturbed set, $\rho_{l+1}$.
Based on the above two subsets, we update $\hat{\pi}_{l+l}(\Vec{\theta})$ by concatenating $\hat{\pi}_{l,A}(\Vec{\theta}|x_{i, obs}, \eta_{i, obs}, x_{i-1}, v_{i-1})$ and $\hat{\pi}_{l,P}(\Vec{\theta}|x_{i, obs}, \eta_{i, obs},x_{i-1}, v_{i-1})$, then set $W_{l+1}^k=\frac{1}{K}$, $k=1\dots K$.\\
\hfill \break
(4) Step 4 (Stopping Criteria Checking): if $\rho_l \gets 0 $, Stop the algorithm, otherwise $l=l+1$ and Return to Step 2. Output $\hat{\pi}_l(\Vec{\theta}|x_{i, obs}, \eta_{i, obs},x_{i-1}, v_{i-1})$.

\section{Calibration Results and Statistical Analysis}
\label{S:3}

In this section, we calibrate the stochastic EAB model for ACCs and HDVs based on ABC-ASMC using the open-source empirical datasets, respectively - ACC data collected by \citet{li2021car}, and NGSIM \citep{NGSIM} and High-D \citep{krajewski2018highd} for HDVs. Based on that, we take a holistic examination of the distribution-wise heterogeneity in CF behavior across HDVs, different ACC vehicle models, engine modes, and speeds.  

\subsection{Empirical Data Description}
\begin{figure*}[h]
 \captionsetup{justification=centering} 
        \centering
       \begin{subfigure}[b]{0.49\textwidth}
            \centering
            \includegraphics[width=\textwidth]{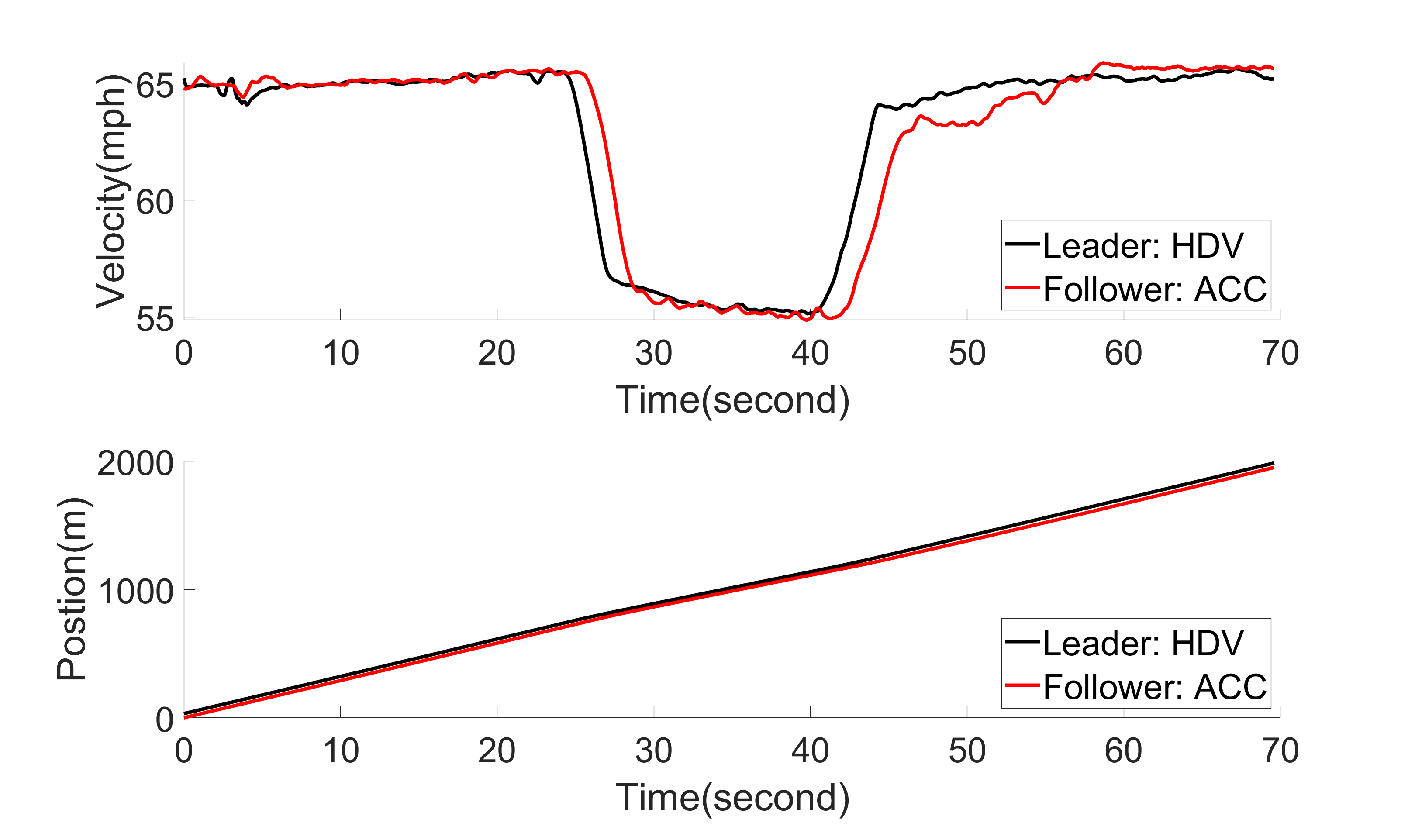}
            \caption*{(a)}%
            {{\small}}    
            \label{}
        \end{subfigure}
        \hfill
        \begin{subfigure}[b]{0.49\textwidth}   \centering\includegraphics[width=\textwidth]{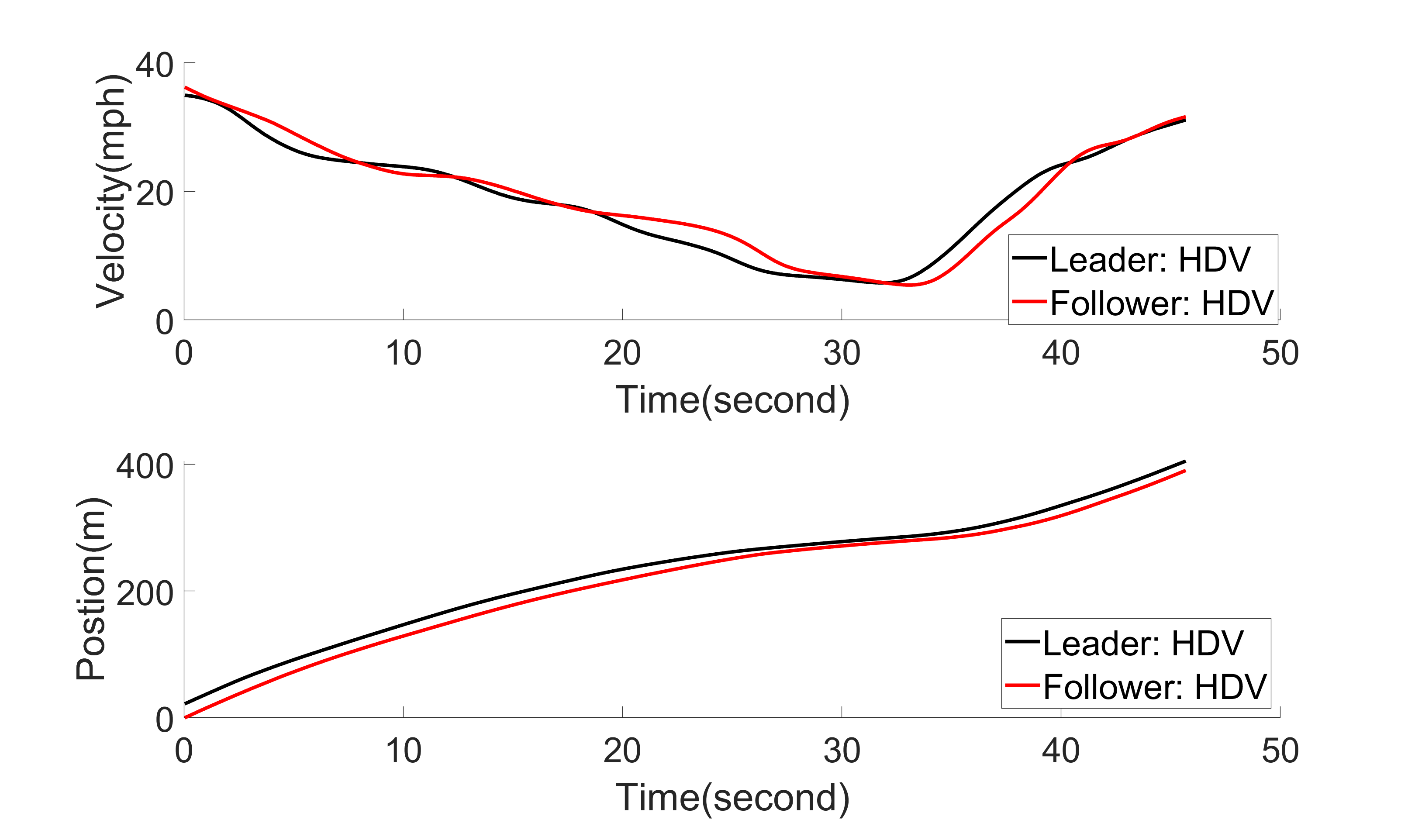}
            \caption*{(b)}%
            {{\small}}    
            \label{}
        \end{subfigure}
        \caption[{Empirical Trajectory around Disturbance}]
        {\small Empirical Trajectory around Disturbance- Time Series Position and Time Series Velocity\\
        (a) ACC: Car Model-X  (b) HDV: HDV-2\\} 
        \label{emp_trajectory}
    \end{figure*}

\citet{li2021car} designed experiments for a three-vehicle platoon consisting of a HDV followed by two ACC vehicles with 1 second headway. To reflect the real-world traffic conditions under disturbances, HDVs are instructed to follow a designed speed profile. HDVs initially travel at a nearly constant speed, then decelerate to a target speed, and finally accelerate to resume the initial constant speed. An illustrative example is provided in Fig. \ref{emp_trajectory} (a).  In the dataset, there are three different controllers, denoted as `Car Model-X', `Car Model-Y', and `Car Model-Z', respectively. The dataset of Car Model-Y is further partitioned into two subsets by different engines (Normal or Sports). The dataset of Car Model-Z has four subsets by different engines (Normal or Power) as well as different speed ranges (Low speed or Median and High speed). To compare the CF behavior of HDVs and ACC vehicles, we extract CF data of HDVs that encountered a disturbance from two datasets (i.e, NGSIM and HighD), as exemplified in Fig. \ref{emp_trajectory}(b). Further, we partitioned the HDV dataset into two subsets by speed ranges. Readers could refer to \citet{li2021car}, \citet{NGSIM} and \citet{krajewski2018highd} for more details. A general description of the selected data is provided in Table \ref{emp_description}.

\begin{table}[!h]
\small
	\caption{General Description of Empirical Trajectories}\label{tab:versions}
	\begin{center}
		\begin{tabular}{c c c c| c c c }\Xhline{1pt}
		\multirow{2}*{Type}&\multirow{2}*{Car Model} & \multirow{2}*{Engine}&\multirow{2}*{Initial Speed Level} &\multirow{2}*{Number of Trajectories} & \multicolumn{2}{c}{Follower Initial Speed (mph)}\\
        	&&&&& mean (mph) & SD (mph)\\ \Xhline{1pt}
        	  \multirow{2}*{HDV} & HDV-1& & Low&16&30.33&1.05\\
        		 & HDV-2& &Median and High&50&48.22&7.52\\
        			\Xhline{0.5pt}
			 \multirow{7}*{ACC}& X &  &Median and High &48 &48.41 &12.64 \\
				 & Y-1 &  Normal &Median and High& 48 & 48.83 &12.73  \\
				 & Y-2 & Sport &Median and High& 48 & 48.45 &12.69  \\
						& Z-1  & Normal &Low & 16 & 35.78 & 1.61  \\
					 	 & Z-2  & Normal &Median and High&32 & 55.87 &10.44  \\
			     		 & Z-3  & Power &Low& 16 & 36.54 & 1.40\\
			     	 	 & Z-4 & Power &Median and High& 32 & 56.36 &10.11
		 \\\Xhline{1pt}
		\end{tabular}
	\end{center}
 \small
 \label{emp_description}
\end{table}

\subsection{Stochastic Calibration Results and Performance}

For each AV controller and HDV subsets, we randomly select 75\% of the empirical trajectories as a training set, $M_1$, and the remaining 25\% as a testing set, $M_2$. We apply ABC-ASMC to calibrate the EAB model using the training set and validate the framework by reproducing trajectories using the estimated posterior joint distributions of model parameters and comparing them with the trajectories in the testing set. As the EAB model is essentially an extension of Newell's simplified CF model, calibration is conducted in two stages, where basic parameters $\tau$ and $\delta$ are first calibrated, and the remaining parameters related to $\eta(t)$ evolution are calibrated in the second stage. We do this to cope with the high dimension of the parameter space and estimate the congestion wave speed, the necessary parameter to measure $\eta(t)$. The first stage calibration is conducted in a deterministic fashion by minimizing the GOF measure, the normalized root mean square error (NRMSE) as below:

\begin{flalign}
GOF(x_{i,obs},x_{i,sim})=\sum_{i=1}^{|M_1|}NRMSE(x_{i,obs},x_{i,sim})=(\sum_{1}^{|M_1|}\frac{\sqrt{\frac{1}{T}\sum_{t=1}^{T}(x_{i,obs}(t)-x_{i, sim}(t))^2}}{\sqrt{\frac{1}{T}\sum_{t=1}^{T}(x_{i,obs}(t))^2}})
\end{flalign}
where $|M_1|$ is the cardinality of training set $M_1$, $x_{i,obs}$ is the observed position and $x_{i, sim}$ is the simulated position. $T$ is the total CF time. 

\begin{table}[!h]
\small
	\caption{Calibrated Parameters in the Newell Model}\label{tab:versions}
	\begin{center}
		\begin{tabular}{c c c c |c c c  }\Xhline{1pt}
    Type & Car Model & Engine & Speed Level & $\tau$ (s) & $\delta$ (m) & $w$ (m/s)\\ \Xhline{1pt}
        	  \multirow{2}*{HDV} & HDV-1&& Low speed&1&9&-9\\
        		 & HDV-2& &Median and high speed&1&6&-6\\
        			\Xhline{0.5pt}
			 \multirow{5}*{ACC}& X & & Median and high &1.1 &10 &-9.09 \\
				 & Y &  Normal &Median and high& 1.1 & 6 & -5.45 \\
				 & Y & Sport &Median and high& 1 & 8 & -8 \\
			   & Z & Normal &Low, median and high& 1 & 12 & -12  \\
				 & Z  & Power &Low, median and high& 1 & 12&-12 \\\Xhline{1pt}
		\end{tabular}
	\end{center}
 \small
 \label{newell}
\end{table}

The first stage calibration result in Table \ref{newell} is consistent with the findings from several empirical studies conducted on HDVs (\citet{chiabaut2010heterogeneous}) and ACCs (\citet{gunter2019model}). It shows that even the basic parameters show some variations across HDVs, different ACC car models and engines. HDV with median and high speed range (HDV-2) appears to be the most aggressive with the smallest $\tau$ and $\delta$. Car Model-Y has different settings for normal and sports engines, whereas Car Model-Z appears to share the same setting. It further shows that Car Model-Y is set to be more aggressive with a smaller $\delta$. Generally, Car Model-Z appears to have a more conservative setting than the other ACC car models. As a result, the estimated congestion wave speed varies across HDVs and ACC car models: it is fastest with Car Model-Y and slowest with Car Model-Z.

For the second stage stochastic calibration, we set $K=20000$ and $\lambda=0.95$ considering the $\Vec{\theta}$ dimension. Based on the maximum and minimum of the empirical reaction pattern $\eta_{obs}$, we set the prior distributions for the model parameters as independent uniform distributions whose marginal distributions are: (1) $\eta^0, \eta^1, \eta^2, \eta^3 \sim Uniform(0.5,1.5)$ for ACCs and $\eta^0, \eta^1, \eta^2, \eta^3 \sim Uniform(0.3,3)$ for HDVs; (2) $\epsilon^0, \epsilon^1, \epsilon^2 \sim Uniform(-0.15,0.15)$; (3) $t_1\sim Uniform(0,25)$.

For our calibration, we aim to reproduce the empirical position $x_{i,obs}$ as well as the reaction pattern $\eta_{i,obs}$.
Hence, we modify the GOF for the second stage calibration as a weighted NRMSE between the empirical ($x_{i,obs}$,$\eta_{i,obs}$, ${\eta_{i,obs}}^{c}$) and calibrated ($x_{i, sim}$, $\eta_{i,sim}$, ${\eta_{i,sim}}^{c}$) as below:
 
\begin{flalign}
GOF
=c_1NRMSE(x_{i,obs}, x_{i, sim})+c_2NRMSE(\eta_{i,obs}, \eta_{i,sim})+c_3NRMSE({\eta_{i,obs}}^{c}, {\eta_{i,sim}}^{c})\label{GOF_EAB}
\end{flalign}
where $c_1$, $c_2$, and $c_3$ are weight coefficients, set as $c_1=0.4$, $c_2=0.4$, and $c_3=0.2$ in our study. $\eta^c$ denotes a set of critical $\eta(t)$ points, consisting of $\eta_{i,obs}^{max}$, $\eta_{i,obs}^{min}$, and the maximum of $|\eta_{i,sim}-\eta_{i,obs}|$, where $\eta_{obs}^{max}$ and $\eta_{obs}^{min}$ are the maximum and minimum points of $\eta_{i,obs}$, respectively.

\begin{figure*}[h]
        \centering
        \begin{subfigure}[b]{0.48\textwidth}
           \centering
       \includegraphics[width=\textwidth]{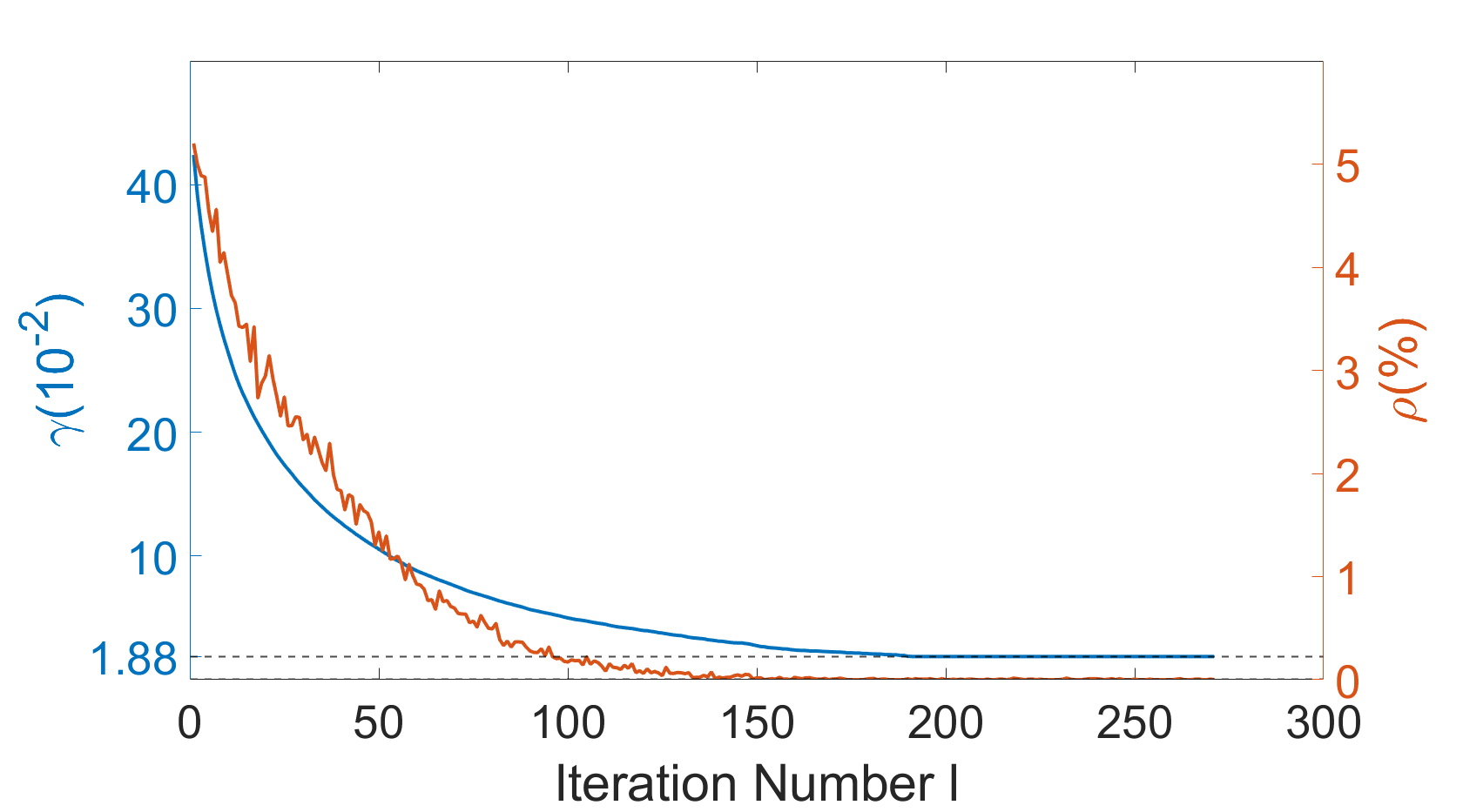}
              \caption*{(a)}%
              \label{fig:a}
            {{\small }}   
        \end{subfigure}
        \begin{subfigure}[b]{0.48\textwidth} 
            \centering             \includegraphics[width=\textwidth]{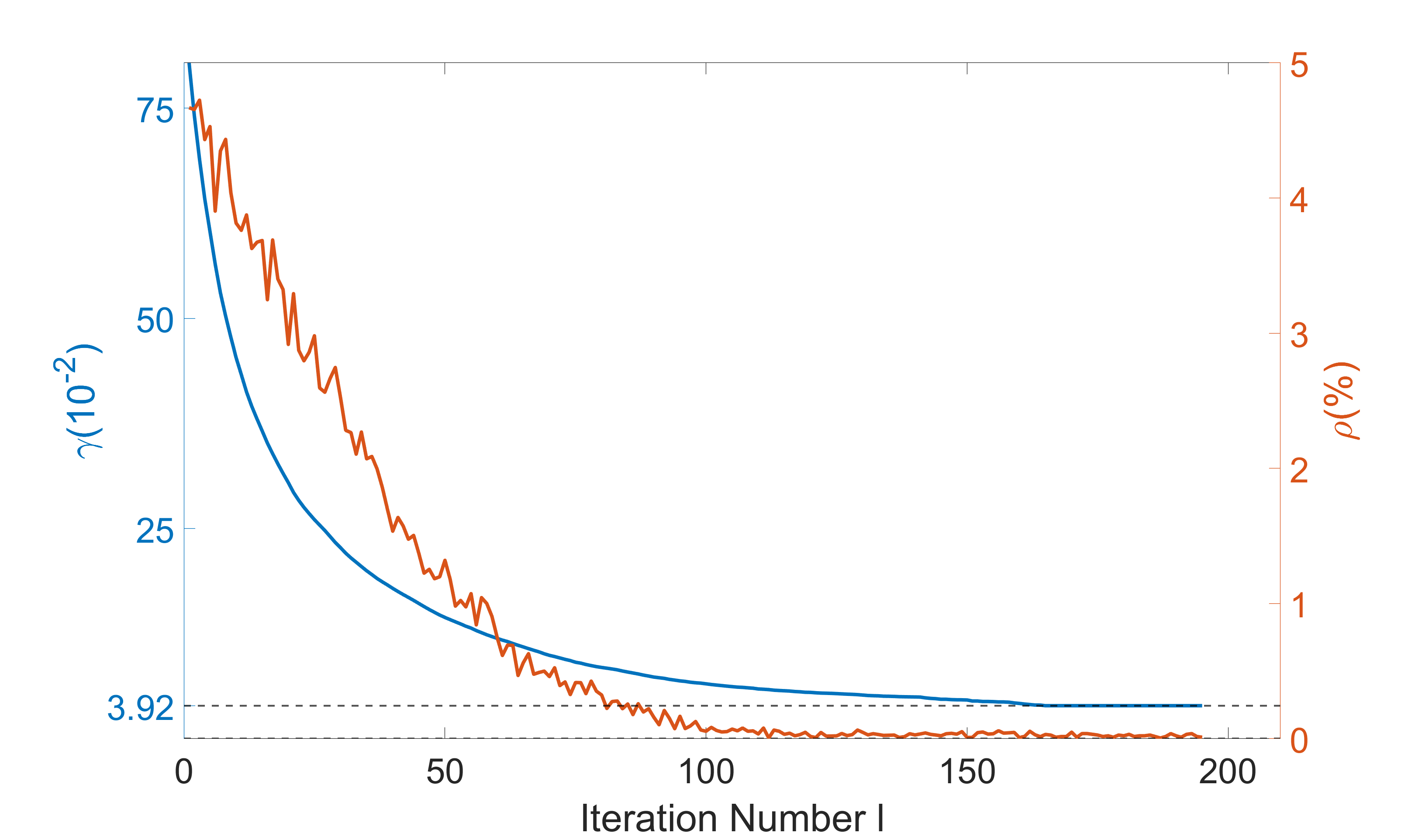}
            \caption*{(b)}%
            {{\small }}    
            \label{fig:b}
        \end{subfigure}
         \caption[Illustrative Example: Convergence of $\gamma$ and $\rho$]%
        {\small Convergence of $\gamma$ and $\rho$  \\
        (a): Car Model-X
        (b): HDV-2
        } 
        \label{convergence}
    \end{figure*}

Fig. \ref{convergence} gives the examples (HDV-2 and Car Model X) illustrating the convergence of $\gamma$ and $\rho$ in ABC-ASMC. The results of other Car Models and HDVs are shown in Appendix A. It shows that $\gamma$ for all cases decreases quickly within the first 80 iterations, demonstrating relatively quick convergence. Further, $\rho$ also decreases significantly, which means that ABC-ASMC becomes more selective in estimating the posterior joint distribution. After 100 iterations, $\gamma$ and $\rho$ both converge, suggesting that further sampling efforts cannot enhance GOF anymore, and the algorithm reaches the converged tolerance, $\gamma$.

\begin{figure*}[!h]
 \captionsetup{justification=centering} 
        \centering
        \begin{subfigure}[b]{0.48\textwidth}
            \centering
       \includegraphics[width=\textwidth]{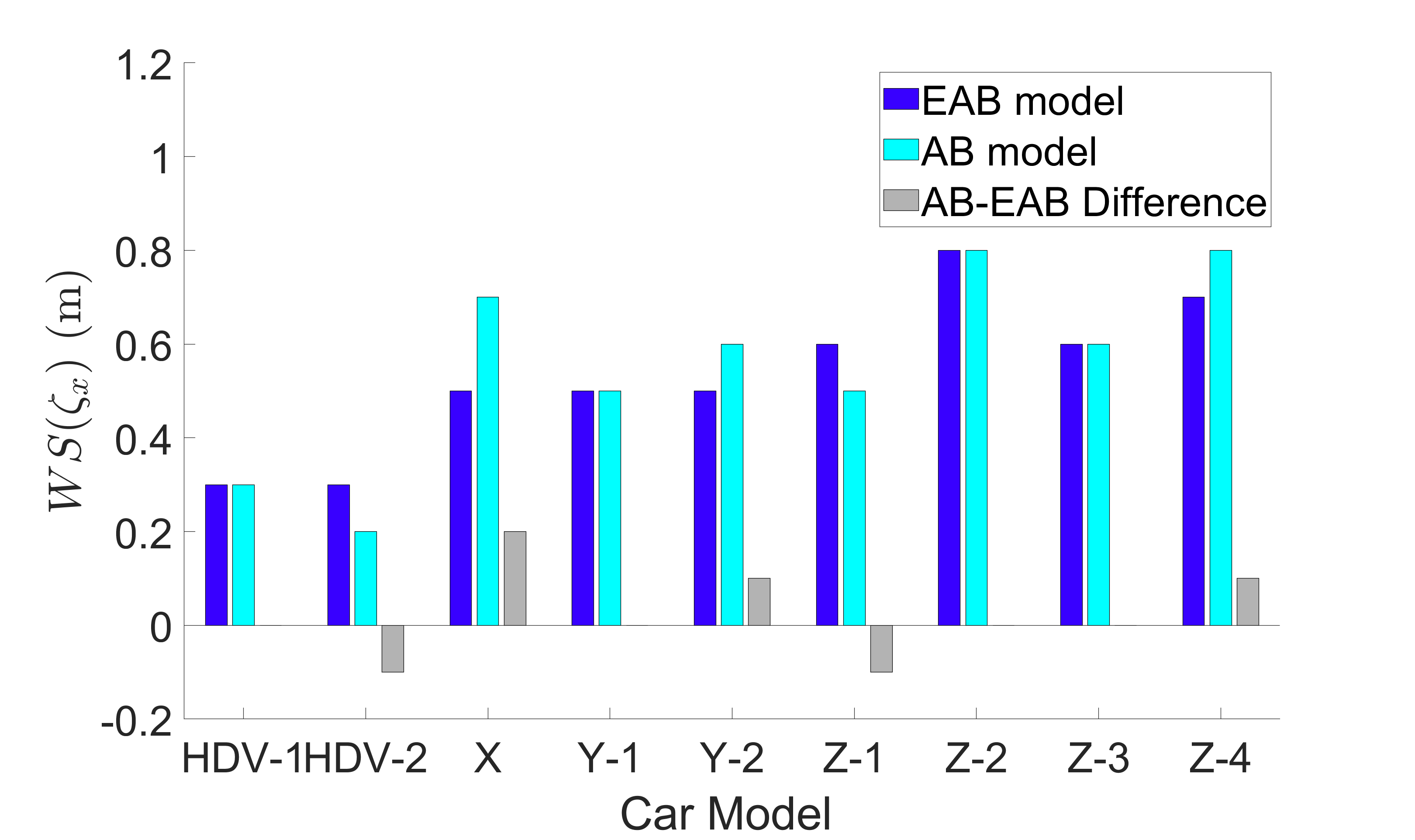}
              \caption*{(a1)}%
            {{\small }}   
        \end{subfigure}
        \begin{subfigure}[b]{0.48\textwidth} 
            \centering 
        \includegraphics[width=\textwidth]{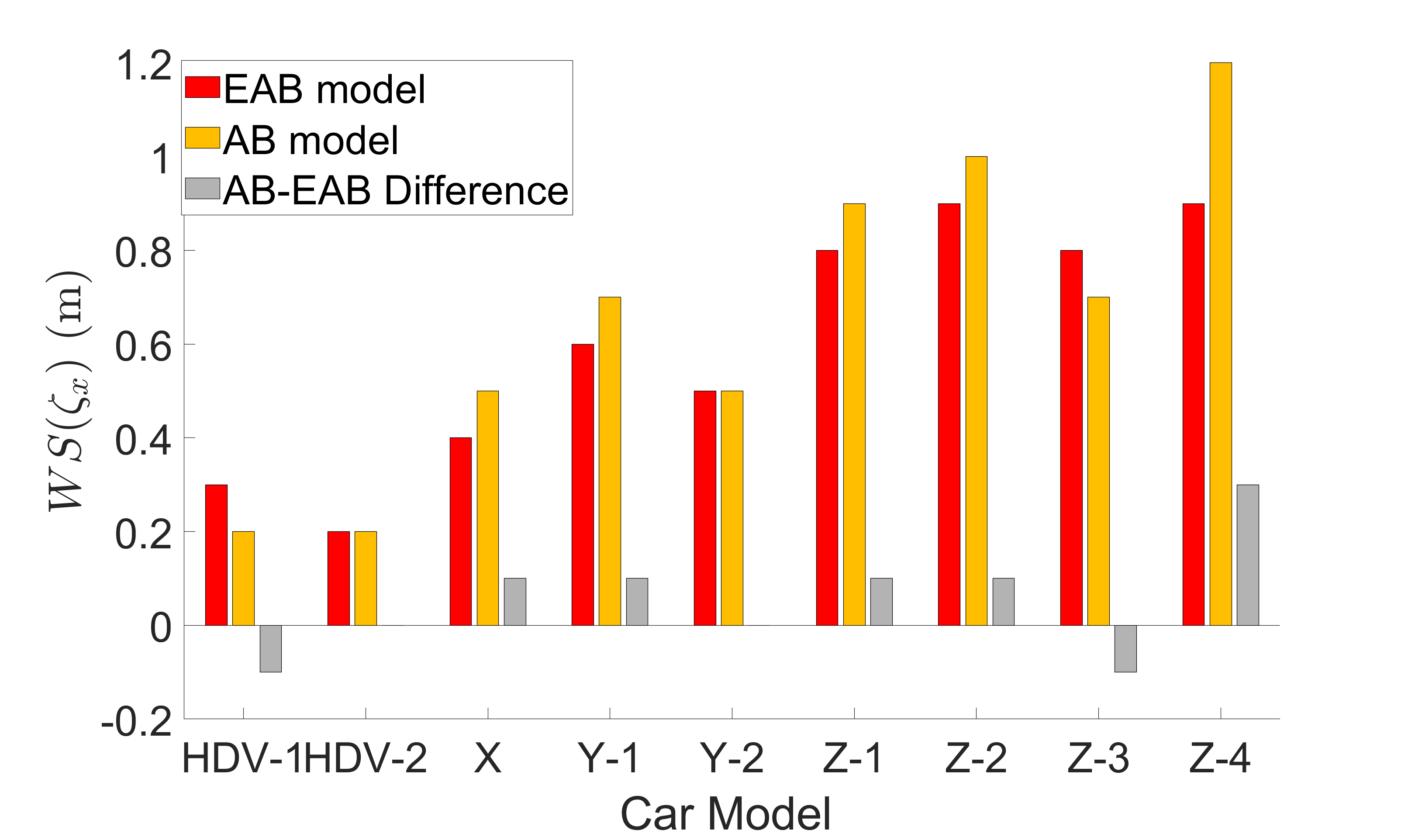}
             \caption*{(b1)}%
            {{\small }}   
        \end{subfigure}
         \begin{subfigure}[b]{0.48\textwidth}
            \centering
  \includegraphics[width=\textwidth]{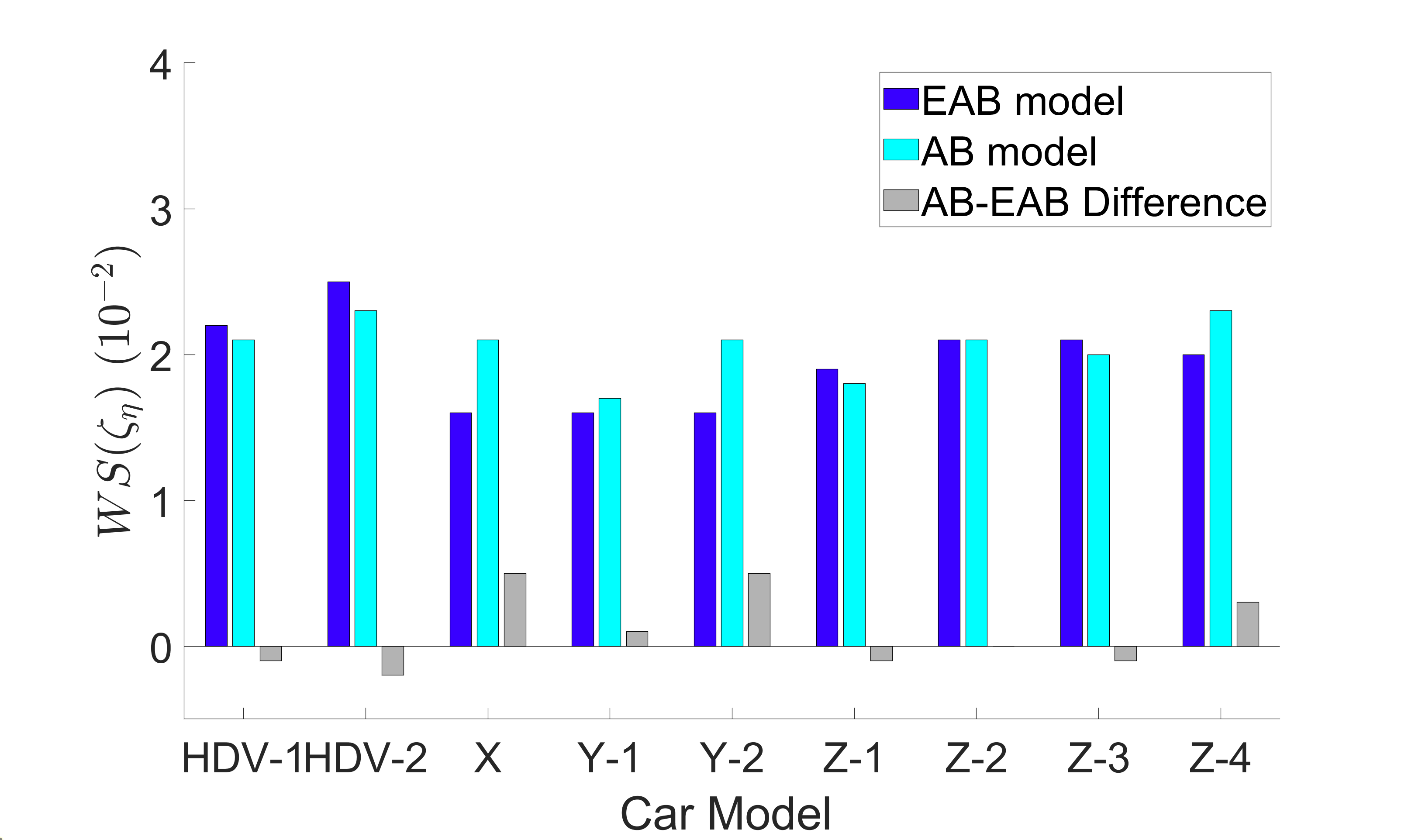}
         \caption*{(a2)}%
            {{\small }}   
        \end{subfigure}
        \hfill
        \begin{subfigure}[b]{0.48\textwidth} 
            \centering 
        \includegraphics[width=\textwidth]{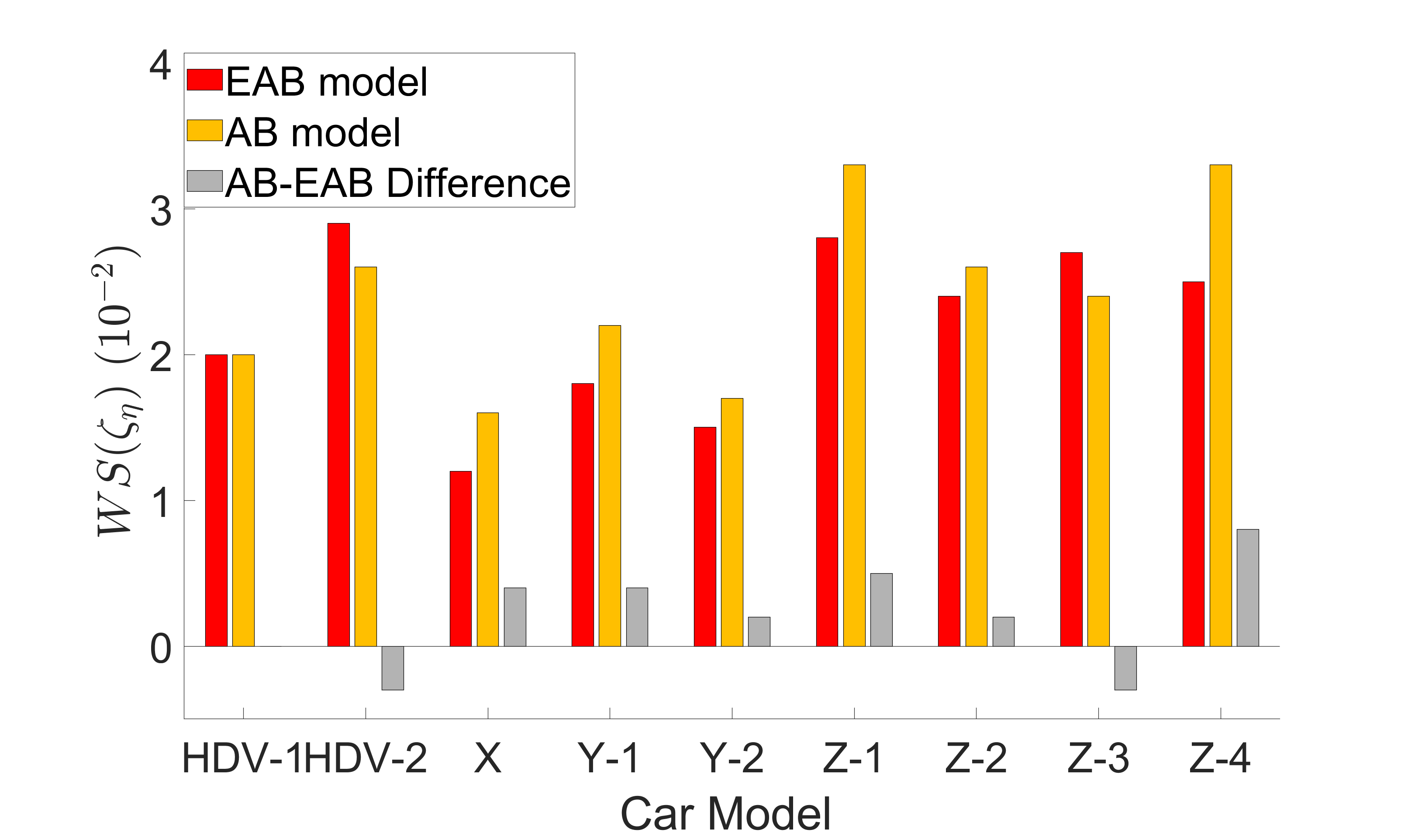}
            \caption*{(b2)}%
            {{\small }}    
            \label{}
        \end{subfigure}
        \begin{subfigure}[b]{0.48\textwidth}
            \centering
       \includegraphics[width=\textwidth]{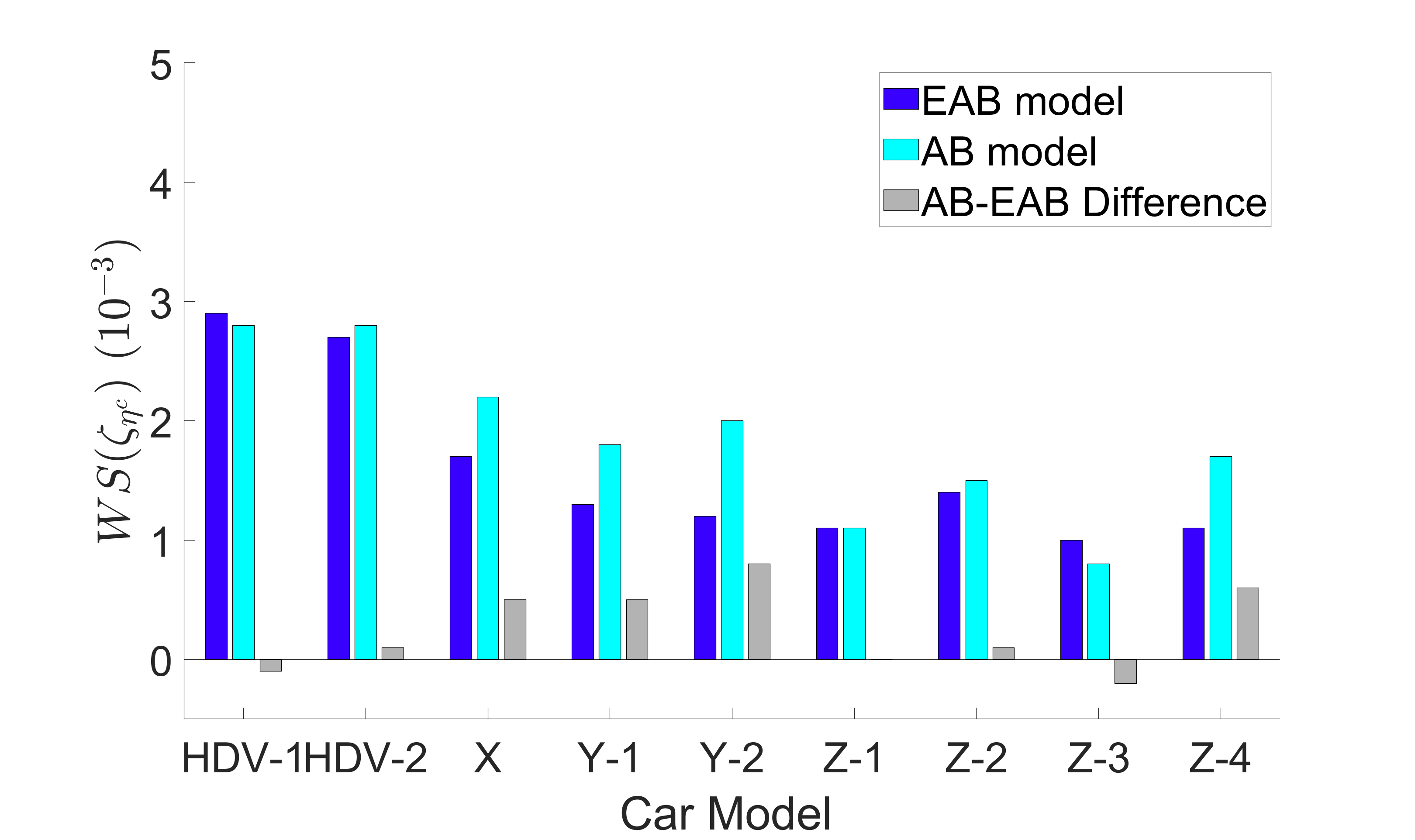}
              \caption*{(a3)}%
            {{\small }}   
        \end{subfigure}
        \hfill
        \begin{subfigure}[b]{0.48\textwidth} 
            \centering 
        \includegraphics[width=\textwidth]{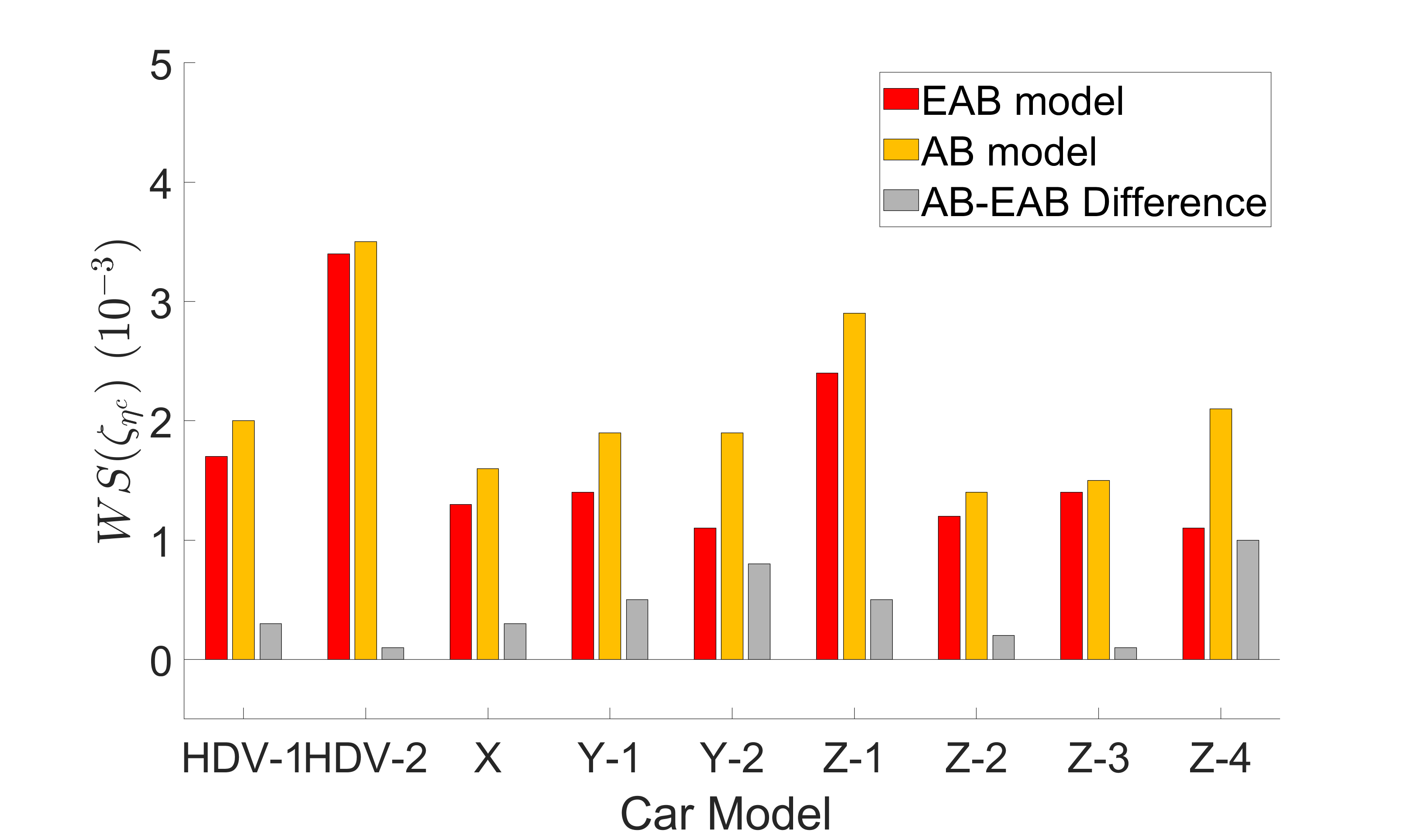}
             \caption*{(b3)}%
            {{\small }}   
        \end{subfigure}
         \caption[Stochastic Calibration Performance Validation using Training and Testing Trajectories]%
        {\small Stochastic Calibration Performance Validation using Training and Testing Trajectories\\
        (a): Training Results
        (b): Testing Results
        } 
        \label{calibration}   
    \end{figure*}

Fig. \ref{calibration} gives the overall calibration and validation results of EAB and AB model quantified by applying the Wasserstein (WS) metric \citep{panaretos2019statistical} both to training set $M_1$ and testing set $M_2$. The WS metric gives the optimal simulated-observed reproduction distance by probabilistically matching a particle $k$ to a car following pair $m$. The objective function (Eq. (\ref{WS})) aims to minimize the distribution-wise errors by determining the optimal weight distribution of particles that best fit CF pairs. 

\begin{flalign} \label{WS}
WS(\zeta)=\frac{1}{|M|}\min_{\kappa}\frac{1}{K}(\sum_{m,k}^{M,K}\kappa_{m,k}\zeta),\zeta\in\{\zeta_x,\zeta_{\eta}, \zeta_{\eta^c}\},\kappa_{m,k}\in \{0,1\}. 
 \end{flalign}
\begin{flalign} \label{WS_Contraint} 
s.t. \sum_{k=1}^{K}\kappa_{m,k}=1 
 \end{flalign}
 \begin{flalign} 
\zeta_x=\sqrt{\frac{1}{T_{m}}\sum_{t=1}^{T_{m}}(x^{m}_{i,obs}(t)-x^{m,k}_{i,sim}(t))^2} \label{WS_feature}  \\
 \zeta_{\eta}=\sqrt{\frac{1}{T_{m}}\sum_{t=1}^{T_{m}}(\eta^{m}_{i,obs}(t)-\eta^{m,k}_{i,sim}(t))^2}\notag \\
 \zeta_{\eta^c}=\sqrt{\frac{1}{T_{m}}\sum_{t=1}^{T_{m}}(\eta^{c,m}_{i,obs}(t)-\eta^{c,m,k}_{i,sim}(t))^2}\notag 
 \end{flalign}
where, $M$ is the trajectory set, and $m$ is the specific CF pair in $M$, $T_{m}$ is the total CF time of $m$. $|M|$ is the cardinality of $M$. $\zeta_x$, $\zeta_{\eta}$ and $\zeta_{\eta^c}$ denote the root mean square error in position $x$, reaction pattern $\eta$, and the critical points of $\eta$, respectively in Eq. \ref{WS_feature}. $\kappa_{m,k}$ is the weight assigned to the particle $k$ fitting the CF pair $m$. Eq. (\ref{WS_Contraint}) governs that the sum of the weights $\kappa$ should be equal to 1. 

The calibration results with training data (Fig. \ref{calibration}(a)) shows that $WS(\zeta_x)$ for all car models calibrated by EAB and AB models are smaller than $1$ m; $WS(\zeta_{\eta})$ are all below $0.03$; and $WS(\zeta_{\eta^c})$ are all smaller than 0.003. The WS values for the ACC vehicles are generally smaller with the EAB model than the AB models, justifying the need for the EAB model framework. Further, the validation errors using the testing data (Fig. 4(b)) are reasonably close to the calibration errors, demonstrating good performance of the EAB model. These findings suggest that EAB model, compared with the AB model, is overall more flexible and accurate in describing the ACC CF behavior.

\subsection{Reaction Pattern Analysis}
\label{SS:Mappings}

We further categorize the reaction patterns, as shown in Table \ref{eta_categorization}. Note that the categorization is determined by a predefined threshold, $\Delta\eta_T=0.09$ for AVs and $\Delta\eta_T=0.18$ for HDVs, based on the difference between the 75th percentile and 25th percentile $\eta_0$ values to prevent the categorization from being overly sensitive to random fluctuations. The higher threshold for HDVs suggests a higher level of stochasticity compared to AVs. We caution that the higher stochasticity could be more attributed to the nature of the data sets than the behavior itself. Notably, the HDV data are from naturalistic driving as opposed to controlled experiments, as is the case for ACC vehicles. 

The results in Table \ref{eta_categorization} show that both single and composite reaction patterns are observed across the board, particularly for Car Model Z. However, when compared to HDVs, the results for ACCs demonstrate a significant divergence. At low speeds, Car Model Z exhibits fewer NE patterns but more concave patterns compared to HDVs. At median and high speeds, ACC vehicles display higher proportions of NE, convex, and non-increasing patterns than HDVs. Early responses are evident in the majority of non-increasing patterns among ACC. Notably, the non-decreasing pattern takes up a significant portion, ranging from 20\% to 36\%. This trend, though in lower incidence than HDVs, is not desirable because it indicates a lower traffic throughput. However, this could be attributed to data limitations, where the recovery from a disturbance may not have been fully captured. A further experimental study is needed to confirm this finding.

Variations are also significant across ACC vehicles. For example, Car Model X exhibits the highest occurrence of NE and non-decreasing patterns, while Car Model Z displays a significant portion of concave patterns. Variations are notable even within the same ACC developers. Notably, the distribution of behavioral patterns between Normal and Sports/Power engines is markedly different at low speed (for Model Z), though the differences are more nuanced at median and high speeds (Model Y and Z). With the same engine at different speeds, there is a higher frequency of concave patterns at low speeds, accompanied by a decrease in the occurrences of convex patterns.
We also provide Jensen-Shannon Distance analysis in the Appendix B to corroborate the categorization remarks. 


Based on the findings, we draw the following conclusions: (1) The behavioral response to disturbance, which directly influences the propagation of disturbances, varies between HDVs and ACC vehicles, and across ACC vehicles. (2) The engine modes and speed have a substantial impact on behavioral patterns. These findings underscore the possibility of highly heterogeneous behavior in mixed traffic. The traffic-level implications of these findings are investigated through the analysis of traffic hysteresis in the following section.

\begin{table}
\small
	\caption{ Categorization: Proportion of Different $\eta$ Patterns under $\Delta\eta_T$}\label{tab:versions}
	\centering
		\begin{tabular}{c c| c |c c| c| c c c c c  }\Xhline{1pt}
   Speed&&\multicolumn{3}{c|}{Low}&\multicolumn{6}{c}{Median and High}\\ \Xhline{1pt} 
			$\Delta\eta_T$ &&0.18&\multicolumn{2}{c|}{0.09}&0.18&\multicolumn{5}{c}{0.09}\\ \Xhline{1pt}
Car Model  &Response & \multirow{2}*{HDV}   &\multicolumn{2}{c|}{Z}  & \multirow{2}*{HDV}  &\multirow{2}*{X}     &\multicolumn{2}{c}{Y}  &\multicolumn{2}{c}{Z}\\ 
Engine   &&  &Normal    &Power    &  &     &Normal    &Sports    &Normal    &Power\\ \Xhline{1pt}
 NE & &0.01& 0.03 & 0 & 0.01 & 0.27 & 0.17 & 0.16 & 0.05 & 0.03 \\\hline
\multirow{2}*{Concave} &  early & 0.13 & 0.03 & 0.04 & 0.01 & 0.01 & 0.03 & 0.04 & 0.04 & 0.02 \\\cline{2-11}
&others&0.14 & 0.44 & 0.60 & 0.24 & 0 & 0.10 & 0.08 & 0.21 & 0.31 \\\hline
\multirow{2}*{Convex} &  early & 0.02& 0.14 & 0.03 & 0.03 & 0.01 & 0.08 & 0.15 & 0.12 & 0.14 \\\cline{2-11}
& others & 0.01& 0 & 0.01 & 0.04 & 0.05 & 0.05 & 0.06 & 0.12 & 0.03 \\\hline
Concave-convex &&0.05& 0.11 & 0.08 & 0.06 & 0.00 & 0.03 & 0.02 & 0.07 & 0.10 \\\hline
Convex-concave&&0.02 & 0.03 & 0.15 & 0.02 & 0.01 & 0& 0.03 & 0.08 & 0.06 \\\hline
\multirow{2}*{Non-decreasing} & early & 0.10& 0.09 & 0 & 0.10 & 0.04 & 0.03 & 0.06 & 0.01 & 0.06 \\\cline{2-11}
&others& 0.51& 0.13 & 0 & 0.48 & 0.36 & 0.31 & 0.22 & 0.22 & 0.20 \\\hline
\multirow{2}*{Non-increasing} & early &0.01 & 0 & 0.08 & 0.01 & 0.24 & 0.15 & 0.15 & 0.05 & 0.05 \\\cline{2-11}
 & others &0 & 0 & 0.01 & 0 & 0.01 & 0.05 & 0.03 & 0.03 & 0 \\
\Xhline{1pt}
\end{tabular}
\small
\label{eta_categorization}
\end{table}


\section{Traffic Hysteresis Evaluation}
\label{S:4}
Traffic Hysteresis, an elliptical movement of flow-density evolution under disturbance, is an important traffic phenomenon linked to the reduction in traffic throughput and traffic flow instability. Studies suggest that traffic hysteresis is directly associated with dynamic CF behavior, characterized by asymmetric deceleration and acceleration during a disturbance \citep{ahn2013method,saifuzzaman2017understanding,chen2012microscopic}. Thus, a good CF model should be able to reproduce traffic hysteresis observed empirically for macroscopic perspectives. In this paper, we focus our investigation on how different CF behaviors of ACC vehicles manifest in traffic hysteresis.

\subsection{Traffic Hysteresis Measurement}
\begin{figure*}[!h]
 \captionsetup{justification=centering} 
        \centering                      
        \begin{subfigure}[b]{0.48\textwidth}
        \centering
          \includegraphics[width=\textwidth]{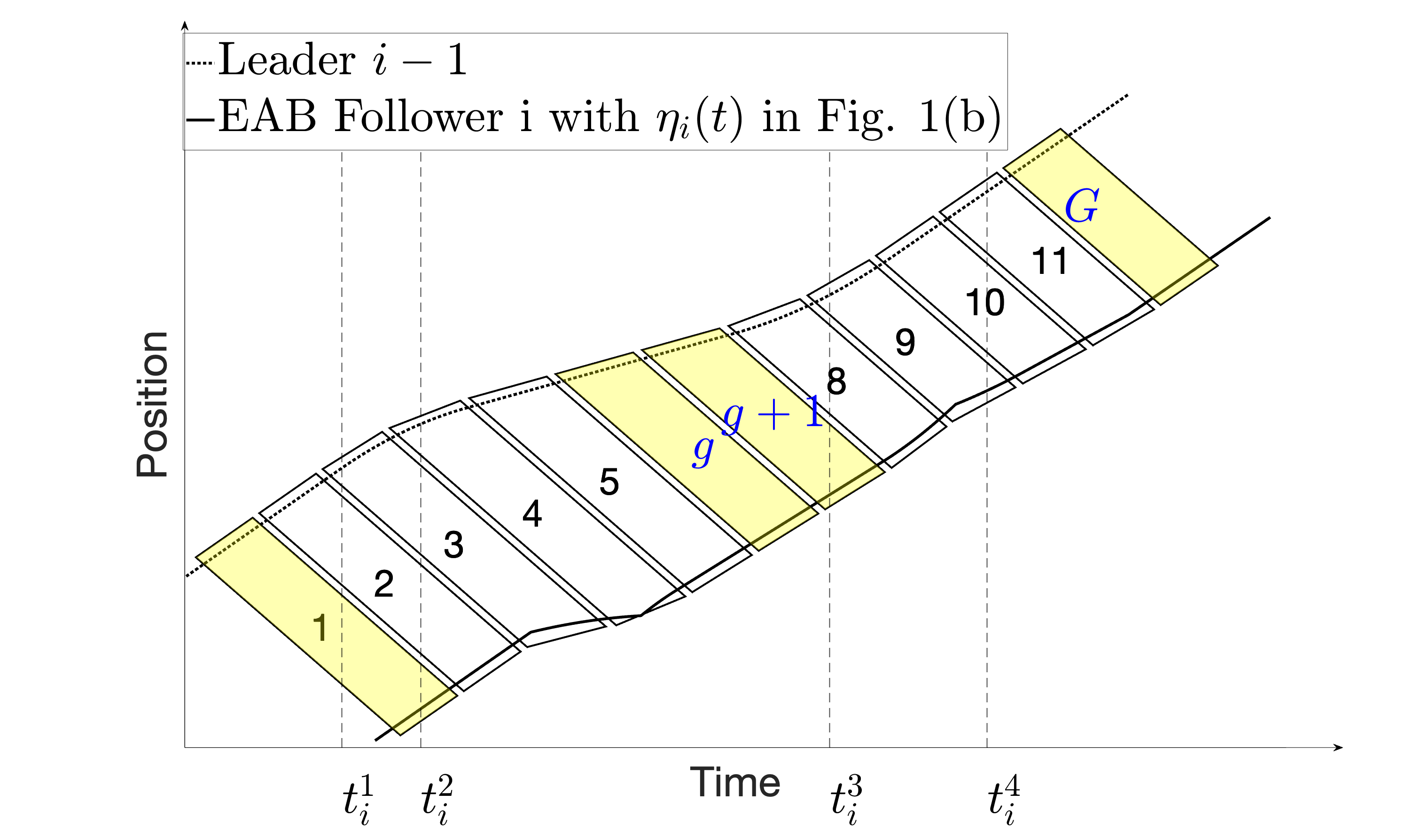}
           \caption[]%
            {{\small }}   
            \label{}
        \end{subfigure}
        \hfill
        \begin{subfigure}[b]{0.48\textwidth} 
            \centering 
        \includegraphics[width=\textwidth]{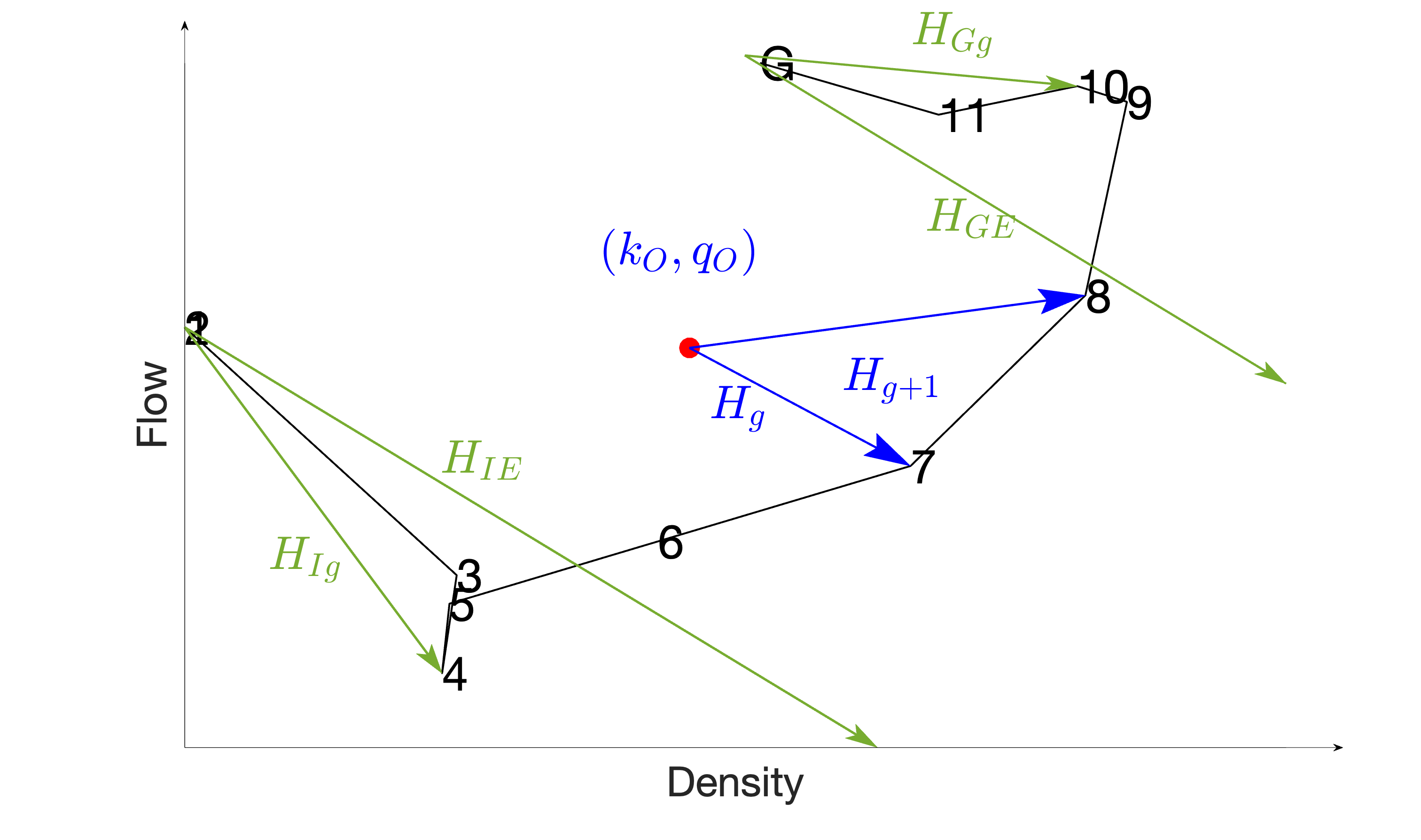}
            \caption[]%
            {{\small }}    
        \end{subfigure}
         \begin{subfigure}[b]{0.48\textwidth}
            \centering
            \includegraphics[width=\textwidth]{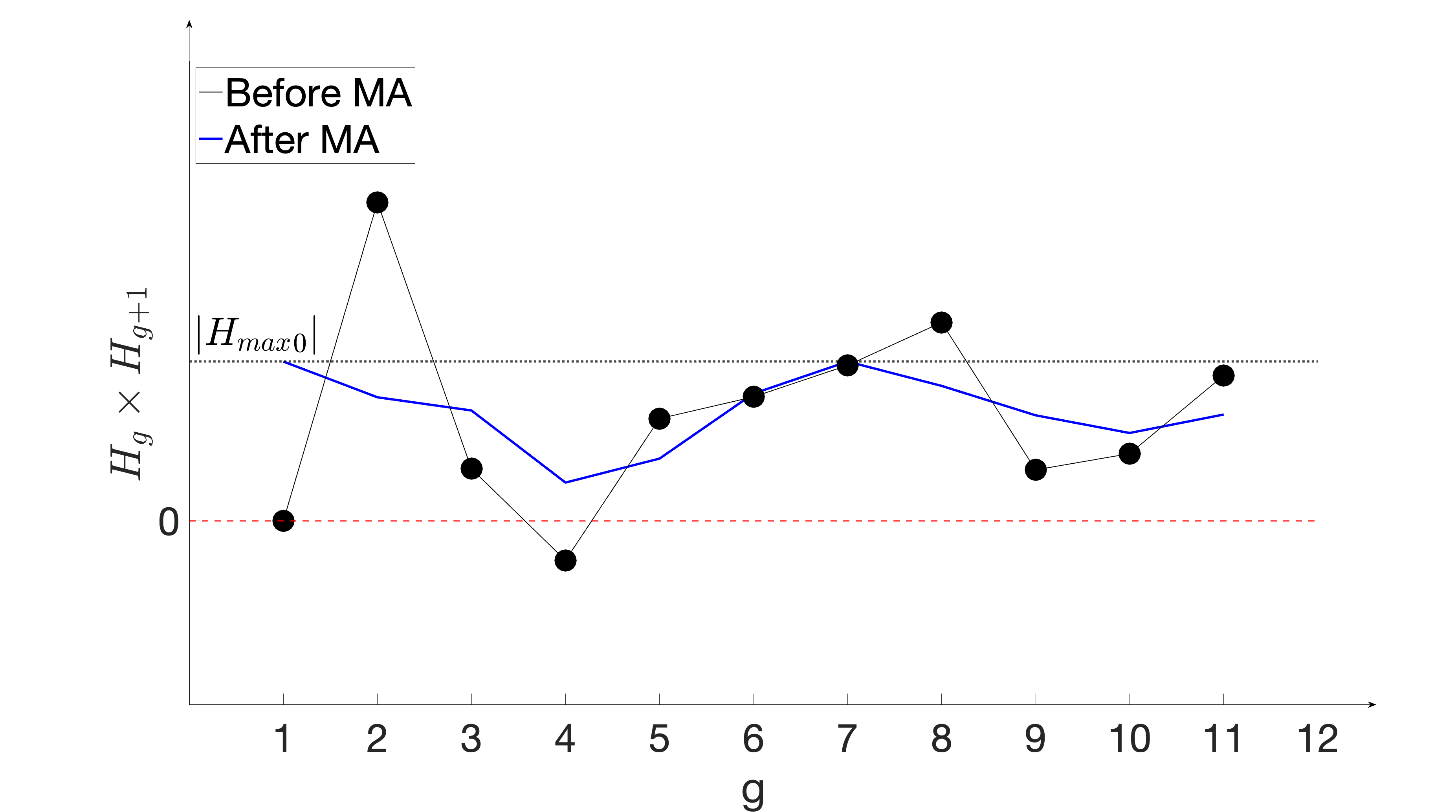}
            \caption[]%
            {{\small}}    
        \end{subfigure}
        \hfill
        \begin{subfigure}[b]{0.48\textwidth} 
            \centering 
        \includegraphics[width=\textwidth]{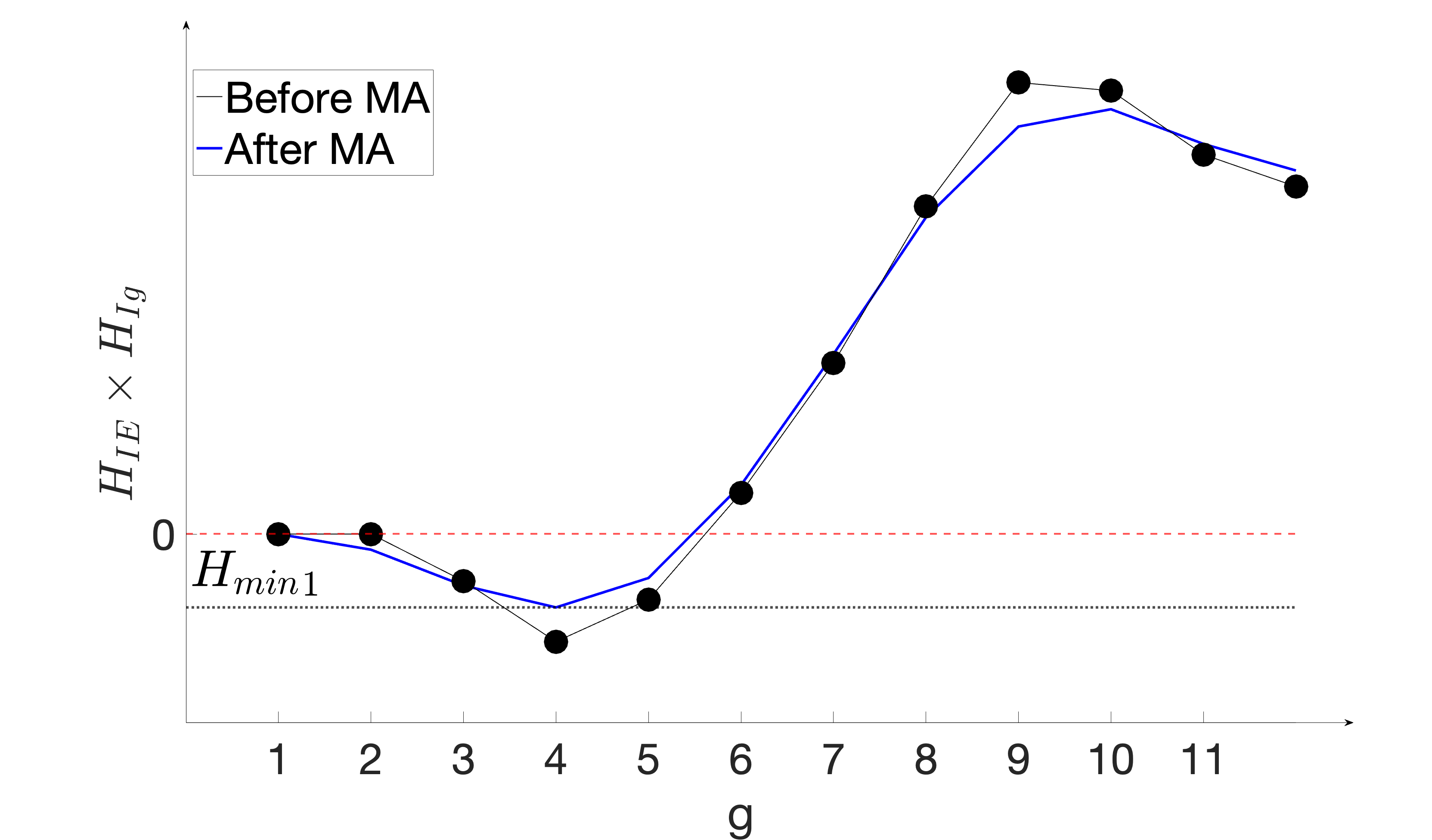}
            \caption[]%
            {{\small }}    
        \end{subfigure}
         \begin{subfigure}[b]{0.48\textwidth}
            \centering
            \includegraphics[width=\textwidth]{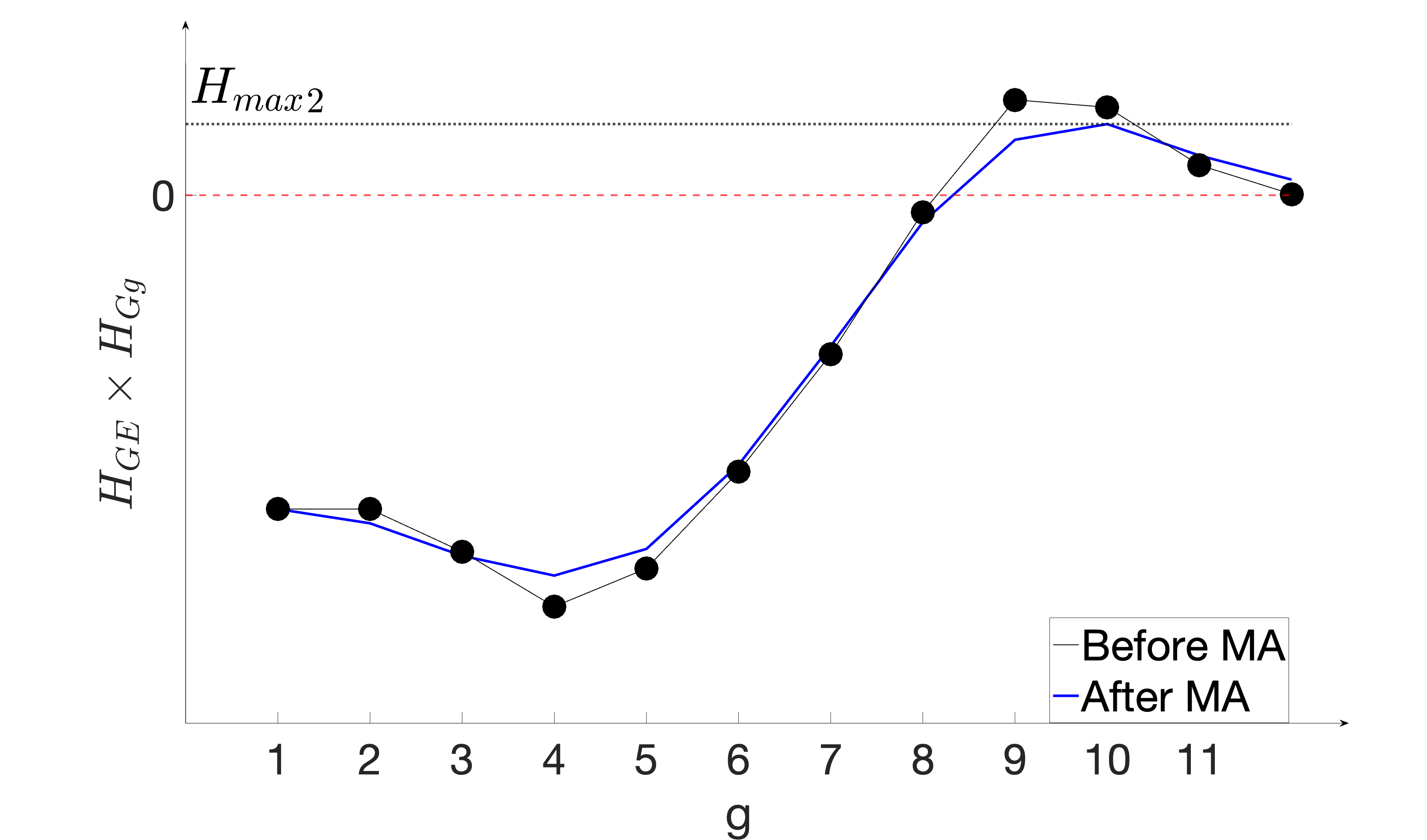}
            \caption[]%
            {{\small}}    
        \end{subfigure}
        \hfill
        \begin{subfigure}[b]{0.48\textwidth}  
            \centering 
            \includegraphics[width=\textwidth]{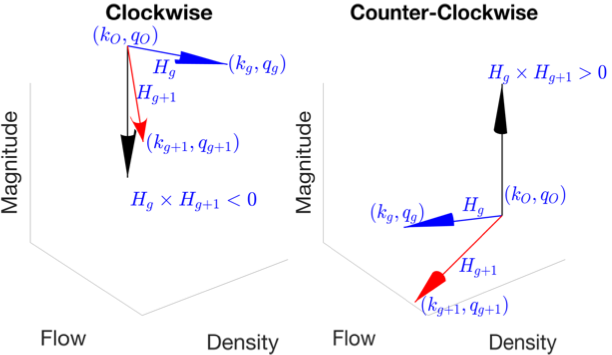}
            \caption[]%
            {{\small }}    
        \end{subfigure}
       \caption[{Determining the Hysteresis Orientation and Magnitude}]
        {\small Schematic Illustration of Hysteresis Measurement\\
        (a) Density-Flow Measurement (b) Hysteresis Orientation Measurement \\(c) Time-varying Hysteresis Magnitude (d) Time-varying Hysteresis Magnitude Relative to Initial Equilibrium\\
        (e) Time-varying Hysteresis Magnitude Relative to New Equilibrium (f) Cross-Product (Right-hand Rule)} 
        \label{hysteresis_measure}
    \end{figure*}

We establish a systematic method to quantify traffic hysteresis. Such method is lacking in the current literature due to the challenge that traffic hysteresis is a directed two-dimensional movement. We first measure the flow and density as vehicles go through a disturbance based on Edie’s generalized definitions \citep{edie1963discussion} and the measurement method by \cite{laval2011hysteresis,shi2021constructing}. Specifically, we define parallelograms rotated by the wave speed, $w$, along the trajectories; see Fig. \ref{hysteresis_measure}(a). Then in each parallelogram, the flow and density are measured according to the generalized definitions: 

 \begin{flalign}
  k(g)=\sum_i^I\frac{\Delta t_i}{|Z_g|}, q(g)=\sum_i^I\frac{\Delta x_i}{|Z_g|}
\end{flalign}\\
where $k$ is the density; $q$ is the flow; $\Delta t_i$ and $\Delta x_i$ are the travel time and distance of the $i^{th}$ vehicle in Zone $g$, respectively; and $|Z_g|$ is the area of Zone $g$, $I$ is the total number of vehicles.  Note that defining the boundary of Zone $g$ is inherently challenging, particularly when $I$ is small (i.e., 2 in the CF pair); a larger boundary leads to the underestimation of flow and density, vice verse. Here, we firstly employ the trajectories of both the leader and the follower to establish the boundary, forming Zone $g$ (yellow shade),  we multiply the area of this region by an additional adjustment parameter $\frac{I}{I-1}$ to obtain an appropriate estimation of $|Z_g|$. Finally, to capture the hysteresis evolution over the zones, the moving average method with window of 3 zones is implemented. 

An example of CCW flow-density evolution is shown in Fig. \ref{hysteresis_measure}(b). To systematically quantify the movement, we utilize four metrics: (1) the center point of flow-density relationship $O$($\frac{\sum_{g=1}^Gk(g)}{G},\frac{\sum_{g=1}^Gq(g)}{G}$), where $G$ is the total number of the zones; (2) the standard deviation (SD) of density and flow;  (3) the time-varying magnitude of hysteresis (Fig. \ref{hysteresis_measure}(c)) and (4) hysteresis patterns. For (3)-(4), we utilize the right-hand rule of the cross-product to establish the hysteresis orientation (i.e., clockwise or counter-clockwise; see Fig. \ref{hysteresis_measure}(f)) and quantify the hysteresis magnitude. Note that the method could be extended to a vehicle platoon with more than 2 vehicles. Specifically, according  to the loop center, and $H_g$, which is the vector connecting center $(k_O, q_O)$ and point $(k_g, q_g)$, the cross-product, $H_g\times H_{g+1}$, represents the dynamic variation from Zone $g$ to Zone $g+1$; see Fig. \ref{hysteresis_measure}(b)-(c). The negative cross product represents the CW while the positive represents the CCW orientation; see Fig. \ref{hysteresis_measure}(d). The absolute cross product value is graphically the area of the parallelogram formed by the sides $H_g$ and $H_{g+1}$.  The value reflects the hysteresis magnitude associated with the transition from Zone $g$ to Zone $g+1$. Further, $H_{IE}$, the vector connecting $(k(1), q(1))$ and $(-\frac{q(1)}{w}+k(1),0)$ represents the initial equilibrium; and  $H_{GE}$ connecting the $(k(G), q(G))$ and $(-\frac{q(G)}{w}+k(G),0)$ represents the new equilibrium. We then apply the cross products $H_{IE}\times H_{Ig}$ (Fig. \ref{hysteresis_measure}(d)) and $H_{GE}\times H_{Gg}$ (Fig. \ref{hysteresis_measure}(e)) to measure the amplification or decay of the disturbance in Zone g relative to the initial and the new equilibrium states in Zone 1 and Zone G.  

We further develop a systematic method to determine the hysteresis patterns. We first identify the maximum absolute average magnitude, $|{H_{max}}_0|$ (Fig. \ref{hysteresis_measure}(c)). The hysteresis orientation is tracked by the signs of ${H_{max}}_0$ (i.e., $-$ for CW type, $+$ for CCW type). Then, we determine whether the new equilibrium is higher or lower than the initial equilibrium by the sign of $H_{IE}\times H_{Ig}$ ($-$, lower; $+$, higher). 
$H_{min_1}$(Fig. \ref{hysteresis_measure}(d)) and $H_{max_2}$ (Fig. \ref{hysteresis_measure}(e)) are identified to monitor the disturbance propagation (i.e., $-$ for amplification, $+$ for decay). By tracking the signs of $H_{IE}\times H_{Ig}$, $H_{max}$, $H_{min_1}$ and $H_{max_2}$, it encapsulates the seven hysteresis patterns detailed in Table \ref{eta_pattern}. The CCW example characterized by ${H_{max}}_0>0$, $H_{min_1}<0$ and $H_{max_2}>0$, implies that the disturbance initially amplifies, followed by partial decay, ultimately resulting in an increased throughput.


\subsection{Hysteresis Stochasticity Analysis}
We first investigate whether the stochastically calibrated EAB model can reproduce the observed traffic hysteresis. Specifically, we select three representative particles from the posterior distributions for comparison: (1) deterministic optimal (baseline), (2) the best-fit particle, and (3) the 5th percentile best-fit particles. (1) represents how typical calibration would be done - by finding the set of parameter values that minimizes the overall fit to all training trajectories. With the stochastic approach, (2) is obtained for each testing trajectory from the estimated joint distribution. (3) gives the distribution-wise sense of validation performance. These particles are used to reproduce hysteresis and compare with the empirical ground truth. The metrics for evaluation are $\overline{d_O}$ (Eq. (\ref{hys_center})): average Euclidean distance between the centers of the observed and simulated hysteresis loops ; $\overline{d_{sd}}$ (Eq. (\ref{hys_sd})): average Euclidean distance between SDs of simulated and observed flow and density; $\overline{NRMSE_H}$: average NRMSE of the simulated and observed hysteresis magnitude (Eq. (\ref{hys_magnitude}));

\begin{flalign}\label{hys_center}
\overline{d_o}=\frac{1}{|M_2|}(\sum_{i=1}^{|M_2|}\sqrt{(\frac{\sum_{g=1}^Gk_{i,obs}(g)}{G}-\frac{\sum_{g=1}^Gk_{i,sim}(g)}{G})^2+(\frac{\sum_{g=1}^Gq_{i,obs}(g)}{G}-\frac{\sum_{g=1}^Gq_{i,sim}(g)}{G})^2}
\end{flalign}
\begin{flalign}\label{hys_sd}
\overline{d_{sd}}=\frac{1}{|M_2|}(\sum_{i=1}^{|M_2|}\sqrt{(SD(k_{i,obs})-SD(k_{i,sim}))^2+(SD(q_{i,obs})-SD(q_{i,sim}))^2}
\end{flalign}
\begin{flalign}\label{hys_magnitude}
\overline{NRMSE_H}=\frac{1}{|M_2|}(\sum_{i=1}^{|M_2|}\frac{\sqrt{\frac{1}{G-1}\sum_{g=1}^{G-1} (H_g^{i,obs}\times H_{g+1}^{i,obs}-H_g^{i,sim}\times H_{g+1}^{i,sim})^2}}{\sqrt{\frac{1}{G-1}\sum_{g=1}^{G-1}(H_g^{i,obs}\times H_{g+1}^{i,obs})^2}})
\end{flalign}
where, $|M_2|$ is the cardinality of the testing set $M_2$.

The results in Table \ref{Hysteresis_summary} show that the stochastic EAB performs better than the deterministic counterpart at reproducing traffic hysteresis, judging by smaller values for the four metrics across the board with few exceptions. 

\begin{table}[!h]
\small
	\caption{Summary Statistics: Hysteresis Stochasticity Analysis based on Testing Trajectories}\label{tab:versions}
	\begin{center}
		\begin{tabular}{c|c |c |c c c }\Xhline{1pt}
Speed &Car Model & Particle &$\overline{d_O}$ & $\overline{d_{sd}}$&$\overline{NRMSE_H}$\\\Xhline{1pt} 
\multirow{9}*{Low} &\multirow{3}*{HDV-1}  &Best Fitted &18.13&28.59&1.17\\
    &	   &5 Percentile&25.91&35.43 &1.10\\
	          	& &Deterministic Optimal&162.83&49.03&1.06\\\cline{2-6}
                 &\multirow{3}*{\makecell[c]{Z-1\\(Normal)}} &Best Fitted &10.61&12.63&0.86\\
                 &	   &5 Percentile&23.29&19.52 &0.84\\
&	   	&Deterministic Optimal&44.69&22.32&0.69\\\cline{2-6}
&	   		\multirow{3}*{\makecell[c]{Z-3\\(Power)}} &Best Fitted &14.21&6.70&0.47\\
&	   &5 Percentile&26.70&10.49&0.78\\
&	   	&Deterministic Optimal&26.20&11.17&0.90\\\hline
\multirow{15}*{Median and High}  & \multirow{3}*{HDV-2} &Best Fitted &14.07&28.23&0.92\\
&	   &5 Percentile&55.20&72.83&1.71\\
	&   	&Deterministic Optimal&165.63&74.52&1.68\\\cline{2-6}
	&   			\multirow{3}*{X} &Best Fitted &14.20&6.24&1.01 \\
	 &  &5 Percentile&17.89&12.05&1.01\\
	&   	&Deterministic Optimal&68.19&12.26&0.99\\\cline{2-6}
	&   		\multirow{3}*{\makecell[c]{Y-1\\(Normal)}} &Best Fitted &12.51&15.05&0.70\\
	 &  &5 Percentile&14.37&26.20 &1.27\\
	&   	&Deterministic Optimal&54.17&26.09&0.86\\\cline{2-6}
	 &  \multirow{3}*{\makecell[c]{Y-2\\(Sports)}} &Best Fitted &10.18&5.16&0.85\\
&	   &5 Percentile&13.48&12.90 &0.97\\
&	   	&Deterministic Optimal&26.17&8.93&0.93\\\cline{2-6}
&	   		\multirow{3}*{\makecell[c]{Z-2\\(Normal)}} &Best Fitted &12.30&6.82&1.58\\
&	   &5 Percentile&22.32&20.50&0.99\\
&	   	&Deterministic Optimal&60.87&16.43&1.84\\\cline{2-6}
&	   	\multirow{3}*{\makecell[c]{Z-4\\(Power)}} &Best Fitted &23.97&12.25&1.37\\
&	   &5 Percentile&28.04&21.38&1.72\\
&	   	&Deterministic Optimal&51.74&17.11&1.23\\\Xhline{1pt}
	\end{tabular}
	\end{center}
 \small
 \label{Hysteresis_summary}
\end{table}

\begin{figure}[htbp]
 \captionsetup{justification=centering} 
        \centering
        \begin{subfigure}[b]{0.45\textwidth}
            \centering
     \includegraphics[width=\textwidth]{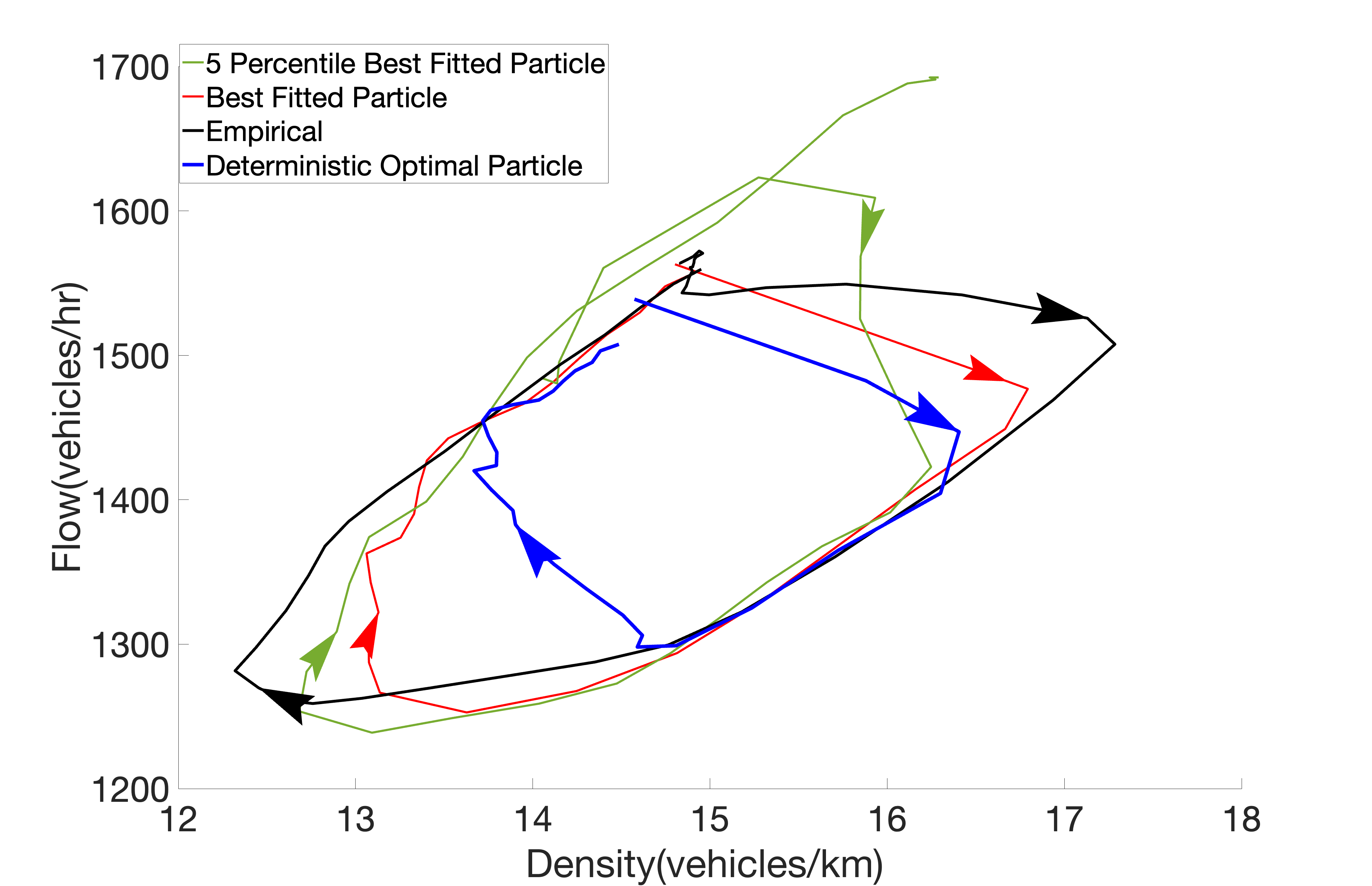}
            \caption[]%
            {{\small}}    
        \end{subfigure}
        \begin{subfigure}[b]{0.45\textwidth}  
            \centering 
    \includegraphics[width=\textwidth]{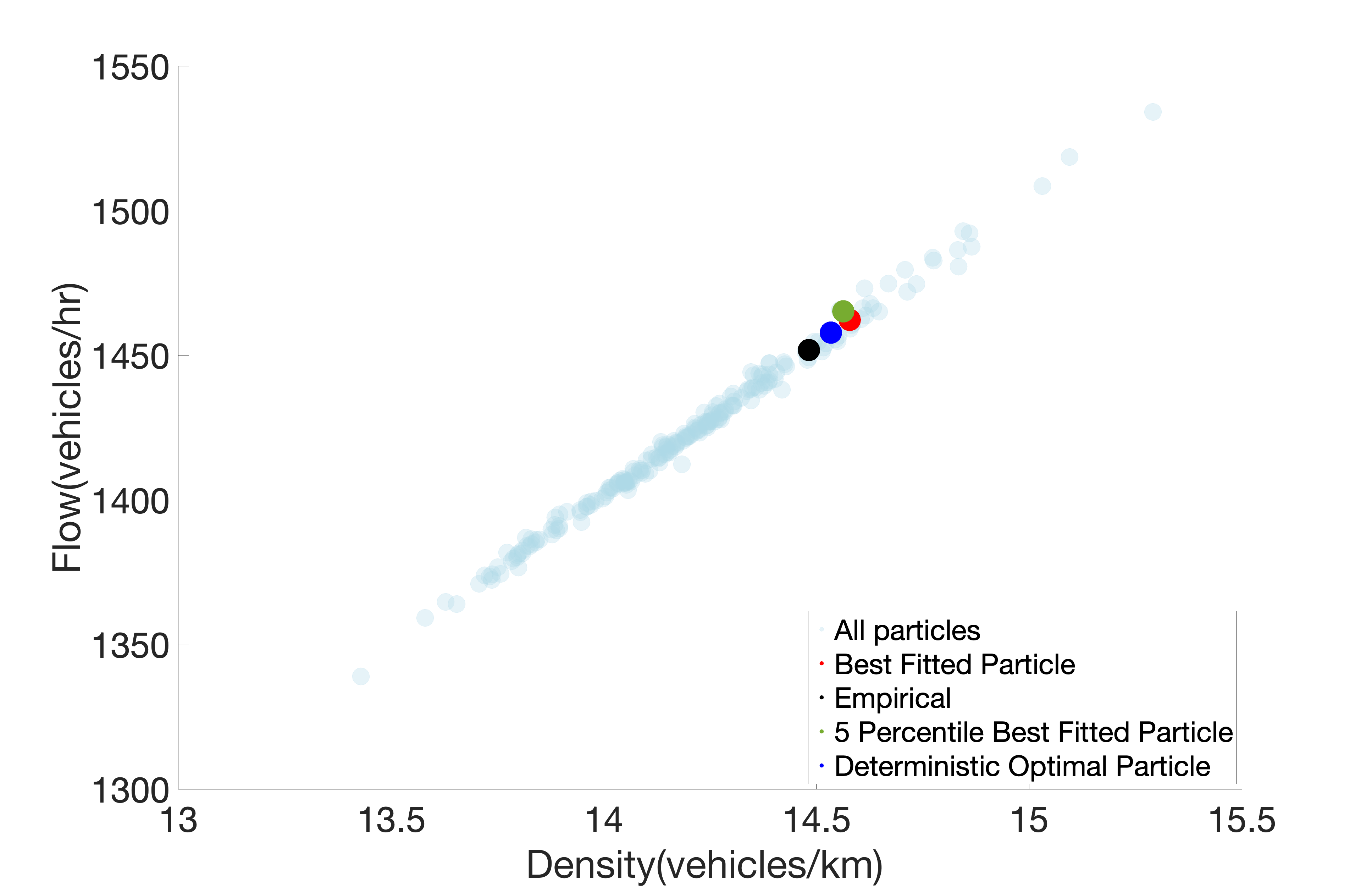}
            \caption[]%
            {{\small }}    
        \end{subfigure}
        \begin{subfigure}[b]{0.45\textwidth} 
            \centering 
        \includegraphics[width=\textwidth]{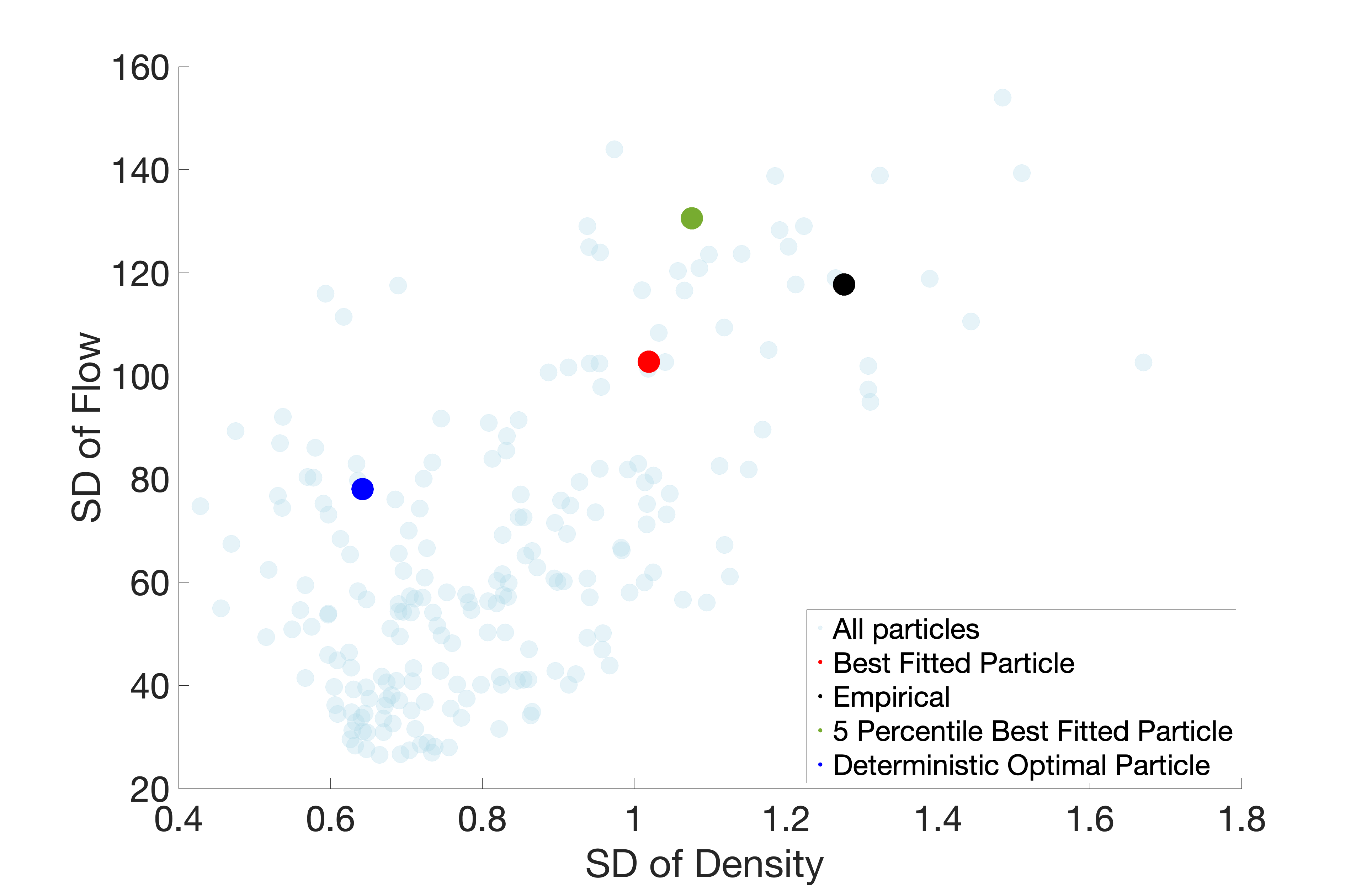}
            \caption[]%
            {{\small}}    
        \end{subfigure}
        \begin{subfigure}[b]{0.45\textwidth}   
            \centering 
            \includegraphics[width=\textwidth]{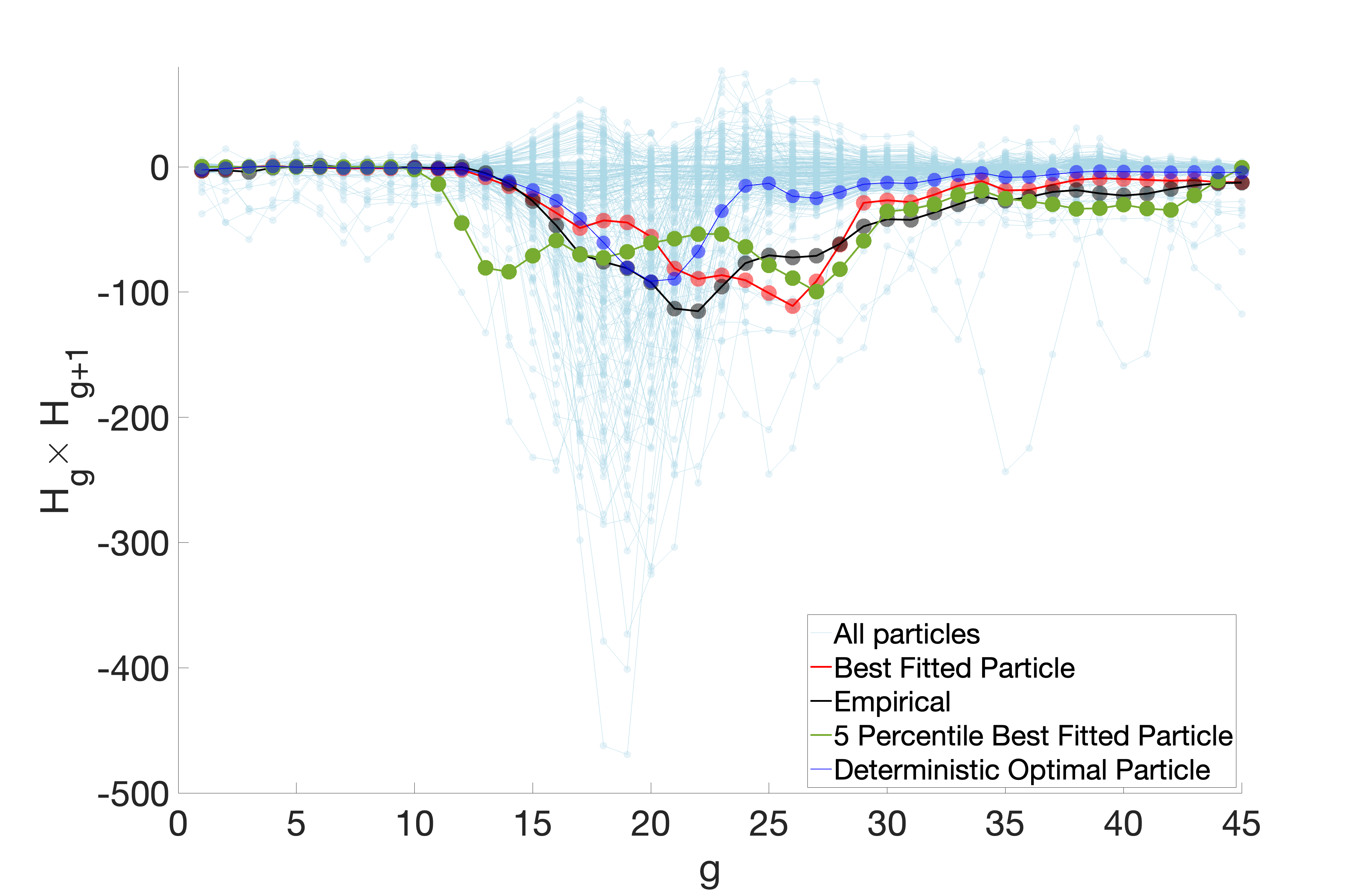}
              \caption[]%
            {{\small}}    
        \end{subfigure}
          \begin{subfigure}[b]{0.45\textwidth}
            \centering
     \includegraphics[width=\textwidth]{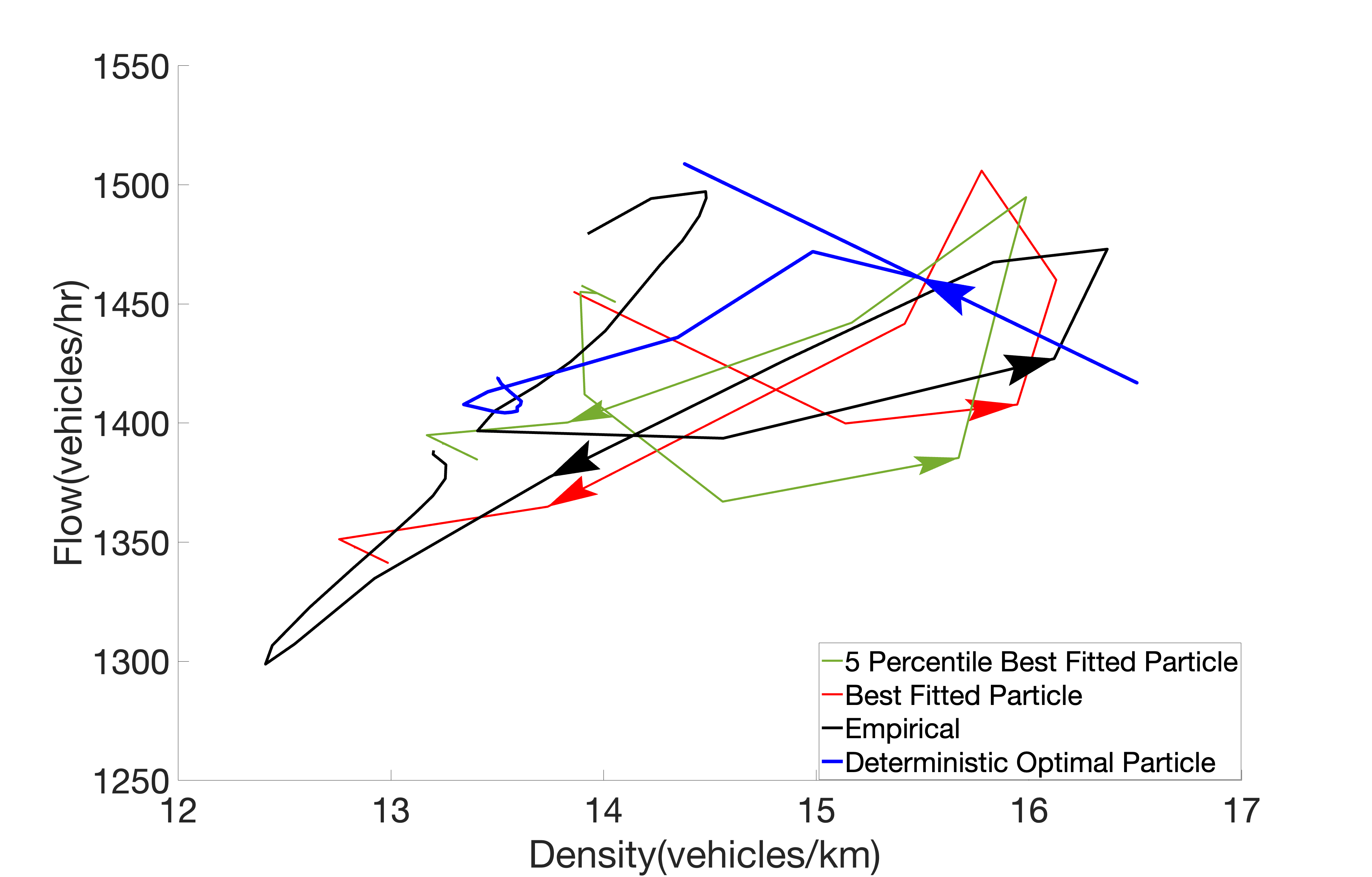}
            \caption[]%
            {{\small}}    
        \end{subfigure}
        \begin{subfigure}[b]{0.45\textwidth}  
            \centering 
    \includegraphics[width=\textwidth]{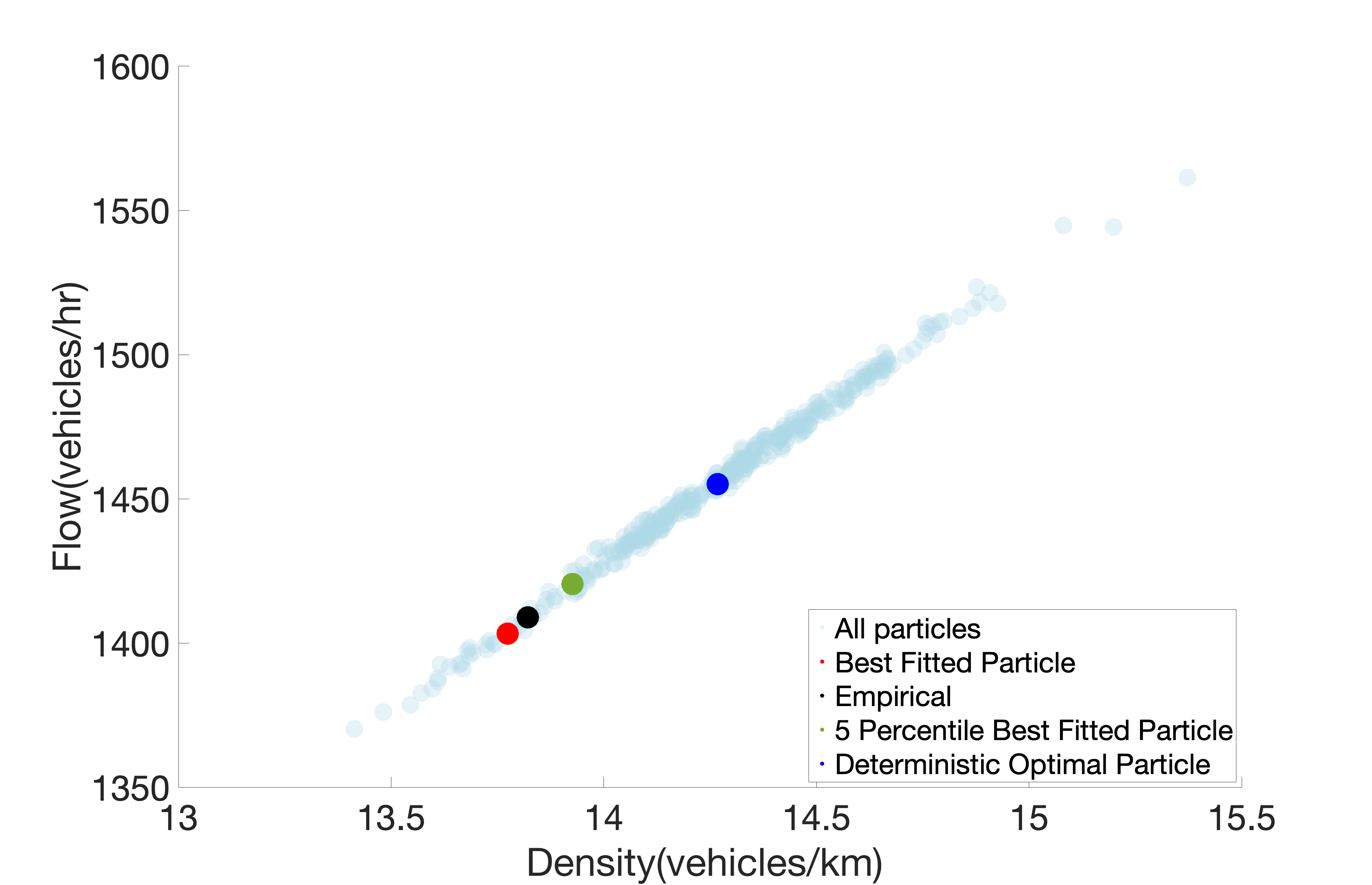}
            \caption[]%
            {{\small }}    
        \end{subfigure}
        \begin{subfigure}[b]{0.45\textwidth} 
            \centering 
        \includegraphics[width=\textwidth]{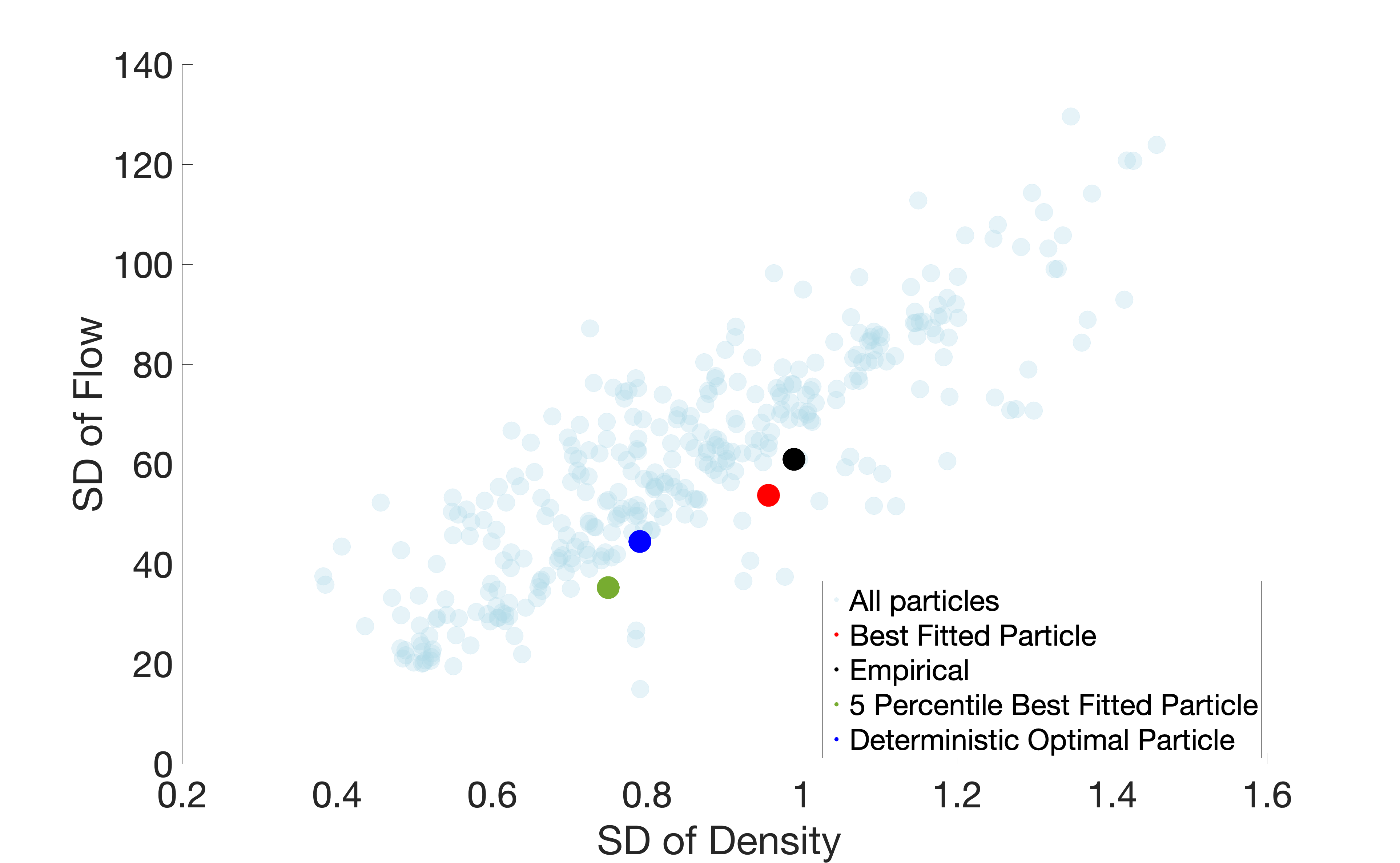}
            \caption[]%
            {{\small}}    
        \end{subfigure}
        \begin{subfigure}[b]{0.45\textwidth}   
            \centering 
            \includegraphics[width=\textwidth]{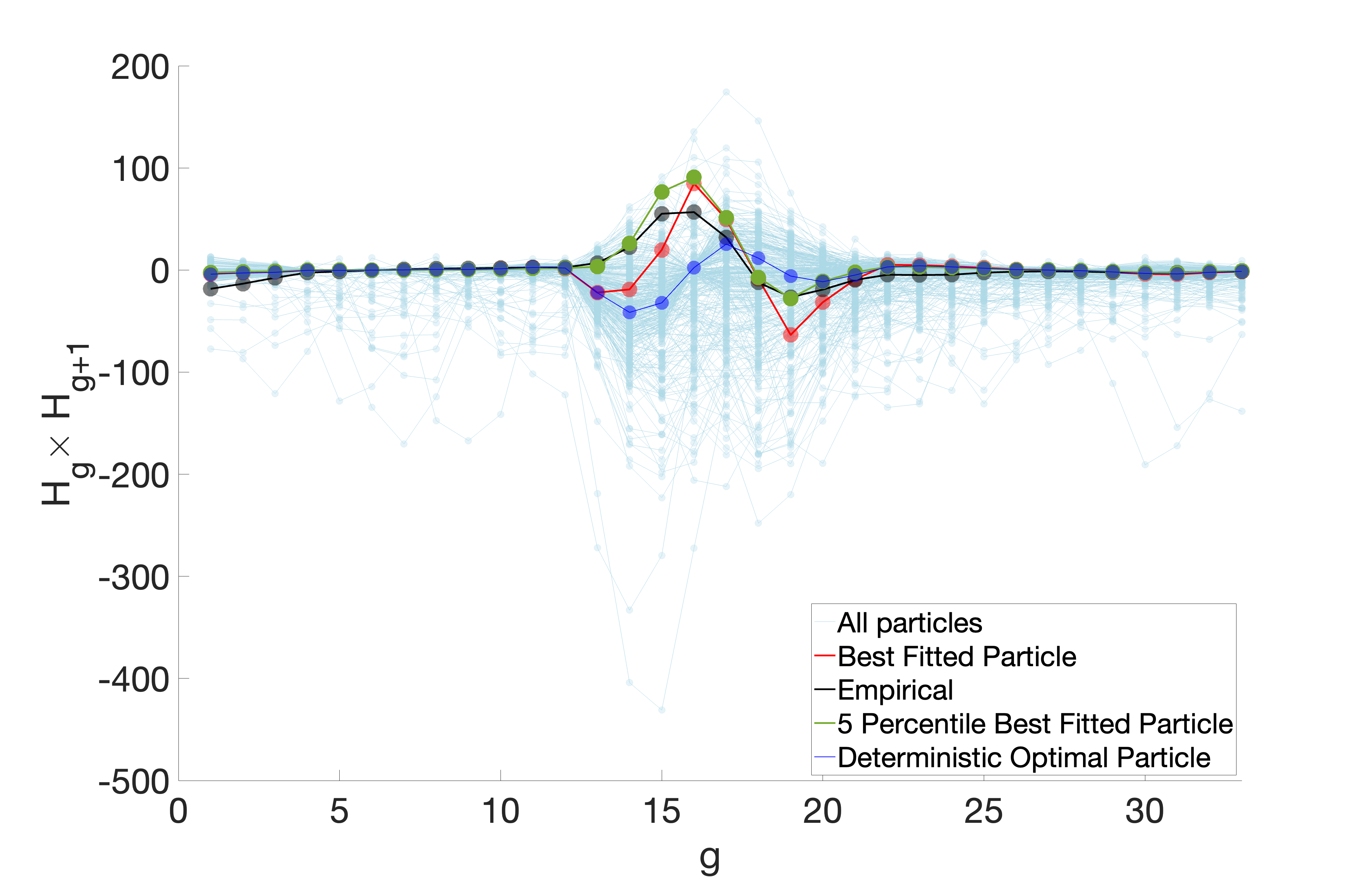}
              \caption[]%
            {{\small}}    
        \end{subfigure}
       \caption[{Hysteresis }]
        {\small Hysteresis Examples\\
       CW-loop: (a) Hysteresis Loop
        (b) Center of Hysteresis 
        (c) SD of Flow and Density
        (d) Cross-product of the Movement;\\
       CCW-loop:
        (e) Hysteresis Loop
        (f) Center of Hysteresis 
        (g) SD of Flow and Density
        (h) Cross-product of the Movement}       
        \label{hysteresis_examples}
      \end{figure}

We then provide two typical examples: CW in Fig. \ref{hysteresis_examples}(a)-(d) and CCW in Fig. \ref{hysteresis_examples}(e)-(h). From Fig. \ref{hysteresis_examples}(a), we can see that all three particles successfully reproduce the CW orientation of the empirical hysteresis (black line). However, the best-fit particle (red) and the 5th percentile best-fit particle (green) produce hysteresis loops that are much closer to the empirical one. The hysteresis loop based on the deterministic particle (blue) is much narrower, signifying underestimation of density variation. Fig. \ref{hysteresis_examples}(e) is a more complicated CCW example. From Fig. \ref{hysteresis_examples}(e), the deterministic optimal particle shows very poor performance at reproducing the dynamic process of the hysteresis formation. This is corroborated in Fig. \ref{hysteresis_examples}(b)-(d) and (f)-(h), where the stochastic estimations perform significantly better than the deterministic one in terms of center (Fig. \ref{hysteresis_examples}(b)(f)), SD of density and flow (Fig. \ref{hysteresis_examples}(c)(g)), and cross product of the movement (Fig. \ref{hysteresis_examples}(d)(h)).  All these findings strongly indicate that the stochastically calibrated EAB model can replicate traffic hysteresis by capturing the stochastic nature of CF dynamics and linking to traffic-level dynamics.

We further examine how the hysteresis patterns differ across HDVs and ACCs in Table \ref{hysteresis_catogrization}. Note that hysteresis $H_g\times H_{g+1}$ is considered significant if its magnitude is greater than a predefined threshold, $H_T$. ${H_T}_0$ and ${H_T}_1$ are used to measure the significance of the disturbance amplification and decay relative to the initial and new equilibrium states (i.e., $H_{IE}\times H_{Ig}$  and  $H_{GE}\times H_{Gg}$). $H_T$, ${H_T}_0$, ${H_T}_1$ are defined based on the difference between 75th and 25th percentiles of $|H_{1}\times H_{2}|$ and $|H_{IE}\times H_{I2}|$ and $|H_{GE}\times H_{G1}|$, respectively. Similar to the results for the behavioral pattern, significant differences between HDVs and ACC vehicles are observed, and to a lesser extent, across ACC vehicles and within the same ACC developer. For HDVs vs. ACCs, differences are particularly notable at medium to high speeds. For instance, HDVs predominantly show $CW^-$ patterns ($>0.42$) and $CCW^-$ ($>0.18$). ACC vehicles display a significantly lower proportion for $CW^-$ and $CCW^-$ patterns. Instead, they display higher proportions for $CW$ patterns. Across ACC vehicles, variations are also notable. For example, Model X exhibits the highest occurrence of NSL pattern and lowest occurrence of $CW$ patterns.  Within the same ACC developers with different engine modes, Model Y and Model Z at median to high speeds exhibit largely similar patterns.  However, some differences with more $CW^-$ and fewer $CW$ patterns are notable at low speeds with Model Z.

\begin{table}[H]
\small
	\caption{ Categorization: Proportion of Empirical Hysteresis Patterns under $H_T$}\label{tab:versions}
	\centering
		\begin{tabular}{c| c |c c| c| c c c c c  }\Xhline{1pt}
   Speed&\multicolumn{3}{c|}{Low}&\multicolumn{6}{c}{Median and High}\\ \Xhline{1pt} 
    $H_T$&400 &\multicolumn{2}{c|}{15}& 400&\multicolumn{5}{c}{15}\\ \Xhline{1pt}
      ${H_T}_0$&21700 &\multicolumn{2}{c|}{4770}& 21700&\multicolumn{5}{c}{4770}\\ \Xhline{1pt}
      ${H_T}_1$&36700&\multicolumn{2}{c|}{8460}& 36700&\multicolumn{5}{c}{8460}\\ \Xhline{1pt}
Car Model   & \multirow{2}*{HDV}   &\multicolumn{2}{c|}{Z}  & \multirow{2}*{HDV}  &\multirow{2}*{X}     &\multicolumn{2}{c}{Y}  &\multicolumn{2}{c}{Z}\\ 
Engine    &  &Normal    &Power    &  &     &Normal    &Sports    &Normal    &Power\\ \Xhline{1pt}
NSL     & 0.19 & 0    & 0    & 0.26   & 0.27 & 0.06 & 0.04 & 0.03 & 0 \\\hline
CW$^+$  & 0.13 & 0.19 & 0.13 & 0.08   & 0.10 & 0.17 & 0.19 & 0.13 & 0.09 \\\hline
CW$^-$  & 0.44 & 0.75 & 0.75 & 0.42   & 0.46 & 0.40 & 0.52 & 0.47 & 0.41 \\\hline
CW      & 0    & 0    & 0.13 & 0.04   & 0.08 & 0.33  & 0.21 & 0.19 & 0.34 \\\hline
CCW$^+$ & 0    & 0    & 0    & 0.02   & 0.02 & 0    & 0.04 & 0.03    & 0.09 \\\hline
CCW$^-$ & 0.25 & 0.06 & 0    & 0.18   & 0.06 & 0.04 & 0    & 0.16 & 0.03 \\\hline
CCW     & 0    & 0    & 0    & 0.02   & 0    & 0    & 0    & 0 & 0.03 
\\\Xhline{1pt}
\end{tabular}
\small
\label{hysteresis_catogrization}
\end{table}

The hysteresis analysis results are consistent with those of the reaction pattern analysis in section 3. From the findings, several conclusions can be drawn: (1) Variations in CF dynamics directly influence the variations in traffic dynamics. (2) ACCs impact traffic differently than HDVs stemming from significantly different CF dynamics.  (3) ACC vehicles exhibit notable heterogeneity among themselves. (4) Even the same ACC vehicle model displays differences in CF and traffic dynamics by engine mode, particularly at low speeds. (5) The speed contributes to heterogeneity in CF and traffic dynamics.

\section{Mixed Platoon Behavior}
In this section, we investigate how the heterogeneity in CF dynamics and hysteresis patterns scale up to the platoon behavior to obtain more direct insight into the mixed traffic dynamics. We investigate (1) the disturbance evolution through a platoon and (2) the accompanying hysteresis characteristics with respect to the ACC penetration rate. 

We simulate a 20-vehicle platoon according to the calibrated stochastic EAB framework with varying ACC market penetration rates ($0-100\%$). The first vehicle trajectory is taken from the real data (i.e, Car Model Z-normal engine for low speed and  Car Model X for high speed). Under the same leading trajectory, 19 followers are simulated to form a mixed platoon using the posterior joint distributions for HDVs and ACC vehicles given the penetration rate. Note that we control for the speed range and engine mode as per our findings in Section 3 that they can induce different CF dynamics.

\begin{figure*}[!h]
 \captionsetup{justification=centering} 
\centering
\includegraphics[width=\textwidth]{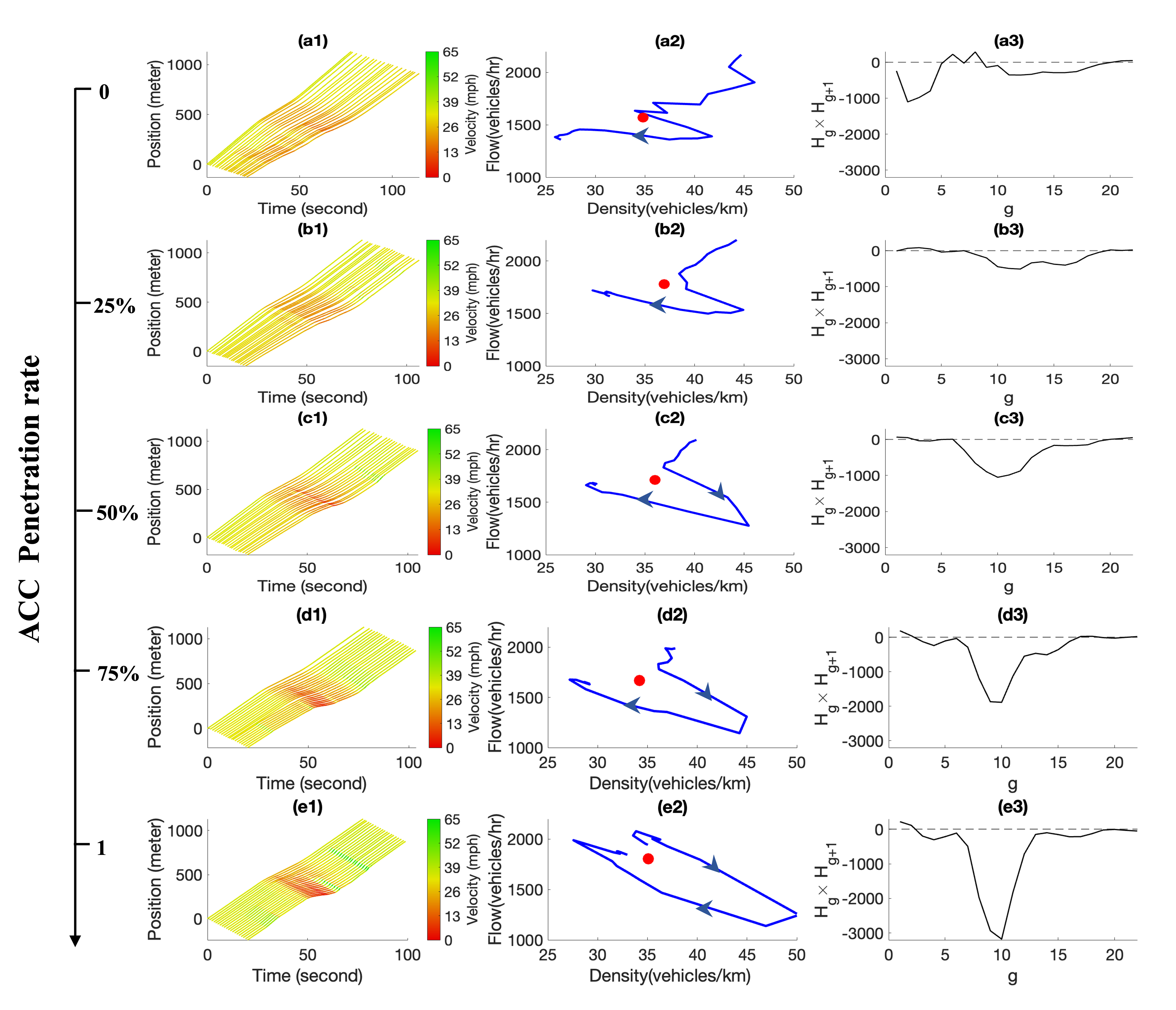}  
\caption[{Mixed Example}]
        {\small Low Speed: Disturbance Propagation and Hysteresis Variation under different ACC penetration rates \\}
        \label{low_mixed_example}
    \end{figure*}

\begin{figure*}[!h]
 \captionsetup{justification=centering} 
\centering
\includegraphics[width=\textwidth]{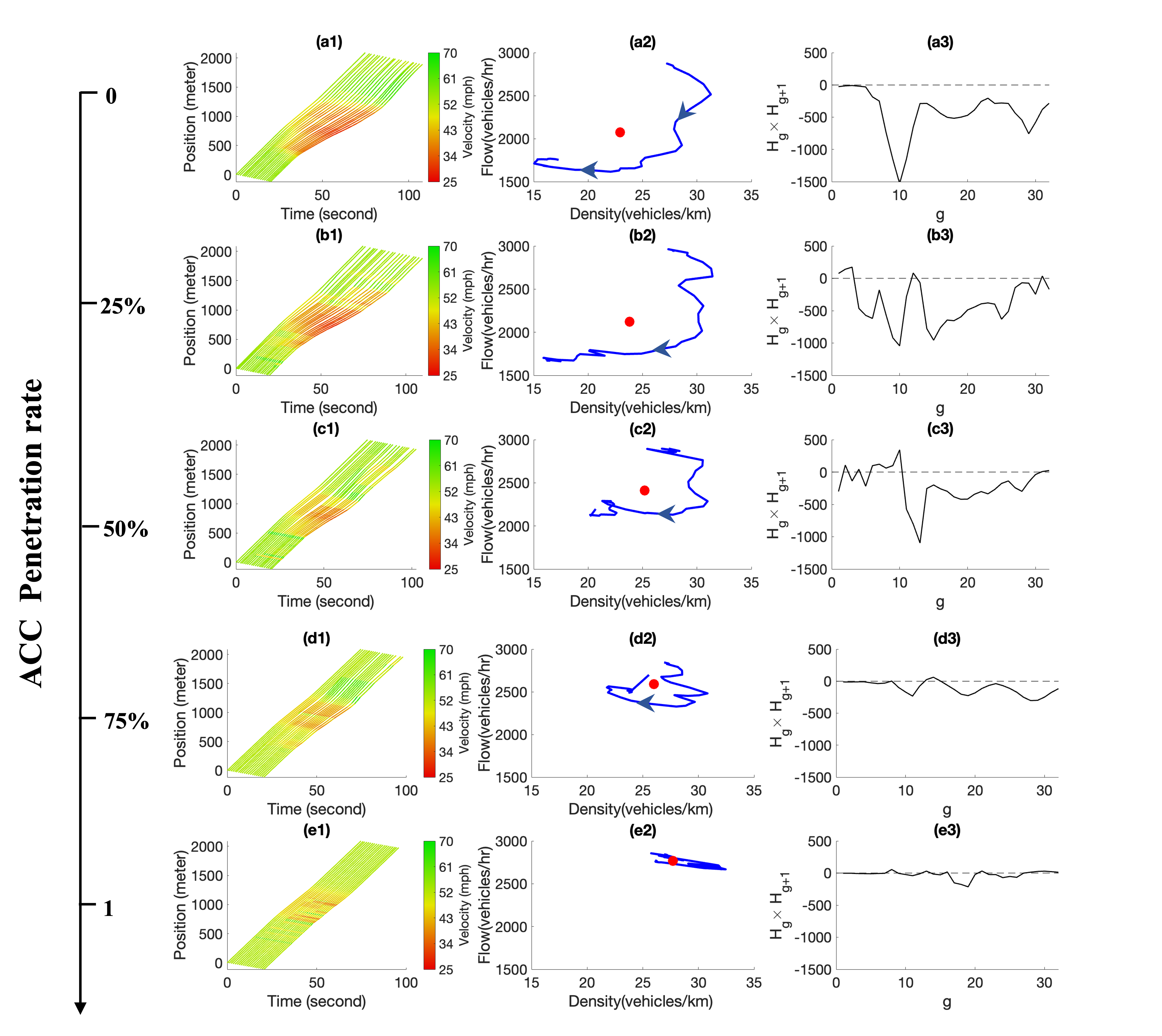}  
\caption[{Mixed Example}]
        {\small Median and high Speed: Disturbance Propagation and Hysteresis Variation under different ACC penetration rates \\}
        \label{mixed_example}
    \end{figure*}

Fig. \ref{low_mixed_example} and \ref{mixed_example} provide two typical examples (Car Model Z - Normal at low speed and Car Model Y-Normal at median and high speed) to investigate the changes in the platoon-level disturbance propagation (the left column, (a1)-(e1)), hysteresis orientation (middle column, (a2)-(e2)), and hysteresis magnitude (right column, (a3)-(e3)) with the ACC penetration rate. We obtained the following observations. 

In low speed (Fig. \ref{low_mixed_example}), the new equilibrium state reaches closer to the original state (i.e., lower incidence of $CW^{-}$) with increasing ACC penetration, albeit at the increasing hysteresis magnitude. This trend can be explained by the higher frequency of non-decreasing patterns in HDVs, which leads to a reduction in traffic throughput. In contrast, ACCs exhibit a higher frequency of concave pattern, combined with a greater deviation of $\eta$. This leads to a substantial deviation from the initial equilibrium but a closer return to the initial equilibrium after a disturbance. This pattern signifies a more pronounced disturbance propagation, where a larger hysteresis magnitude indicates a notable reduction in the average speed for ACCs during the disturbance.

In median and high speed, a significant reduction of disturbance is notable with higher ACC penetration; see the left column of Fig. \ref{mixed_example}. Further, the hysteresis loop becomes much smaller and complete with increasing penetration of ACC vehicles (the middle and right columns of the figure), indicating lower throughput reduction and disturbance magnitude. This is attributed to the higher incidence of convex and non-increasing patterns in ACC vehicles.

\begin{figure}[!h]
\captionsetup{justification=centering} 
        \centering
        \begin{subfigure}[b]{0.45\textwidth}  
            \centering 
    \includegraphics[width=\textwidth]{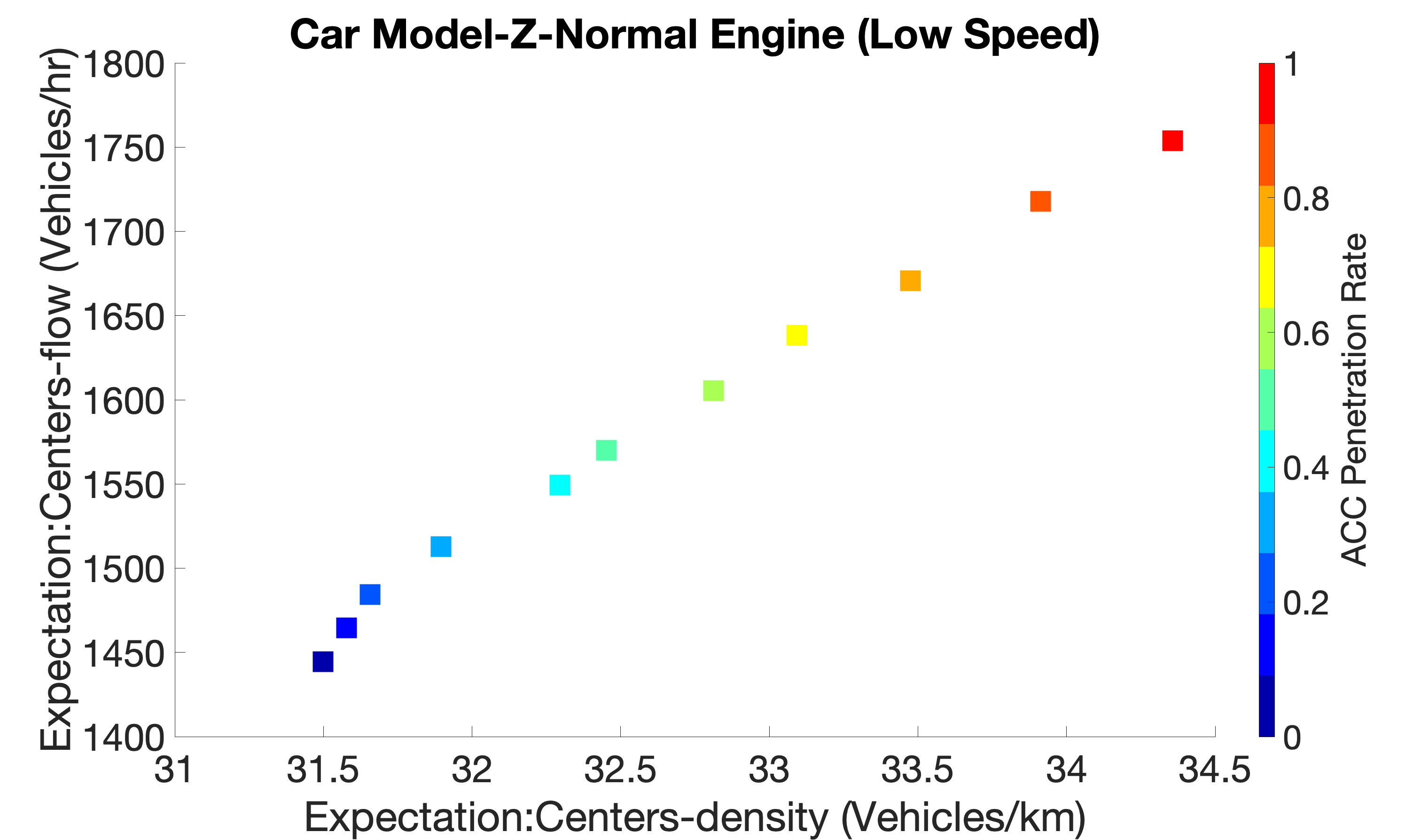}
           \caption[*(a1)]%
            {{\small }}    
        \end{subfigure} 
          \centering
        \begin{subfigure}[b]{0.45\textwidth}  
            \centering 
    \includegraphics[width=\textwidth]{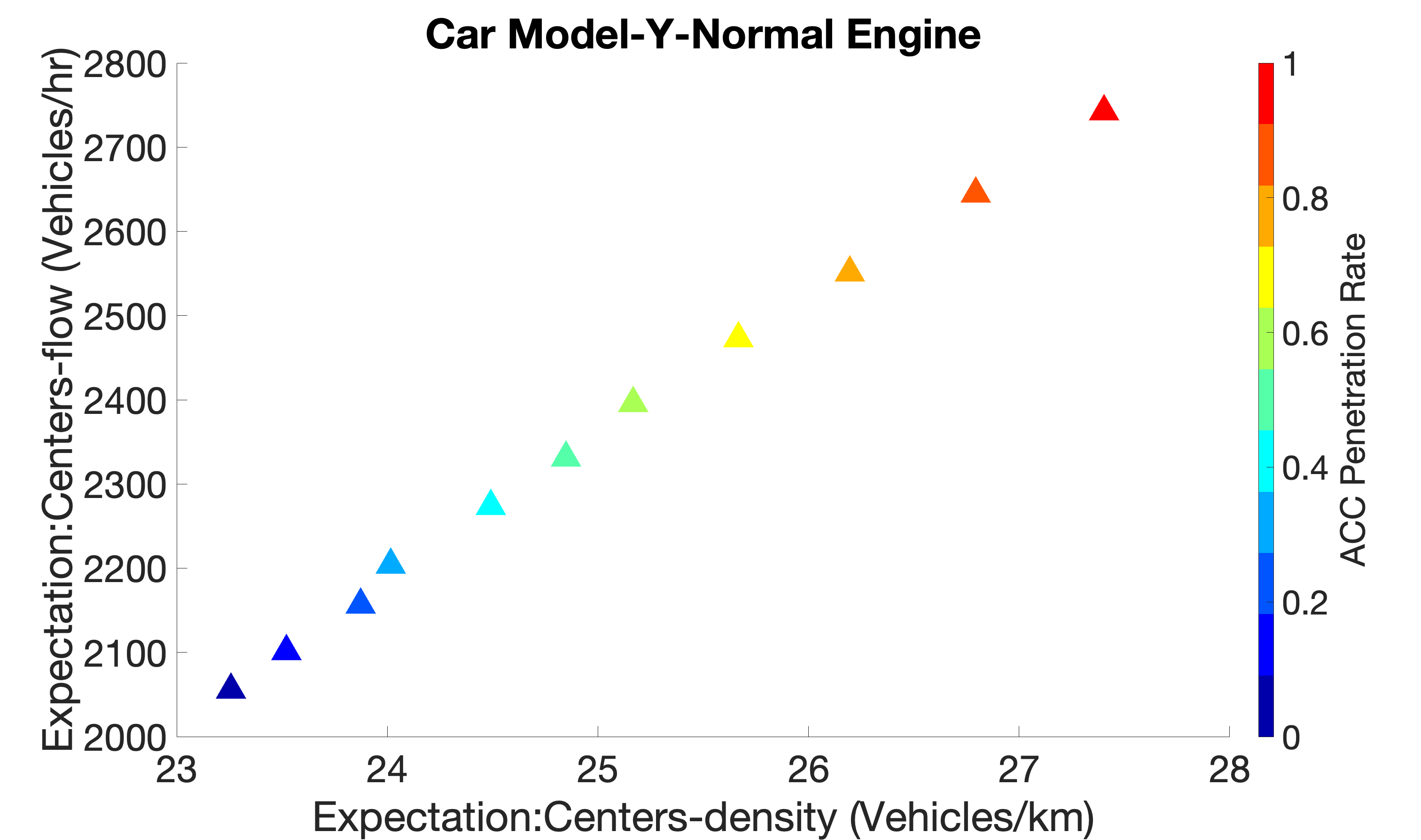}
           \caption[(a2)]%
            {{\small }}    
        \end{subfigure} 
        \begin{subfigure}[b]{0.45\textwidth}
            \centering
     \includegraphics[width=\textwidth]{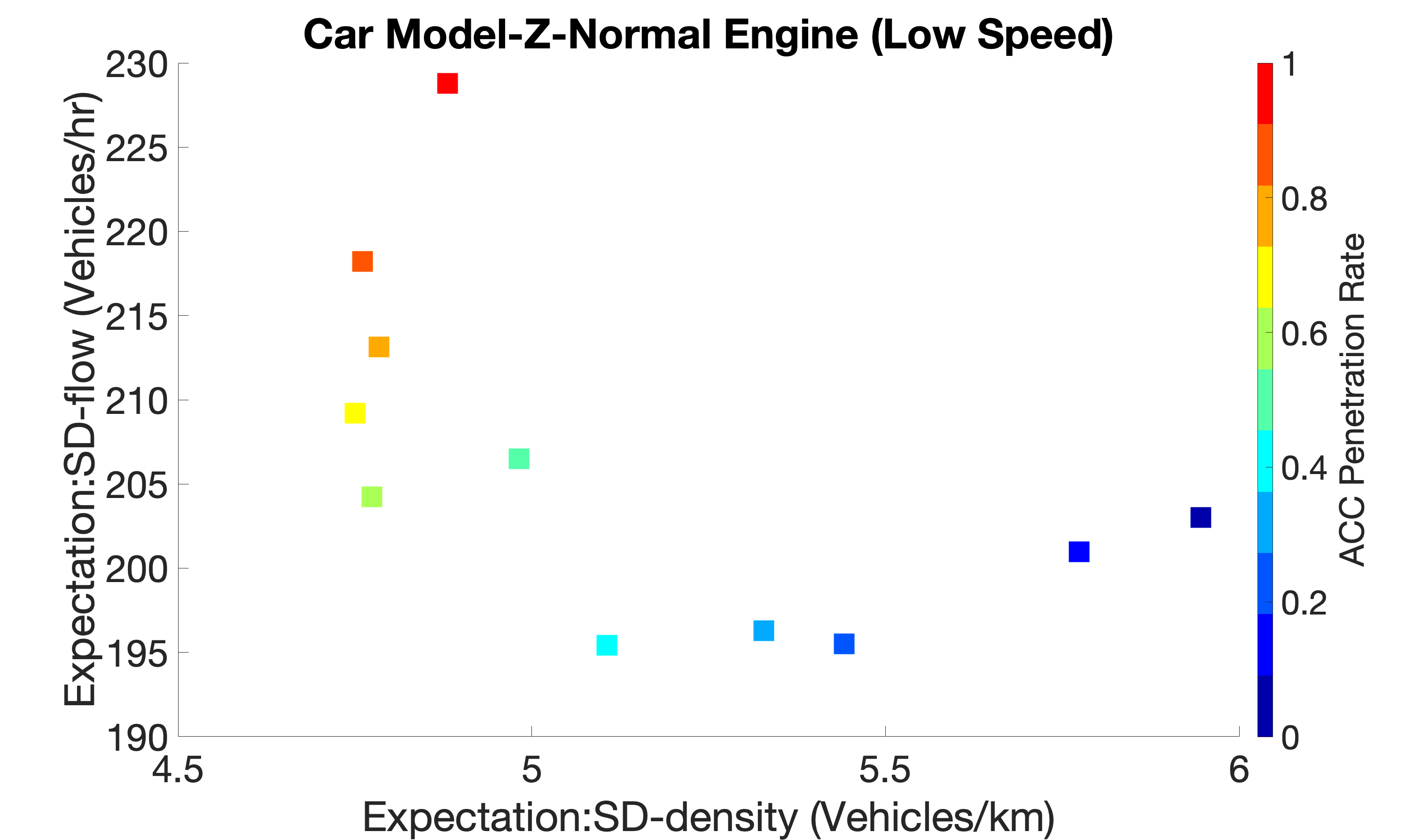}
            \caption[(b1)]%
            {{\small}}    
        \end{subfigure}   
          \begin{subfigure}[b]{0.45\textwidth}
            \centering
     \includegraphics[width=\textwidth]{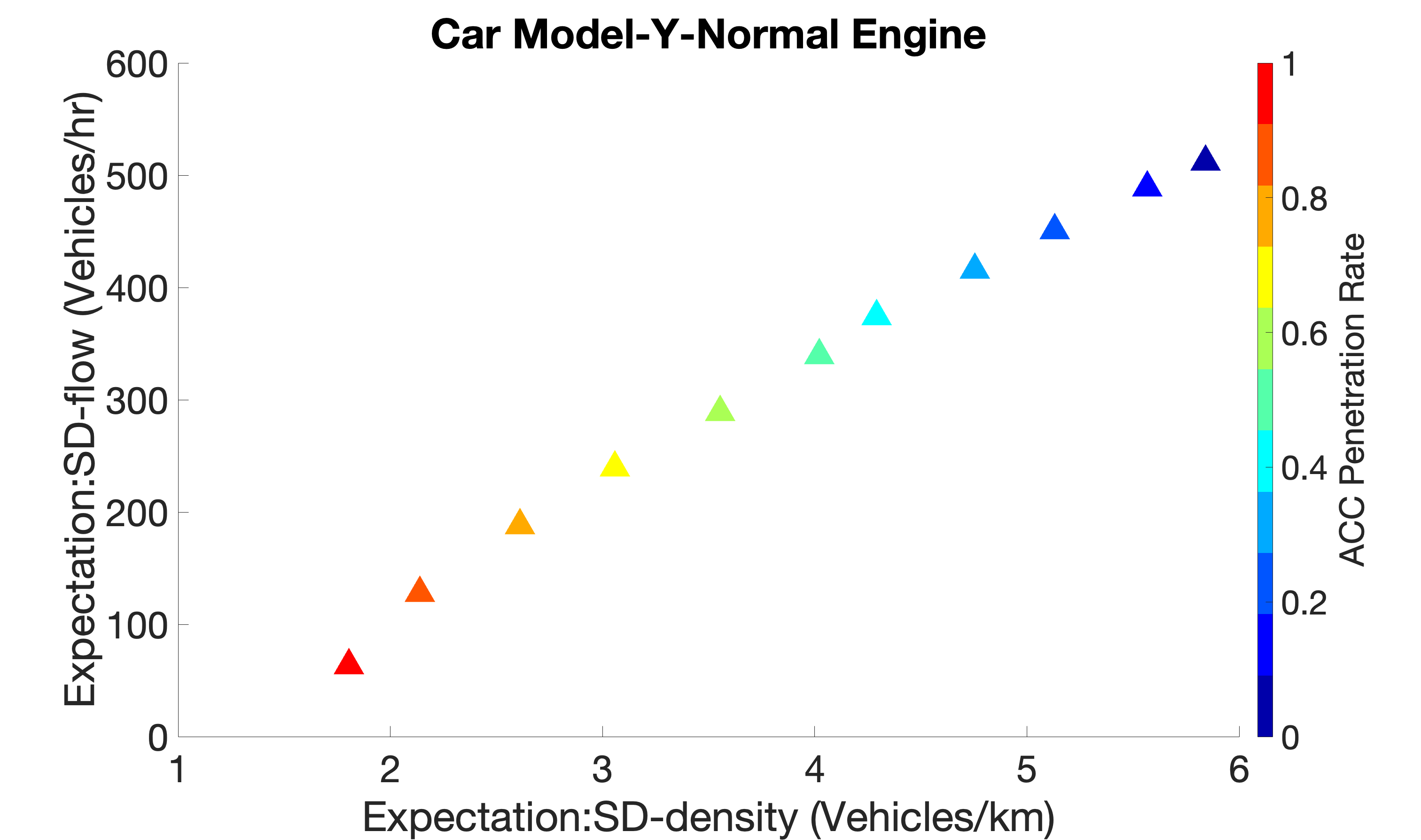}
            \caption[(b2)]%
            {{\small}}    
        \end{subfigure} 
         \begin{subfigure}[b]{0.45\textwidth}  
           \centering 
    \includegraphics[width=\textwidth]{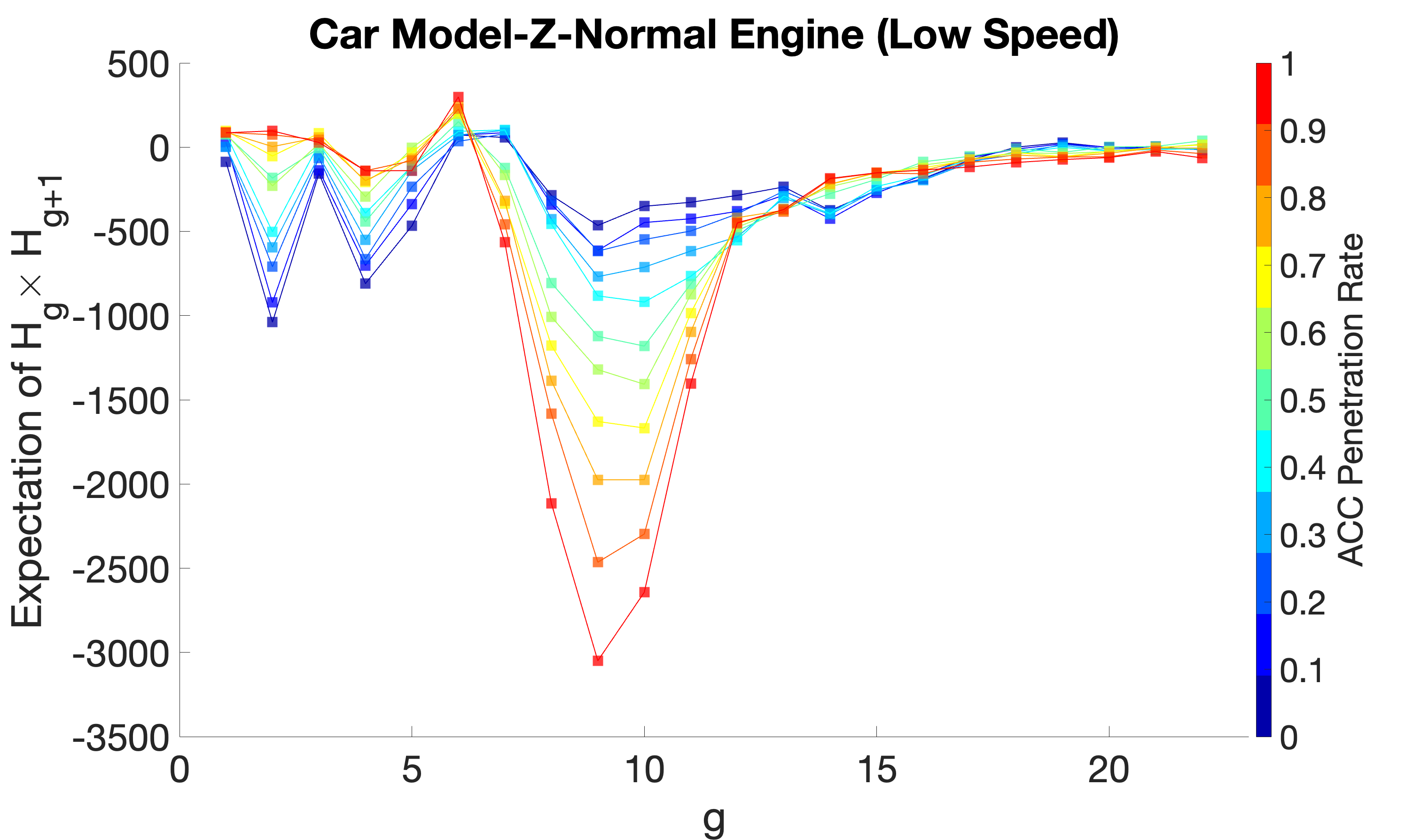}
            \caption[]%
            {{\small }}    
        \end{subfigure} 
        \begin{subfigure}[b]{0.45\textwidth}
            \centering
     \includegraphics[width=\textwidth]{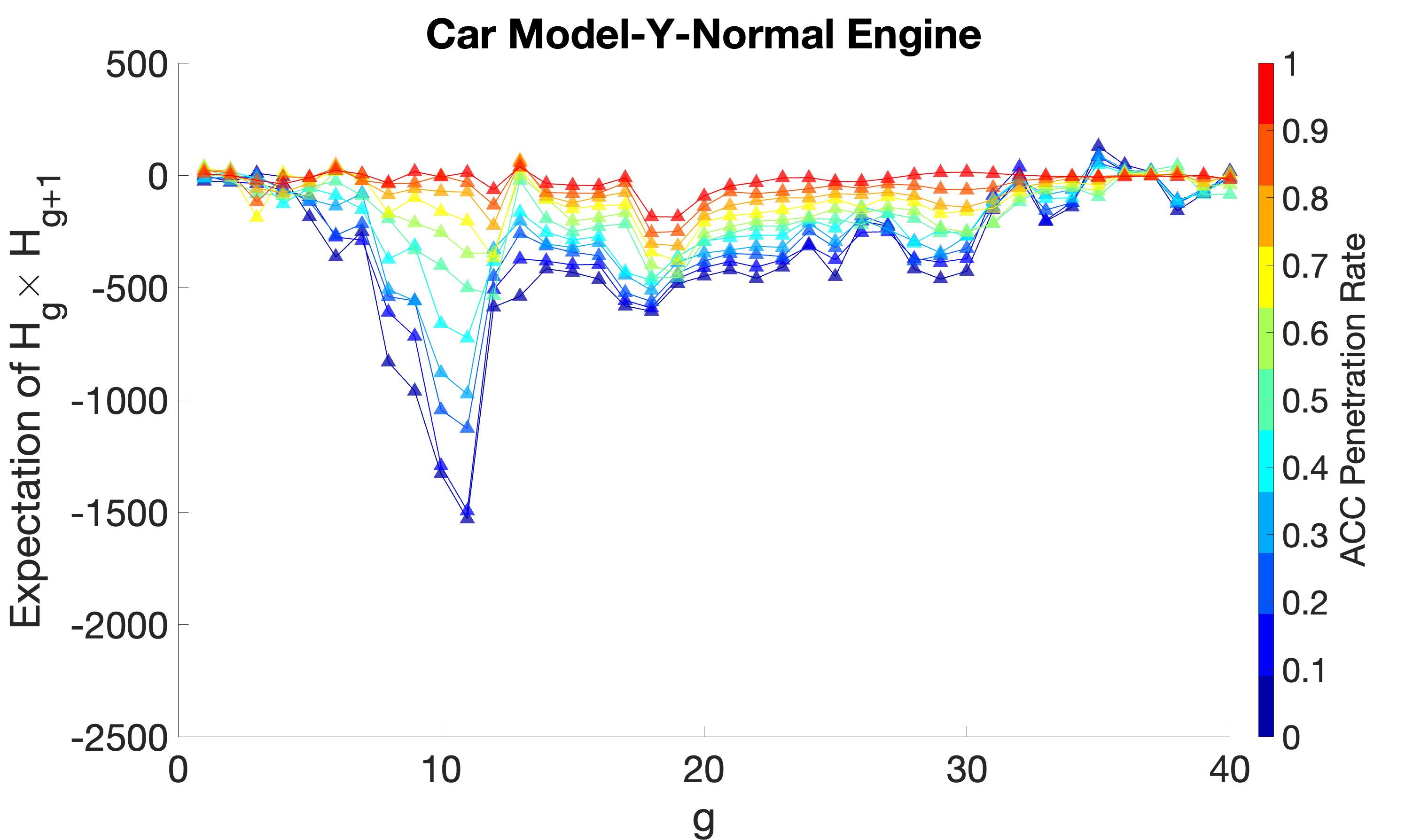}
            \caption[]%
            {{\small}}    
        \end{subfigure}   
        \caption[{Expectation}]
       {\small Hysteresis Expectation-ACC penetration rates over 200 simulations \\ 
      Low speed: Car Model Z - Normal Engine: (a) Centers (c) SD  (e) Magnitude \\
      Median and high speed: Car Model Y - Normal Engine: (b) Centers (d) SD  (f) Magnitude} 
        \label{expectation}
      \end{figure}

The resulting hysteresis characteristics, specifically the expectation of centers, SDs, hysteresis magnitude and hysteresis loop over 200 simulations, with respect to the ACC penetration rate are presented in Fig. \ref{expectation}. Again, Model Z-Normal Engine and Model Y- Normal Engine are presented as the representative examples for the low speed range and the median and high speed range, respectively, as the trends are qualitatively consistent across different ACCs within the same speed range. The hysteresis center shows a consistently increasing trend for both speeds. But the SD of flows when at the low speed shows an increasing trend in Fig. \ref{expectation}(c), resulting in a larger hysteresis magnitude in Fig. \ref{expectation}(e).

From the findings, several conclusions can be drawn: (1) the heterogeneity and stochasticity observed from CF dynamics and traffic dynamics will ultimately result in a reduced throughput. (2) The speed is a significant contributor to the ACC heterogeneity in the platoon-level performance. (3) With increasing penetration of ACC vehicles, the mixed platoon exhibits more complete hysteresis loops compared to HDVs, indicating lower throughput reduction.

\section{Conclusion}
This paper developed a stochastic unifying behavioral CF model, EAB model, to approximate the CF behaviors of commercial ACC vehicles. The proposed approach is developed from a more interpretable, CF behavioral perspective, rather than a control theory-based approach. Specifically, the main advantages are that it provides (1) a common platform to compare different ACC vehicles and discern behavioral differences from HDVs and among ACC developers, engine modes, and speeds, and (2) a direct link between the CF behavior of ACC vehicles and traffic-level features such as traffic hysteresis. 

Further, the stochastic treatment of the EAB model enables characterization of uncertainties originating from vehicle dynamics, model mismatch, potential switch of control logic, etc. Specifically, we applied ABC-ASMC that enables efficient estimation of the joint distribution of the EAB model parameters without having to specify a likelihood function. The calibration results demonstrated that our algorithm is quickly-converging and robust, and can reproduce vehicle trajectories and reaction patterns to disturbances well. The results also suggested that the stochastic treatment is more descriptive for behavior patterns than the deterministic approach. Further, the EAB model can capture composite behavior patterns (i.e., convex-concave and convex-concave), which are not captured by the AB model. We found that there are notable distribution-wise differences in the behavioral patterns between ACC vehicles and HDVs, as well as across different ACC developers, engines, and speed ranges, albeit to a lesser degree.  

Connecting directly to the traffic-level dynamics, we investigated how heterogeneous CF dynamics manifests itself in throughput reduction and traffic hysteresis, important traffic phenomena linked to traffic throughput and stability. To this end, we established a systematic framework to determine the hysteresis orientation and quantify its magnitude. We found that our stochastic EAB approach is capable of reproducing empirical traffic hysteresis. The stochastic treatment overwhelmingly outperforms the deterministic one, further justifying our approach.


As expected, significant heterogeneity in CF dynamics leads to heterogeneity in traffic hysteresis in terms of orientation and magnitude. In comparison to HDVs, the hysteresis loop is more complete with ACC vehicles, implying lower throughput reduction. This was corroborated in mixed traffic simulation experiments, where throughput reduction decreased with higher penetration of ACC vehicles. Likewise, the hysteresis magnitude decreased with higher penetration of ACC vehicles in the median-high speed range; however, the trend was opposite in low speed. 

Some future studies are desired. The current work is limited to a specific traffic scenario (e.g., one stop-and-go disturbance) and should be expanded to a wide range of scenarios (e.g., multiple disturbances) consistent with real-world traffic systems. Further, the findings related to the variations in CF dynamics and traffic hysteresis are specific to the data used in this study and can thus change in the future as the technology develops. The proposed analysis framework can lead to new insights as more data become available in the future.

\section*{Acknowledgements}
This research was sponsored by the US National Science Foundation (Award CMMI 1932932 and 1932921).

\bibliographystyle{elsarticle-harv}
\biboptions{authoryear}
\bibliography{references.bib}

\begin{thebibliography}{43}
\expandafter\ifx\csname natexlab\endcsname\relax\def\natexlab#1{#1}\fi
\providecommand{\url}[1]{\texttt{#1}}
\providecommand{\href}[2]{#2}
\providecommand{\path}[1]{#1}
\providecommand{\DOIprefix}{doi:}
\providecommand{\ArXivprefix}{arXiv:}
\providecommand{\URLprefix}{URL: }
\providecommand{\Pubmedprefix}{pmid:}
\providecommand{\doi}[1]{\href{http://dx.doi.org/#1}{\path{#1}}}
\providecommand{\Pubmed}[1]{\href{pmid:#1}{\path{#1}}}
\providecommand{\bibinfo}[2]{#2}
\ifx\xfnm\relax \def\xfnm[#1]{\unskip,\space#1}\fi
\bibitem[{Ahn et~al.(2013)Ahn, Vadlamani and Laval}]{ahn2013method}
\bibinfo{author}{Ahn, S.}, \bibinfo{author}{Vadlamani, S.},
  \bibinfo{author}{Laval, J.}, \bibinfo{year}{2013}.
\newblock \bibinfo{title}{A method to account for non-steady state conditions
  in measuring traffic hysteresis}.
\newblock \bibinfo{journal}{Transportation Research Part C: Emerging
  Technologies} \bibinfo{volume}{34}, \bibinfo{pages}{138--147}.
\bibitem[{Beaumont et~al.(2009)Beaumont, Cornuet, Marin and
  Robert}]{beaumont2009adaptive}
\bibinfo{author}{Beaumont, M.A.}, \bibinfo{author}{Cornuet, J.M.},
  \bibinfo{author}{Marin, J.M.}, \bibinfo{author}{Robert, C.P.},
  \bibinfo{year}{2009}.
\newblock \bibinfo{title}{Adaptive approximate bayesian computation}.
\newblock \bibinfo{journal}{Biometrika} \bibinfo{volume}{96},
  \bibinfo{pages}{983--990}.
\bibitem[{Besselink and Johansson(2017)}]{besselink2017string}
\bibinfo{author}{Besselink, B.}, \bibinfo{author}{Johansson, K.H.},
  \bibinfo{year}{2017}.
\newblock \bibinfo{title}{String stability and a delay-based spacing policy for
  vehicle platoons subject to disturbances}.
\newblock \bibinfo{journal}{IEEE Transactions on Automatic Control}
  \bibinfo{volume}{62}, \bibinfo{pages}{4376--4391}.
\bibitem[{Chen et~al.(2012a)Chen, Laval, Zheng and Ahn}]{chen2012behavioral}
\bibinfo{author}{Chen, D.}, \bibinfo{author}{Laval, J.},
  \bibinfo{author}{Zheng, Z.}, \bibinfo{author}{Ahn, S.},
  \bibinfo{year}{2012}a.
\newblock \bibinfo{title}{A behavioral car-following model that captures
  traffic oscillations}.
\newblock \bibinfo{journal}{Transportation research part B: methodological}
  \bibinfo{volume}{46}, \bibinfo{pages}{744--761}.
\bibitem[{Chen et~al.(2012b)Chen, Laval, Ahn and Zheng}]{chen2012microscopic}
\bibinfo{author}{Chen, D.}, \bibinfo{author}{Laval, J.A.},
  \bibinfo{author}{Ahn, S.}, \bibinfo{author}{Zheng, Z.},
  \bibinfo{year}{2012}b.
\newblock \bibinfo{title}{Microscopic traffic hysteresis in traffic
  oscillations: A behavioral perspective}.
\newblock \bibinfo{journal}{Transportation Research Part B: Methodological}
  \bibinfo{volume}{46}, \bibinfo{pages}{1440--1453}.
\bibitem[{Chiabaut et~al.(2010)Chiabaut, Leclercq and
  Buisson}]{chiabaut2010heterogeneous}
\bibinfo{author}{Chiabaut, N.}, \bibinfo{author}{Leclercq, L.},
  \bibinfo{author}{Buisson, C.}, \bibinfo{year}{2010}.
\newblock \bibinfo{title}{From heterogeneous drivers to macroscopic patterns in
  congestion}.
\newblock \bibinfo{journal}{Transportation Research Part B: Methodological}
  \bibinfo{volume}{44}, \bibinfo{pages}{299--308}.
\bibitem[{Csill{\'e}ry et~al.(2010)Csill{\'e}ry, Blum, Gaggiotti and
  Fran{\c{c}}ois}]{csillery2010approximate}
\bibinfo{author}{Csill{\'e}ry, K.}, \bibinfo{author}{Blum, M.G.},
  \bibinfo{author}{Gaggiotti, O.E.}, \bibinfo{author}{Fran{\c{c}}ois, O.},
  \bibinfo{year}{2010}.
\newblock \bibinfo{title}{Approximate bayesian computation (abc) in practice}.
\newblock \bibinfo{journal}{Trends in ecology \& evolution}
  \bibinfo{volume}{25}, \bibinfo{pages}{410--418}.
\bibitem[{Del~Moral et~al.(2012)Del~Moral, Doucet and Jasra}]{del2012adaptive}
\bibinfo{author}{Del~Moral, P.}, \bibinfo{author}{Doucet, A.},
  \bibinfo{author}{Jasra, A.}, \bibinfo{year}{2012}.
\newblock \bibinfo{title}{An adaptive sequential monte carlo method for
  approximate bayesian computation}.
\newblock \bibinfo{journal}{Statistics and computing} \bibinfo{volume}{22},
  \bibinfo{pages}{1009--1020}.
\bibitem[{Edie et~al.(1963)}]{edie1963discussion}
\bibinfo{author}{Edie, L.C.}, et~al., \bibinfo{year}{1963}.
\newblock \bibinfo{title}{Discussion of traffic stream measurements and
  definitions}.
\newblock \bibinfo{publisher}{Port of New York Authority New York}.
\bibitem[{Fuglede and Topsoe(2004)}]{fuglede2004jensen}
\bibinfo{author}{Fuglede, B.}, \bibinfo{author}{Topsoe, F.},
  \bibinfo{year}{2004}.
\newblock \bibinfo{title}{Jensen-shannon divergence and hilbert space
  embedding}, in: \bibinfo{booktitle}{International symposium onInformation
  theory, 2004. ISIT 2004. Proceedings.}, \bibinfo{organization}{IEEE}.
  p.~\bibinfo{pages}{31}.
\bibitem[{Gunter et~al.(2020)Gunter, Gloudemans, Stern, McQuade, Bhadani,
  Bunting, Delle~Monache, Lysecky, Seibold, Sprinkle
  et~al.}]{gunter2020commercially}
\bibinfo{author}{Gunter, G.}, \bibinfo{author}{Gloudemans, D.},
  \bibinfo{author}{Stern, R.E.}, \bibinfo{author}{McQuade, S.},
  \bibinfo{author}{Bhadani, R.}, \bibinfo{author}{Bunting, M.},
  \bibinfo{author}{Delle~Monache, M.L.}, \bibinfo{author}{Lysecky, R.},
  \bibinfo{author}{Seibold, B.}, \bibinfo{author}{Sprinkle, J.}, et~al.,
  \bibinfo{year}{2020}.
\newblock \bibinfo{title}{Are commercially implemented adaptive cruise control
  systems string stable?}
\newblock \bibinfo{journal}{IEEE Transactions on Intelligent Transportation
  Systems} \bibinfo{volume}{22}, \bibinfo{pages}{6992--7003}.
\bibitem[{Gunter et~al.(2019)Gunter, Janssen, Barbour, Stern and
  Work}]{gunter2019model}
\bibinfo{author}{Gunter, G.}, \bibinfo{author}{Janssen, C.},
  \bibinfo{author}{Barbour, W.}, \bibinfo{author}{Stern, R.E.},
  \bibinfo{author}{Work, D.B.}, \bibinfo{year}{2019}.
\newblock \bibinfo{title}{Model-based string stability of adaptive cruise
  control systems using field data}.
\newblock \bibinfo{journal}{IEEE Transactions on Intelligent Vehicles}
  \bibinfo{volume}{5}, \bibinfo{pages}{90--99}.
\bibitem[{He et~al.(2020)He, Makridis, Fontaras, Mattas, Xu and
  Ciuffo}]{he2020energy}
\bibinfo{author}{He, Y.}, \bibinfo{author}{Makridis, M.},
  \bibinfo{author}{Fontaras, G.}, \bibinfo{author}{Mattas, K.},
  \bibinfo{author}{Xu, H.}, \bibinfo{author}{Ciuffo, B.}, \bibinfo{year}{2020}.
\newblock \bibinfo{title}{The energy impact of adaptive cruise control in
  real-world highway multiple-car-following scenarios}.
\newblock \bibinfo{journal}{European Transport Research Review}
  \bibinfo{volume}{12}, \bibinfo{pages}{1--11}.
\bibitem[{Jiang et~al.(2022)Jiang, Xie, Evans, Wen, Li and
  Chen}]{jiang2022reinforcement}
\bibinfo{author}{Jiang, L.}, \bibinfo{author}{Xie, Y.}, \bibinfo{author}{Evans,
  N.G.}, \bibinfo{author}{Wen, X.}, \bibinfo{author}{Li, T.},
  \bibinfo{author}{Chen, D.}, \bibinfo{year}{2022}.
\newblock \bibinfo{title}{Reinforcement learning based cooperative longitudinal
  control for reducing traffic oscillations and improving platoon stability}.
\newblock \bibinfo{journal}{Transportation Research Part C: Emerging
  Technologies} \bibinfo{volume}{141}, \bibinfo{pages}{103744}.
\bibitem[{Kontar et~al.(2021)Kontar, Li, Srivastava, Zhou, Chen and
  Ahn}]{kontar2021multi}
\bibinfo{author}{Kontar, W.}, \bibinfo{author}{Li, T.},
  \bibinfo{author}{Srivastava, A.}, \bibinfo{author}{Zhou, Y.},
  \bibinfo{author}{Chen, D.}, \bibinfo{author}{Ahn, S.}, \bibinfo{year}{2021}.
\newblock \bibinfo{title}{On multi-class automated vehicles: Car-following
  behavior and its implications for traffic dynamics}.
\newblock \bibinfo{journal}{Transportation research part C: emerging
  technologies} \bibinfo{volume}{128}, \bibinfo{pages}{103166}.
\bibitem[{Krajewski et~al.(2018)Krajewski, Bock, Kloeker and
  Eckstein}]{krajewski2018highd}
\bibinfo{author}{Krajewski, R.}, \bibinfo{author}{Bock, J.},
  \bibinfo{author}{Kloeker, L.}, \bibinfo{author}{Eckstein, L.},
  \bibinfo{year}{2018}.
\newblock \bibinfo{title}{The highd dataset: A drone dataset of naturalistic
  vehicle trajectories on german highways for validation of highly automated
  driving systems}, in: \bibinfo{booktitle}{2018 21st International Conference
  on Intelligent Transportation Systems (ITSC)}, \bibinfo{organization}{IEEE}.
  pp. \bibinfo{pages}{2118--2125}.
\bibitem[{Laval(2011)}]{laval2011hysteresis}
\bibinfo{author}{Laval, J.A.}, \bibinfo{year}{2011}.
\newblock \bibinfo{title}{Hysteresis in traffic flow revisited: An improved
  measurement method}.
\newblock \bibinfo{journal}{Transportation Research Part B: Methodological}
  \bibinfo{volume}{45}, \bibinfo{pages}{385--391}.
\bibitem[{Laval and Leclercq(2010)}]{laval2010mechanism}
\bibinfo{author}{Laval, J.A.}, \bibinfo{author}{Leclercq, L.},
  \bibinfo{year}{2010}.
\newblock \bibinfo{title}{A mechanism to describe the formation and propagation
  of stop-and-go waves in congested freeway traffic}.
\newblock \bibinfo{journal}{Philosophical Transactions of the Royal Society A:
  Mathematical, Physical and Engineering Sciences} \bibinfo{volume}{368},
  \bibinfo{pages}{4519--4541}.
\bibitem[{Li et~al.(2021)Li, Chen, Zhou, Laval and Xie}]{li2021car}
\bibinfo{author}{Li, T.}, \bibinfo{author}{Chen, D.}, \bibinfo{author}{Zhou,
  H.}, \bibinfo{author}{Laval, J.}, \bibinfo{author}{Xie, Y.},
  \bibinfo{year}{2021}.
\newblock \bibinfo{title}{Car-following behavior characteristics of adaptive
  cruise control vehicles based on empirical experiments}.
\newblock \bibinfo{journal}{Transportation research part B: methodological}
  \bibinfo{volume}{147}, \bibinfo{pages}{67--91}.
\bibitem[{Makridis et~al.(2021)Makridis, Mattas, Anesiadou and
  Ciuffo}]{makridis2021openacc}
\bibinfo{author}{Makridis, M.}, \bibinfo{author}{Mattas, K.},
  \bibinfo{author}{Anesiadou, A.}, \bibinfo{author}{Ciuffo, B.},
  \bibinfo{year}{2021}.
\newblock \bibinfo{title}{Openacc. an open database of car-following
  experiments to study the properties of commercial acc systems}.
\newblock \bibinfo{journal}{Transportation research part C: emerging
  technologies} \bibinfo{volume}{125}, \bibinfo{pages}{103047}.
\bibitem[{Makridis et~al.(2020)Makridis, Mattas, Ciuffo, Re, Kriston, Minarini
  and Rognelund}]{makridis2020empirical}
\bibinfo{author}{Makridis, M.}, \bibinfo{author}{Mattas, K.},
  \bibinfo{author}{Ciuffo, B.}, \bibinfo{author}{Re, F.},
  \bibinfo{author}{Kriston, A.}, \bibinfo{author}{Minarini, F.},
  \bibinfo{author}{Rognelund, G.}, \bibinfo{year}{2020}.
\newblock \bibinfo{title}{Empirical study on the properties of adaptive cruise
  control systems and their impact on traffic flow and string stability}.
\newblock \bibinfo{journal}{Transportation research record}
  \bibinfo{volume}{2674}, \bibinfo{pages}{471--484}.
\bibitem[{Marin et~al.(2012)Marin, Pudlo, Robert and
  Ryder}]{marin2012approximate}
\bibinfo{author}{Marin, J.M.}, \bibinfo{author}{Pudlo, P.},
  \bibinfo{author}{Robert, C.P.}, \bibinfo{author}{Ryder, R.J.},
  \bibinfo{year}{2012}.
\newblock \bibinfo{title}{Approximate bayesian computational methods}.
\newblock \bibinfo{journal}{Statistics and Computing} \bibinfo{volume}{22},
  \bibinfo{pages}{1167--1180}.
\bibitem[{Milan{\'e}s et~al.(2013)Milan{\'e}s, Shladover, Spring, Nowakowski,
  Kawazoe and Nakamura}]{milanes2013cooperative}
\bibinfo{author}{Milan{\'e}s, V.}, \bibinfo{author}{Shladover, S.E.},
  \bibinfo{author}{Spring, J.}, \bibinfo{author}{Nowakowski, C.},
  \bibinfo{author}{Kawazoe, H.}, \bibinfo{author}{Nakamura, M.},
  \bibinfo{year}{2013}.
\newblock \bibinfo{title}{Cooperative adaptive cruise control in real traffic
  situations}.
\newblock \bibinfo{journal}{IEEE Transactions on intelligent transportation
  systems} \bibinfo{volume}{15}, \bibinfo{pages}{296--305}.
\bibitem[{Newell(2002)}]{newell2002simplified}
\bibinfo{author}{Newell, G.F.}, \bibinfo{year}{2002}.
\newblock \bibinfo{title}{A simplified car-following theory: a lower order
  model}.
\newblock \bibinfo{journal}{Transportation Research Part B: Methodological}
  \bibinfo{volume}{36}, \bibinfo{pages}{195--205}.
\bibitem[{Panaretos and Zemel(2019)}]{panaretos2019statistical}
\bibinfo{author}{Panaretos, V.M.}, \bibinfo{author}{Zemel, Y.},
  \bibinfo{year}{2019}.
\newblock \bibinfo{title}{Statistical aspects of wasserstein distances}.
\newblock \bibinfo{journal}{Annual review of statistics and its application}
  \bibinfo{volume}{6}, \bibinfo{pages}{405--431}.
\bibitem[{Ploeg et~al.(2011)Ploeg, Scheepers, Van~Nunen, Van~de Wouw and
  Nijmeijer}]{ploeg2011design}
\bibinfo{author}{Ploeg, J.}, \bibinfo{author}{Scheepers, B.T.},
  \bibinfo{author}{Van~Nunen, E.}, \bibinfo{author}{Van~de Wouw, N.},
  \bibinfo{author}{Nijmeijer, H.}, \bibinfo{year}{2011}.
\newblock \bibinfo{title}{Design and experimental evaluation of cooperative
  adaptive cruise control}, in: \bibinfo{booktitle}{2011 14th International
  IEEE Conference on Intelligent Transportation Systems (ITSC)},
  \bibinfo{organization}{IEEE}. pp. \bibinfo{pages}{260--265}.
\bibitem[{Ploeg et~al.(2013)Ploeg, Shukla, Van De~Wouw and
  Nijmeijer}]{ploeg2013controller}
\bibinfo{author}{Ploeg, J.}, \bibinfo{author}{Shukla, D.P.},
  \bibinfo{author}{Van De~Wouw, N.}, \bibinfo{author}{Nijmeijer, H.},
  \bibinfo{year}{2013}.
\newblock \bibinfo{title}{Controller synthesis for string stability of vehicle
  platoons}.
\newblock \bibinfo{journal}{IEEE Transactions on Intelligent Transportation
  Systems} \bibinfo{volume}{15}, \bibinfo{pages}{854--865}.
\bibitem[{Qu et~al.(2020)Qu, Yu, Zhou, Lin and Wang}]{qu2020jointly}
\bibinfo{author}{Qu, X.}, \bibinfo{author}{Yu, Y.}, \bibinfo{author}{Zhou, M.},
  \bibinfo{author}{Lin, C.T.}, \bibinfo{author}{Wang, X.},
  \bibinfo{year}{2020}.
\newblock \bibinfo{title}{Jointly dampening traffic oscillations and improving
  energy consumption with electric, connected and automated vehicles: a
  reinforcement learning based approach}.
\newblock \bibinfo{journal}{Applied Energy} \bibinfo{volume}{257},
  \bibinfo{pages}{114030}.
\bibitem[{Saifuzzaman et~al.(2017)Saifuzzaman, Zheng, Haque and
  Washington}]{saifuzzaman2017understanding}
\bibinfo{author}{Saifuzzaman, M.}, \bibinfo{author}{Zheng, Z.},
  \bibinfo{author}{Haque, M.M.}, \bibinfo{author}{Washington, S.},
  \bibinfo{year}{2017}.
\newblock \bibinfo{title}{Understanding the mechanism of traffic hysteresis and
  traffic oscillations through the change in task difficulty level}.
\newblock \bibinfo{journal}{Transportation Research Part B: Methodological}
  \bibinfo{volume}{105}, \bibinfo{pages}{523--538}.
\bibitem[{Shi et~al.(2021)Shi, Zhou, Wu, Wang, Lin and Ran}]{shi2021connected}
\bibinfo{author}{Shi, H.}, \bibinfo{author}{Zhou, Y.}, \bibinfo{author}{Wu,
  K.}, \bibinfo{author}{Wang, X.}, \bibinfo{author}{Lin, Y.},
  \bibinfo{author}{Ran, B.}, \bibinfo{year}{2021}.
\newblock \bibinfo{title}{Connected automated vehicle cooperative control with
  a deep reinforcement learning approach in a mixed traffic environment}.
\newblock \bibinfo{journal}{Transportation Research Part C: Emerging
  Technologies} \bibinfo{volume}{133}, \bibinfo{pages}{103421}.
\bibitem[{Shi and Li(2021)}]{shi2021constructing}
\bibinfo{author}{Shi, X.}, \bibinfo{author}{Li, X.}, \bibinfo{year}{2021}.
\newblock \bibinfo{title}{Constructing a fundamental diagram for traffic flow
  with automated vehicles: Methodology and demonstration}.
\newblock \bibinfo{journal}{Transportation Research Part B: Methodological}
  \bibinfo{volume}{150}, \bibinfo{pages}{279--292}.
\bibitem[{Shladover et~al.(2015)Shladover, Nowakowski, Lu and
  Ferlis}]{shladover2015cooperative}
\bibinfo{author}{Shladover, S.E.}, \bibinfo{author}{Nowakowski, C.},
  \bibinfo{author}{Lu, X.Y.}, \bibinfo{author}{Ferlis, R.},
  \bibinfo{year}{2015}.
\newblock \bibinfo{title}{Cooperative adaptive cruise control: Definitions and
  operating concepts}.
\newblock \bibinfo{journal}{Transportation Research Record}
  \bibinfo{volume}{2489}, \bibinfo{pages}{145--152}.
\bibitem[{Sisson et~al.(2018)Sisson, Fan and Beaumont}]{sisson2018handbook}
\bibinfo{author}{Sisson, S.A.}, \bibinfo{author}{Fan, Y.},
  \bibinfo{author}{Beaumont, M.}, \bibinfo{year}{2018}.
\newblock \bibinfo{title}{Handbook of approximate Bayesian computation}.
\newblock \bibinfo{publisher}{CRC Press}.
\bibitem[{Srivastava et~al.(2021)Srivastava, Chen and
  Ahn}]{srivastava2021modeling}
\bibinfo{author}{Srivastava, A.}, \bibinfo{author}{Chen, D.},
  \bibinfo{author}{Ahn, S.}, \bibinfo{year}{2021}.
\newblock \bibinfo{title}{Modeling and control using connected and automated
  vehicles with chained asymmetric driver behavior under stop-and-go
  oscillations}.
\newblock \bibinfo{journal}{Transportation research record}
  \bibinfo{volume}{2675}, \bibinfo{pages}{342--355}.
\bibitem[{Swaroop and Hedrick(1999)}]{swaroop1999constant}
\bibinfo{author}{Swaroop, D.}, \bibinfo{author}{Hedrick, J.K.},
  \bibinfo{year}{1999}.
\newblock \bibinfo{title}{Constant spacing strategies for platooning in
  automated highway systems} .
\bibitem[{USDOT(2007)}]{NGSIM}
\bibinfo{author}{USDOT, D.o.T.}, \bibinfo{year}{2007}.
\newblock \bibinfo{title}{Next generation simulation (ngsim)}.
\newblock \bibinfo{journal}{http://www.ngsim.fhwa.dot.gov} .
\bibitem[{Wang et~al.(2014)Wang, Daamen, Hoogendoorn and van
  Arem}]{wang2014rolling}
\bibinfo{author}{Wang, M.}, \bibinfo{author}{Daamen, W.},
  \bibinfo{author}{Hoogendoorn, S.P.}, \bibinfo{author}{van Arem, B.},
  \bibinfo{year}{2014}.
\newblock \bibinfo{title}{Rolling horizon control framework for driver
  assistance systems. part i: Mathematical formulation and non-cooperative
  systems}.
\newblock \bibinfo{journal}{Transportation research part C: emerging
  technologies} \bibinfo{volume}{40}, \bibinfo{pages}{271--289}.
\bibitem[{Wang et~al.(2018)Wang, Hoogendoorn, Daamen, van Arem, Shyrokau and
  Happee}]{wang2018delay}
\bibinfo{author}{Wang, M.}, \bibinfo{author}{Hoogendoorn, S.P.},
  \bibinfo{author}{Daamen, W.}, \bibinfo{author}{van Arem, B.},
  \bibinfo{author}{Shyrokau, B.}, \bibinfo{author}{Happee, R.},
  \bibinfo{year}{2018}.
\newblock \bibinfo{title}{Delay-compensating strategy to enhance string
  stability of adaptive cruise controlled vehicles}.
\newblock \bibinfo{journal}{Transportmetrica B: Transport Dynamics}
  \bibinfo{volume}{6}, \bibinfo{pages}{211--229}.
\bibitem[{Yu et~al.(2019)Yu, Sun, Liu and Yang}]{yu2019managing}
\bibinfo{author}{Yu, C.}, \bibinfo{author}{Sun, W.}, \bibinfo{author}{Liu,
  H.X.}, \bibinfo{author}{Yang, X.}, \bibinfo{year}{2019}.
\newblock \bibinfo{title}{Managing connected and automated vehicles at isolated
  intersections: From reservation-to optimization-based methods}.
\newblock \bibinfo{journal}{Transportation research part B: methodological}
  \bibinfo{volume}{122}, \bibinfo{pages}{416--435}.
\bibitem[{Zhong et~al.(2023)Zhong, Zhou and Ahn}]{xinzhi}
\bibinfo{author}{Zhong, X.}, \bibinfo{author}{Zhou, Y.}, \bibinfo{author}{Ahn,
  S.}, \bibinfo{year}{2023}.
\newblock \bibinfo{title}{Quantifying the relation between the car-following
  and traffic dynamics: A refined generalized definition approach} .
\bibitem[{Zhou et~al.(2017)Zhou, Ahn, Chitturi and Noyce}]{zhou2017rolling}
\bibinfo{author}{Zhou, Y.}, \bibinfo{author}{Ahn, S.},
  \bibinfo{author}{Chitturi, M.}, \bibinfo{author}{Noyce, D.A.},
  \bibinfo{year}{2017}.
\newblock \bibinfo{title}{Rolling horizon stochastic optimal control strategy
  for acc and cacc under uncertainty}.
\newblock \bibinfo{journal}{Transportation Research Part C: Emerging
  Technologies} \bibinfo{volume}{83}, \bibinfo{pages}{61--76}.
\bibitem[{Zhou et~al.(2020)Zhou, Ahn, Wang and
  Hoogendoorn}]{zhou2020stabilizing}
\bibinfo{author}{Zhou, Y.}, \bibinfo{author}{Ahn, S.}, \bibinfo{author}{Wang,
  M.}, \bibinfo{author}{Hoogendoorn, S.}, \bibinfo{year}{2020}.
\newblock \bibinfo{title}{Stabilizing mixed vehicular platoons with connected
  automated vehicles: An h-infinity approach}.
\newblock \bibinfo{journal}{Transportation Research Part B: Methodological}
  \bibinfo{volume}{132}, \bibinfo{pages}{152--170}.
\bibitem[{Zhou et~al.(2022)Zhou, Jafarsalehi, Jiang, Wang, Ahn and Lee}]{Zhou}
\bibinfo{author}{Zhou, Y.}, \bibinfo{author}{Jafarsalehi, G.},
  \bibinfo{author}{Jiang, J.}, \bibinfo{author}{Wang, X.},
  \bibinfo{author}{Ahn, S.}, \bibinfo{author}{Lee, J.D.}, \bibinfo{year}{2022}.
\newblock \bibinfo{title}{Stochastic calibration of automated vehicle
  car-following control: An approximate bayesian computation approach}.
\newblock \bibinfo{journal}{Available at SSRN:
  https://ssrn.com/abstract=4084970}
  \DOIprefix\doi{http://dx.doi.org/10.2139/ssrn.4084970}.

\end{thebibliography}

\section*{Appendix}
\subsection*{A: EAB Calibration with ABC-ASMC}
\begin{figure*}[!h]
 \captionsetup{justification=centering} 
        \centering
        \begin{subfigure}[b]{0.48\textwidth}
            \centering
\includegraphics[width=\textwidth]{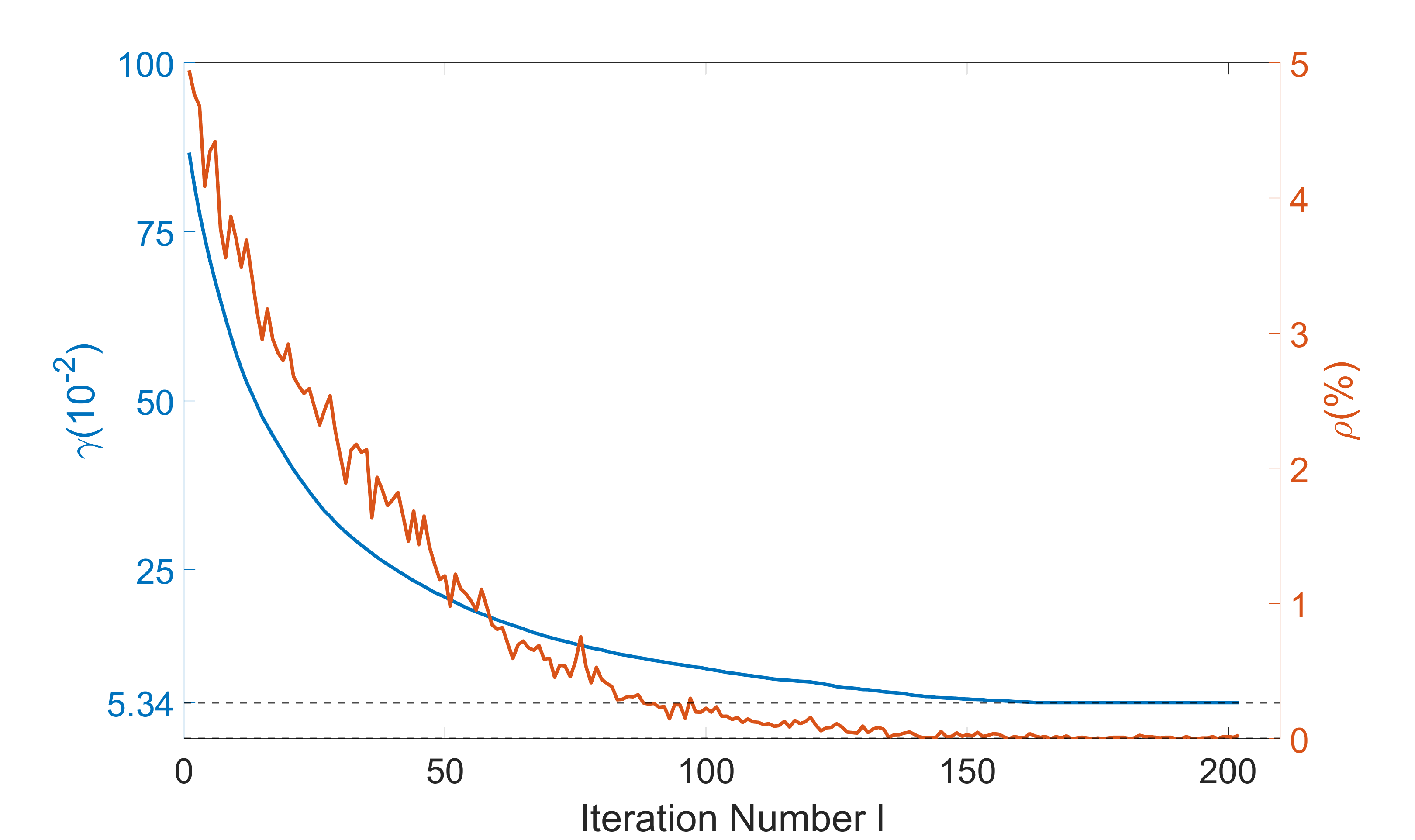}
              \caption[]%
            {{\small }}   
        \end{subfigure}
         \hfill
        \begin{subfigure}[b]{0.48\textwidth} 
            \centering 
            \includegraphics[width=\textwidth]{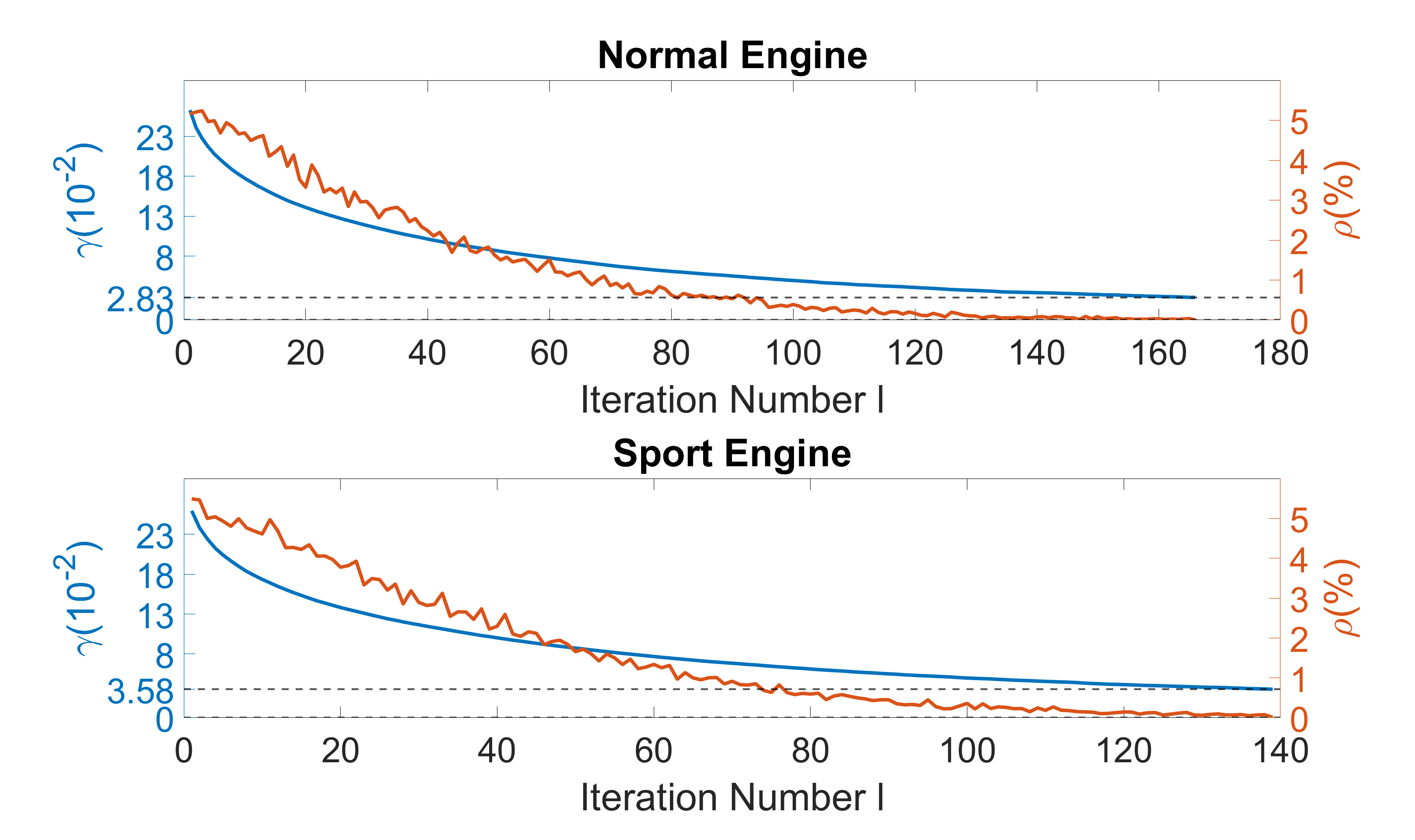}
                 \caption[]%
            {{\small }}   
        \end{subfigure}
         \begin{subfigure}[b]{0.48\textwidth}
            \centering
\includegraphics[width=\textwidth]{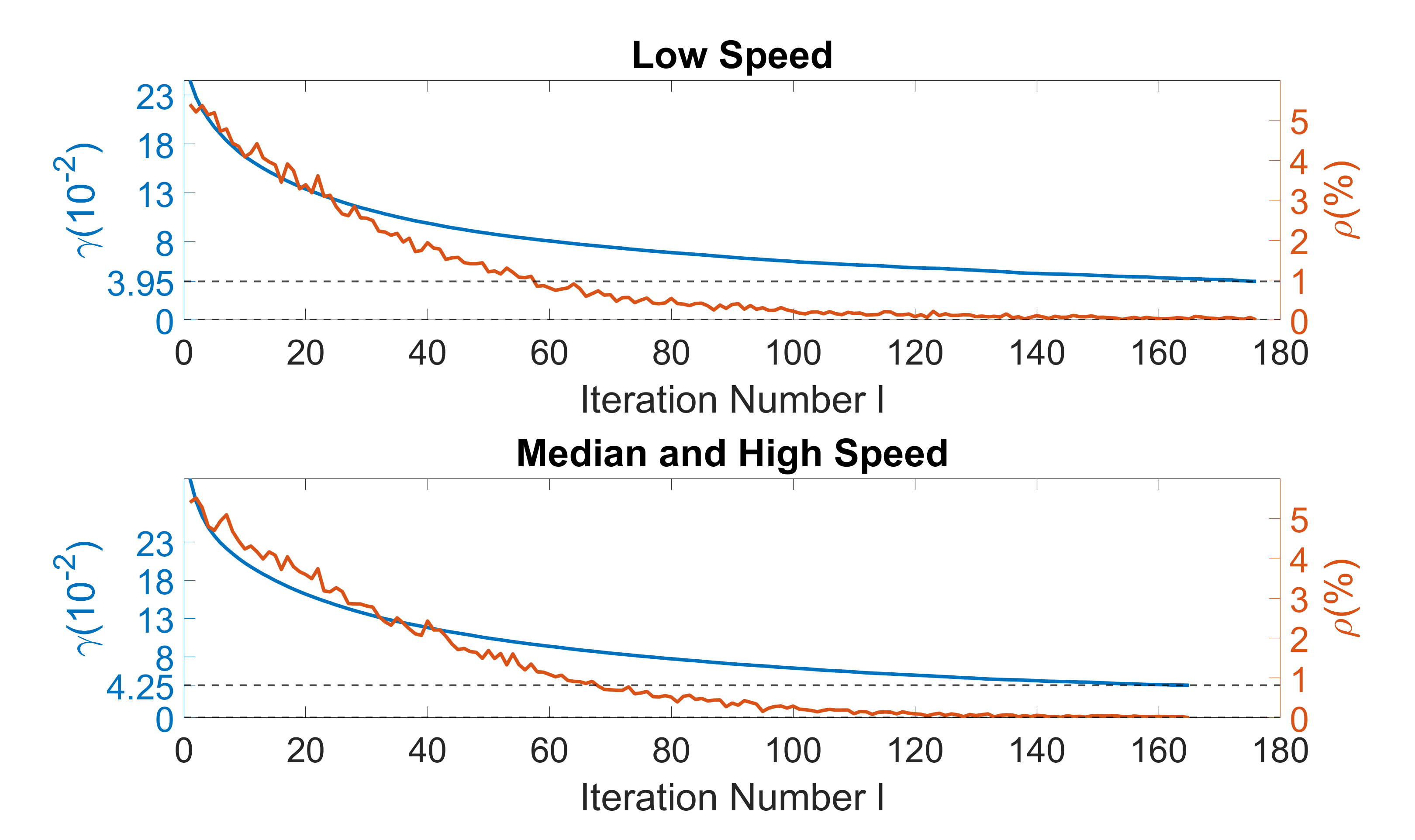}
         \caption[]%
            {{\small }}   
        \end{subfigure}
        \hfill
        \begin{subfigure}[b]{0.48\textwidth} 
            \centering 
\includegraphics[width=\textwidth]{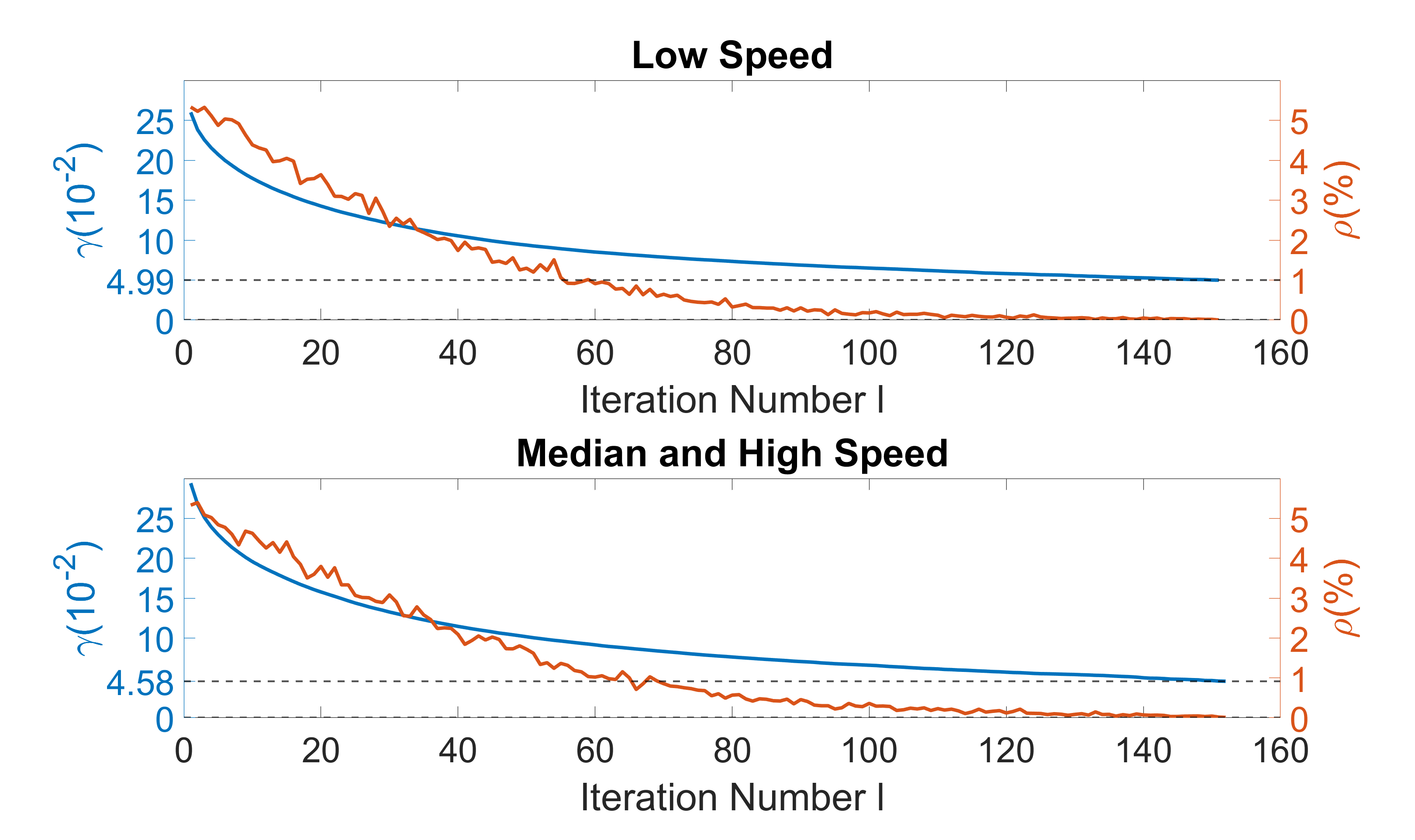}
            \caption[]%
            {{\small }}    
            \label{}
        \end{subfigure}
         \caption[Convergence of $\gamma$ and $\rho$ ]%
        {\small Convergence of $\gamma$ and $\rho$  \\
        HDV: (a) HDV-1
        ACC: (b) Car Model-Y (c) Car Model-Z-Normal Engine (d) Car Model-Z-Power Engine 
        } 
        \label{convergence_others}
    \end{figure*}

Fig. \ref{convergence_others} depicts the convergence behavior of the ABC-ASMC algorithm as it calibrates EAB models for HDVs and ACCs. The quick convergence indicates that the algorithm effectively identifies the optimal posterior joint distributions for the EAB models, enabling accurate characterization of CF behaviors of HDVs and ACCs.

\subsection*{B: Reaction Pattern Analysis based on Jensen-Shannon Distance}
\begin{figure*}[!ht]
 \captionsetup{justification=centering} 
        \centering
        \begin{subfigure}[b]{\textwidth}
           \centering
       \includegraphics[width=0.8\textwidth]{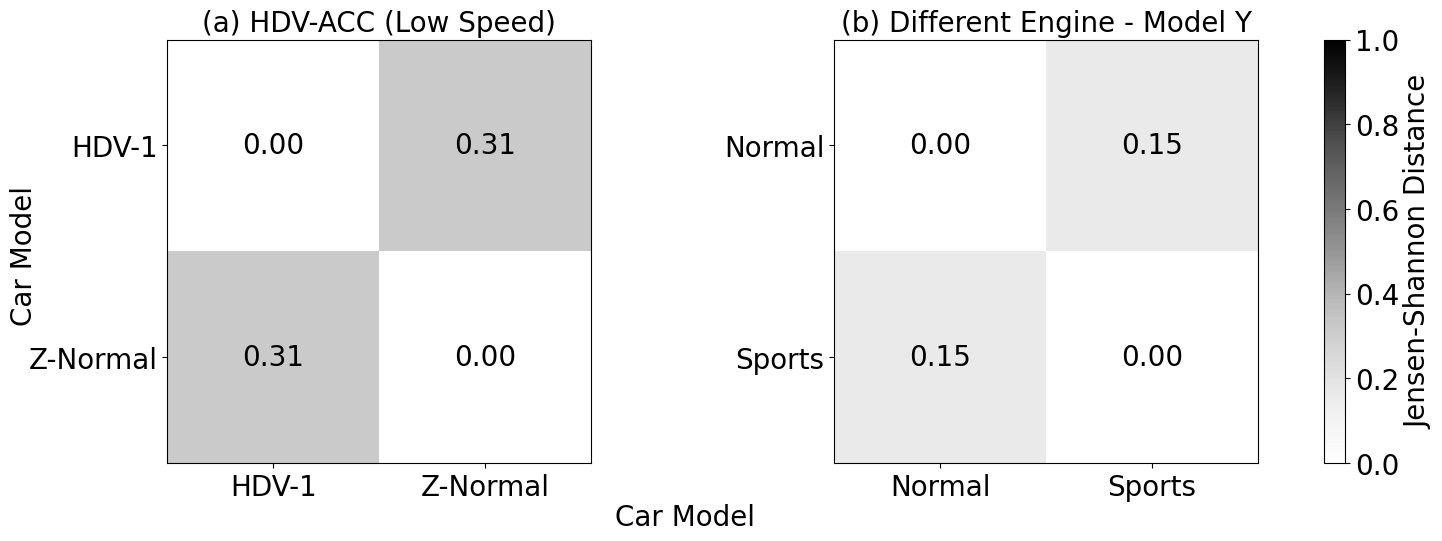}
            {{\small }}   
        \end{subfigure}
        \hfill
        \begin{subfigure}[b]{\textwidth} 
            \centering             \includegraphics[width=\textwidth]{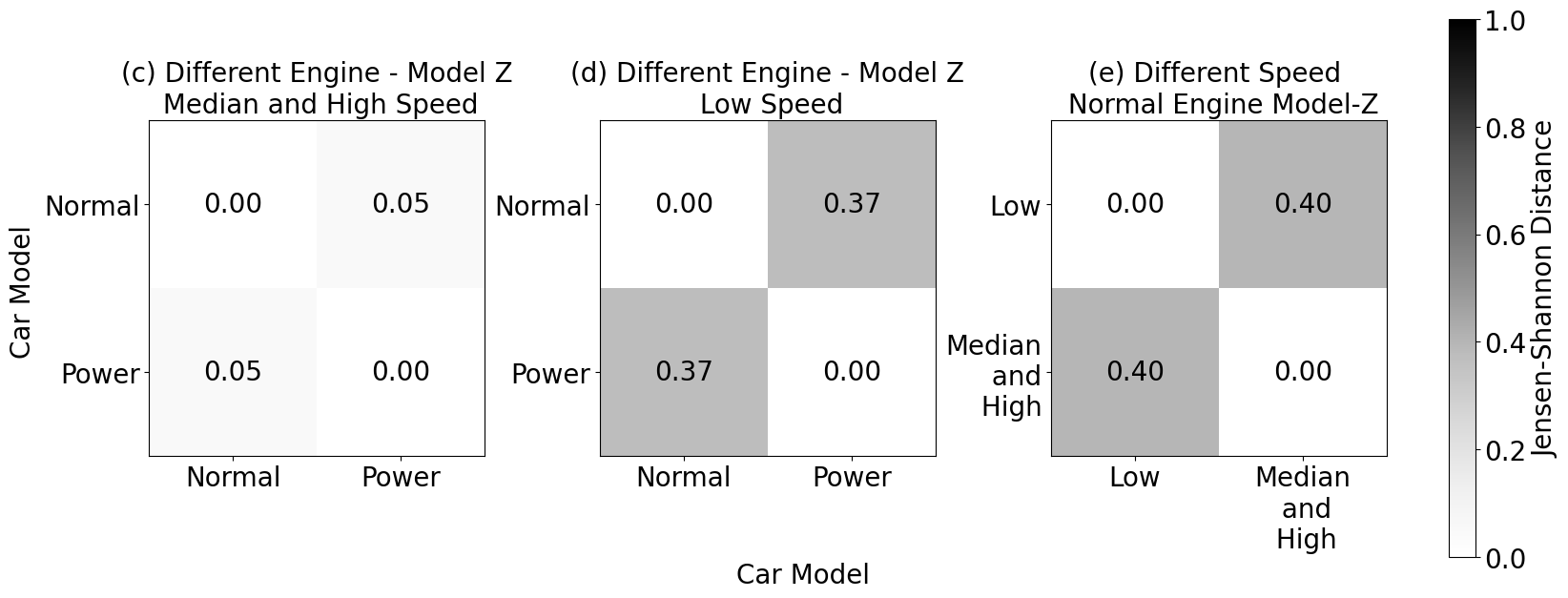}
            {{\small }}    
            \label{}
        \end{subfigure}
        \hfill
        \begin{subfigure}[b]{0.6\textwidth} 
            \centering             \includegraphics[width=\textwidth]{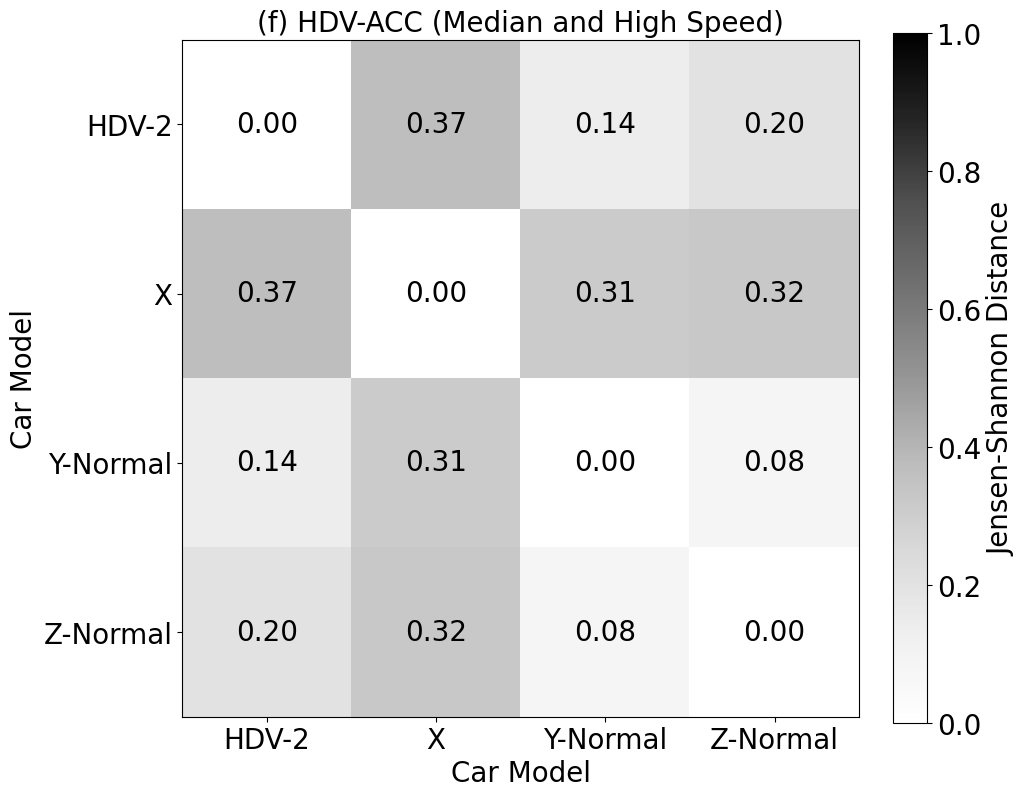}
            {{\small }}    
            \label{}
        \end{subfigure}
         \caption[JSD Metrics of Joint Distributions]%
        {\small JSD Metrics of Joint Distributions  \\
        } 
        \label{JSD}
    \end{figure*}

To evaluate the variability in the reaction pattern across HDVs and ACC vehicles, we adopt the Jensen–Shannon distance (JSD) to compare their calibrated posterior joint distributions (see Fig. \ref{JSD})\citep{fuglede2004jensen}. The JSD serves as a symmetric metric to quantify the difference between two joint distributions. Its value increases from 0 to 1 as the dissimilarity between the distributions grows. In Fig. \ref{JSD}, the pair-wise JSD ranges from 0.05 to 0.40, indicating some variation in the level of dissimilarity among different HDVs and ACCs. Fig. \ref{JSD}(a) and Fig. \ref{JSD}(f) show that JSD values exceed 0.14 between HDV-ACC pairs, indicating significant differences, at low speed and median and high speeds. Notably high levels of dissimilarity are observed for the HDV-Z pair at low speed (Fig. \ref{JSD}(a)) and HDV-X at median and high speeds (Fig. \ref{JSD}(f)). This finding suggests marked differences in $\eta$ evolution between HDVs and ACCs in general. Further, differences are also notable among different ACC developers (Fig. \ref{JSD}(f)), particularly between Car Model X and the other two (Y and Z). Car Model Y and Z appear to share some similar CF characteristics.

For the same ACC developers, JSD values are observed in Fig. \ref{JSD}(b)-(c) for different engines is much lower than in Fig. \ref{JSD}(d), suggesting the difference between different engines is more distinct when at low speeds than at median and high speed. However, when examining Car Model Z at low speed (Fig. \ref{JSD}(d)),  When controlled for the engine and the model, a remarkable difference in the $\eta$ evolution is observed between the two speed categories (Fig. \ref{JSD}(e)). 


\end{document}